\documentclass[12pt]{article}
\usepackage{epsfig}
\usepackage{amssymb}
\usepackage{cite}
\usepackage{axodraw}
\usepackage{color}
\definecolor{GreenYellow}   {cmyk}{0.15,0,0.69,0}
\definecolor{Yellow}        {cmyk}{0,0,1,0}
\definecolor{Goldenrod}     {cmyk}{0,0.10,0.84,0}
\definecolor{Dandelion}     {cmyk}{0,0.29,0.84,0}
\definecolor{Apricot}       {cmyk}{0,0.32,0.52,0}
\definecolor{Peach}         {cmyk}{0,0.50,0.70,0}
\definecolor{Melon}         {cmyk}{0,0.46,0.50,0}
\definecolor{YellowOrange}  {cmyk}{0,0.42,1,0}
\definecolor{Orange}        {cmyk}{0,0.61,0.87,0}
\definecolor{BurntOrange}   {cmyk}{0,0.51,1,0}
\definecolor{Bittersweet}   {cmyk}{0,0.75,1,0.24}
\definecolor{RedOrange}     {cmyk}{0,0.77,0.87,0}
\definecolor{Mahogany}      {cmyk}{0,0.85,0.87,0.35}
\definecolor{Maroon}        {cmyk}{0,0.87,0.68,0.32}
\definecolor{BrickRed}      {cmyk}{0,0.89,0.94,0.28}
\definecolor{Red}           {cmyk}{0,1,1,0}
\definecolor{OrangeRed}     {cmyk}{0,1,0.50,0}
\definecolor{RubineRed}     {cmyk}{0,1,0.13,0}
\definecolor{WildStrawberry}{cmyk}{0,0.96,0.39,0}
\definecolor{Salmon}        {cmyk}{0,0.53,0.38,0}
\definecolor{CarnationPink} {cmyk}{0,0.63,0,0}
\definecolor{Magenta}       {cmyk}{0,1,0,0}
\definecolor{VioletRed}     {cmyk}{0,0.81,0,0}
\definecolor{Rhodamine}     {cmyk}{0,0.82,0,0}
\definecolor{Mulberry}      {cmyk}{0.34,0.90,0,0.02}
\definecolor{RedViolet}     {cmyk}{0.07,0.90,0,0.34}
\definecolor{Fuchsia}       {cmyk}{0.47,0.91,0,0.08}
\definecolor{Lavender}      {cmyk}{0,0.48,0,0}
\definecolor{Thistle}       {cmyk}{0.12,0.59,0,0}
\definecolor{Orchid}        {cmyk}{0.32,0.64,0,0}
\definecolor{DarkOrchid}    {cmyk}{0.40,0.80,0.20,0}
\definecolor{Purple}        {cmyk}{0.45,0.86,0,0}
\definecolor{Plum}          {cmyk}{0.50,1,0,0}
\definecolor{Violet}        {cmyk}{0.79,0.88,0,0}
\definecolor{RoyalPurple}   {cmyk}{0.75,0.90,0,0}
\definecolor{BlueViolet}    {cmyk}{0.86,0.91,0,0.04}
\definecolor{Periwinkle}    {cmyk}{0.57,0.55,0,0}
\definecolor{CadetBlue}     {cmyk}{0.62,0.57,0.23,0}
\definecolor{CornflowerBlue}{cmyk}{0.65,0.13,0,0}
\definecolor{MidnightBlue}  {cmyk}{0.98,0.13,0,0.43}
\definecolor{NavyBlue}      {cmyk}{0.94,0.54,0,0}
\definecolor{RoyalBlue}     {cmyk}{1,0.50,0,0}
\definecolor{Blue}          {cmyk}{1,1,0,0}
\definecolor{Cerulean}      {cmyk}{0.94,0.11,0,0}
\definecolor{Cyan}          {cmyk}{1,0,0,0}
\definecolor{ProcessBlue}   {cmyk}{0.96,0,0,0}
\definecolor{SkyBlue}       {cmyk}{0.62,0,0.12,0}
\definecolor{Turquoise}     {cmyk}{0.85,0,0.20,0}
\definecolor{TealBlue}      {cmyk}{0.86,0,0.34,0.02}
\definecolor{Aquamarine}    {cmyk}{0.82,0,0.30,0}
\definecolor{BlueGreen}     {cmyk}{0.85,0,0.33,0}
\definecolor{Emerald}       {cmyk}{1,0,0.50,0}
\definecolor{JungleGreen}   {cmyk}{0.99,0,0.52,0}
\definecolor{SeaGreen}      {cmyk}{0.69,0,0.50,0}
\definecolor{Green}         {cmyk}{1,0,1,0}
\definecolor{ForestGreen}   {cmyk}{0.91,0,0.88,0.12}
\definecolor{PineGreen}     {cmyk}{0.92,0,0.59,0.25}
\definecolor{LimeGreen}     {cmyk}{0.50,0,1,0}
\definecolor{YellowGreen}   {cmyk}{0.44,0,0.74,0}
\definecolor{SpringGreen}   {cmyk}{0.26,0,0.76,0}
\definecolor{OliveGreen}    {cmyk}{0.64,0,0.95,0.40}
\definecolor{RawSienna}     {cmyk}{0,0.72,1,0.45}
\definecolor{Sepia}         {cmyk}{0,0.83,1,0.70}
\definecolor{Brown}         {cmyk}{0,0.81,1,0.60}
\definecolor{Tan}           {cmyk}{0.14,0.42,0.56,0}
\definecolor{Gray}          {cmyk}{0,0,0,0.50}
\definecolor{Black}         {cmyk}{0,0,0,1}
\definecolor{White}         {cmyk}{0,0,0,0}

\newcommand{\bra}[1]{\langle #1|}
\newcommand{\ket}[1]{|#1\rangle}

\newcommand{\lsim}{\raisebox{-0.13cm}{~\shortstack{$<$ \\[-0.07cm] $\sim$}}~}

\def\theequation{\arabic{section}.\arabic{equation}}

\begin{document}
\tolerance=100000

\begin{flushright}
CERN-PH-TH/2007-136\\[-1mm]
MAN/HEP/2007/13\\[-1mm]
{\tt arXiv:0708.2079}\\[-1mm]
August 2007 
\end{flushright}

\bigskip

\begin{center}
{\Large {\bf {\boldmath $B$}-Meson Observables in the Maximally
CP-Violating}}\\[0.3cm]  
{\Large \bf MSSM  with Minimal Flavour Violation}\\[1.7cm]
{\large John Ellis}$^{\, a}$, 
{\large Jae Sik Lee}$^{\, b}$ and 
{\large Apostolos Pilaftsis}$^{\, a,c}$\\[0.5cm]
{\it $^a$Theory Division, Physics Department, CERN, CH-1211 Geneva 23,
Switzerland}\\[2mm]
{\it $^b$Center for Theoretical Physics, School of Physics, 
Seoul National University,}\\
{\it Seoul 151-747, Korea}\\[2mm]
{\it $^c$School of Physics and Astronomy, University of Manchester}\\
{\it Manchester M13 9PL, United Kingdom}
\end{center}

\vspace*{0.8cm}\centerline{\bf  ABSTRACT}
  \noindent
Additional sources  of CP violation  in the MSSM may  affect $B$-meson
mixings and  decays, even in scenarios with  minimal flavour violation
(MFV).   We   formulate  the  maximally   CP-violating  and  minimally
flavour-violating  (MCPMFV)   variant  of  the  MSSM,   which  has  19
parameters, including  6 phases  that violate CP.   We then  develop a
manifestly   flavour-covariant  effective  Lagrangian   formalism  for
calculating  Higgs-mediated  FCNC observables  in  the  MSSM at  large
$\tan\beta$, and  analyze within the  MCPMFV framework FCNC  and other
processes involving  $B$ mesons.  We  include a new class  of dominant
subleading  contributions   due  to  non-decoupling   effects  of  the
third-generation  quarks.  We  present illustrative  numerical results
that  include effects  of the  CP-odd MCPMFV  parameters on  Higgs and
sparticle masses,  the $B_s$  and $B_d$ mass  differences, and  on the
decays  $B_s  \to \mu^+  \mu^-$,  $B_u  \to \tau  \nu$  and  $b \to  s
\gamma$. We  use these results  to derive illustrative  constraints on
the MCPMFV parameters imposed by D0, CDF, BELLE and BABAR measurements
of $B$ mesons, demonstrating how a potentially observable contribution
to the  CP asymmetry in the  $b \to s  \gamma$ decay may arise  in the
MSSM with MCPMFV.

\medskip


\vspace*{\fill}
\newpage

\setcounter{equation}{0}
\section{Introduction}
  \label{sec:intro}

Models  incorporating  supersymmetry   (SUSY),  such  as  the  Minimal
Supersymmetric Standard Model (MSSM), contain many possible sources of
flavour and CP violation. In particular, the soft SUSY-breaking sector
in general  introduces many new  sources of flavour and  CP violation,
giving  rise to  effects that  may exceed  the experimental  limits by
several    orders    of    magnitude.     The   unitarity    of    the
Cabibbo--Kobayashi--Maskawa  (CKM)   quark  mixing  matrix  suppresses
flavour-changing-neutral  currents (FCNC)  and CP  violation somewhat,
thanks to  the Glashow--Iliopoulos--Maiani (GIM) mechanism~\cite{GIM},
to  the   extent  that  the  soft  SUSY-breaking   scalar  masses  are
universal. One possible solution to  the flavour and CP problems is to
ensure that  the soft SUSY-breaking  sector is fully protected  by the
GIM mechanism.  This can be achieved within the so-called framework of
minimal flavour violation (MFV), where  all flavour and CP effects are
mediated  by  the  superpotential  interactions corresponding  to  the
ordinary Yukawa couplings  of the Higgs bosons to  quarks and leptons.
In this  framework, FCNC and  CP-violating observables depend  only on
the fermion masses and their  mixings, and hence the CKM mixing matrix
${\bf V}$~\cite{CKM}.  In  such a scenario, all FCNC  and CP violation
observables would  vanish in the MSSM  if ${\bf V}$ were  equal to the
unit matrix {\bf 1}.

A minimal realization of MFV in  the MSSM is obtained by assuming that
all soft SUSY-breaking bilinear  masses for the scalar particles, such
as squarks,  sleptons and  Higgs bosons, are  equal to a  common value
$m_0$  at the gauge  coupling unification  point $M_{\rm  GUT}$, where
$M_{\rm GUT}$ might be the threshold for some underlying grand unified
theory  (GUT) based,  e.g., on  SU(5) or  SO(10).  Likewise,  the soft
masses  of  the fermionic  SUSY  partners  of  the gauge  fields,  the
gauginos, might also  be equal to a common  value $m_{1/2}$ at $M_{\rm
GUT}$ and, in the same  spirit, all soft trilinear Yukawa couplings of
the Higgs bosons to squarks and  sleptons could be real and equal to a
universal      parameter     $A$      times      the     corresponding
Higgs-fermion-antifermion couplings.   The Higgs supermultiplet mixing
parameter $\mu$  and the corresponding soft  SUSY-breaking term $B\mu$
introduce  two  additional  mass   scales  in  the  theory.   However,
minimization  conditions  on  the  Higgs  potential  can  be  used  to
eliminate  these two  last mass  scales in  favour of  the electroweak
scale $M_Z$  and $\tan\beta \equiv  v_u/v_d$, where $v_{u,d}$  are the
vacuum expectation  values (VEVs) of the two  Higgs doublets $H_{u,d}$
in the MSSM.

It is well  known that a minimal expansion of  the above MFV framework
is to allow  the soft SUSY-breaking mass parameters  $m_{1/2}$ and $A$
to be  complex with CP-odd phases, thereby  introducing two additional
sources  of  CP  violation in  the  theory.  In  this case,  all  FCNC
observables,  whether CP-conserving or  not, still  depend on  the CKM
mixing matrix~${\bf V}$ in such a way that they vanish if ${\bf V}$ is
assumed to be  diagonal, i.e., equal to the  unit matrix. However, the
two  new   phases  introduce  the  possibility  of   CP  violation  in
flavour-conserving processes even if ${\bf V}$ is real, and in general
CP violation in FCNC processes may differ from CKM predictions.

Here we go one step further,  and ask the following question.  What is
the  maximal number  of additional  CP-violating parameters  and extra
flavour-singlet mass  scales that  could be present  in the  MSSM, for
which  the above  notion of  MFV remains  still valid,  i.e.,~all FCNC
effects vanish  in the limit  of a diagonal  ${\bf V}$?  We  call this
scenario  the   maximally  CP-violating  MSSM   with  minimal  flavour
violation,  or in  short, the  MSSM with  MCPMFV.  As  we will  see in
Section~\ref{sec:MMFV}, there are a total of 19 parameters in the MSSM
with  MCPMFV,  including  6  CP-violating  phases  and  13  real  mass
parameters.  The purposes of this paper are to formulate the MSSM with
MCPMFV, calculate the most relevant $B$-meson observables, and explore
the  experimental constraints  on the  MCPMFV  theoretical parameters,
exploiting   a  manifestly   flavour-covariant   effective  Lagrangian
formalism  for calculating  Higgs-mediated FCNC  observables  at large
$\tan\beta$ that we develop here.

At large values  of $\tan\beta$, e.g.~$\tan\beta \stackrel{>}{{}_\sim}
40$,  one-loop  threshold   effects  on  Higgs-boson  interactions  to
down-type  quarks get enhanced~\cite{TBanks,RH,CGNW},  and so  play an
important role  in FCNC processes, such as  the $K^0$-$\bar{K}^0$ mass
difference, $B_s$-$\bar{B_s}$  and $B_d$-$\bar{B_d}$ mixings,  and the
decays  $B \to  X_s  \gamma$, $B  \to  K l^+l^-$,  $B_{s,d} \to  \mu^+
\mu^-$~\cite{FCNC,FCNCCP,Frank,ADKT,Ambrosio,Buras,DP,Bphases,LR,Carena},
and $B \to  \tau \nu$~\cite{Akeroyd:2003zr,Itoh:2004ye}.  We present in  this paper a
manifestly   flavour-covariant  effective  Lagrangian   formalism  for
calculating  FCNC processes that  follows the  lines of  the effective
Lagrangian approach given in~\cite{DP}.   In addition, we include here
the dominant  subleading contributions to  the one-loop Higgs-mediated
FCNC interactions due  to non-decoupling large Yukawa-coupling effects
of the third-generation quarks.   Based on this improved formalism, we
compute FCNC  observables in constrained  versions of the  MSSM, where
MFV has  been imposed on the  soft SUSY-breaking mass  parameters as a
boundary condition  at the scale $M_{\rm GUT}$.   We present numerical
results  for  $B$-meson  observables  in  one example  of  the  MCPMFV
framework,   from  which   illustrative  constraints   on   the  basic
theoretical  parameters are  derived, after  incorporating  the recent
experimental results from D0 and CDF~\cite{CDFD0}.

The paper  is organized  as follows: in  Section~\ref{sec:MMFV}, after
briefly reviewing the  MFV framework, we derive the  maximal number of
flavour-singlet mass parameters  that can be present in  the MSSM with
MCPMFV  at the GUT  scale.  All  relevant one-loop  RGEs are  given in
Appendix~\ref{sec:RGE}.   In   Section~\ref{sec:EPF},  we  present  an
effective  Lagrangian formalism  for Higgs-mediated  FCNC interactions
that  respects  flavour  covariance.   We also  discuss  the  dominant
subleading  effects at  large  $\tan\beta$, due  to  the large  Yukawa
couplings of the third generation.  Useful relations which result from
Ward identities (WIs) that  involve the $Z$ and $W$-boson interactions
to             quarks            are             derived            in
Appendix~\ref{sec:WIs}. Section~\ref{sec:FCNC} summarizes all relevant
analytic  results   pertinent  to  FCNC   $B$-meson  observables.   In
Section~\ref{sec:num} we  exhibit numerical estimates  and predictions
for  various  FCNC  processes,  including  the  $B_s$-$\bar{B_s}$  and
$B_d$-$\bar{B_d}$ mixings,  and the decays $B_{s,d}  \to \mu^+ \mu^-$,
$B \to  X_s \gamma$,  and $B  \to \tau \nu$.   We also  illustrate the
combined  constraints on  the theoretical  parameters imposed  by data
from  D0,  CDF,  BELLE and  BABAR  in  one  sample MCPMFV  model.   We
summarize our conclusions in Section~\ref{sec:concl}.

\setcounter{equation}{0} 
\section{Maximal CP and Minimal Flavour Violation}
  \label{sec:MMFV}

In this section we derive  the maximal number of CP-violating and real
flavour-singlet   mass  parameters   that  can   be  present   in  the
CP-violating MSSM and satisfy the  property of MFV as described in the
Introduction.

The superpotential defining  the flavour structure of the  MSSM may be
written as
\begin{equation}
  \label{Wpot}
W_{\rm MSSM}\ =\ \widehat{U}^C {\bf h}_u \widehat{Q} \widehat{H}_u\:
+\:   \widehat{D}^C {\bf h}_d \widehat{H}_d \widehat{Q}  \: +\: 
\widehat{E}^C {\bf h}_e \widehat{H}_d \widehat{L} \: +\: \mu
\widehat{H}_u \widehat{H}_d\ ,
\end{equation}
where $\widehat{H}_{u,d}$  are the  two Higgs chiral  superfields, and
$\widehat{Q}$,  $\widehat{L}$,  $\widehat{U}^C$,  $\widehat{D}^C$  and
$\widehat{E}^C$ are the left-  and right-handed superfields related to
up- and  down-type quarks and  charged leptons.  The  Yukawa couplings
${\bf  h}_{u,d,e}$ are  $3\times  3$ complex  matrices describing  the
charged-lepton   and   quark    masses   and   their   mixings.    The
superpotential~(\ref{Wpot})  contains one  mass  parameter, the  $\mu$
parameter that mixes the Higgs supermultiplets, which has to be of the
electroweak order for a natural realization of the Higgs mechanism.

In an  unconstrained version of the  MSSM, there is a  large number of
different  mass  parameters   present  in  the  soft  SUSY-breaking
Lagrangian
\begin{eqnarray} 
  \label{Lsoft}
-{\cal L}_{\rm soft} &=& \frac{1}{2}\, \Big(\, M_1\,
 \widetilde{B}\widetilde{B}\: +\: M_2\, \widetilde{W}^i\widetilde{W}^i\: 
+\: M_3\, \tilde{g}^a\tilde{g}^a\,  \ +\ {\rm h.c.}\Big)\: +\:
 \widetilde{Q}^\dagger 
\widetilde{\bf M}^2_Q \widetilde{Q}\: +\:  \widetilde{L}^\dagger
\widetilde{\bf M}^2_L \widetilde{L}\: +\: \widetilde{U}^\dagger
\widetilde{\bf M}^2_U \widetilde{U}\nonumber\\
&&+\: \widetilde{D}^\dagger \widetilde{\bf M}^2_D \widetilde{D}\: 
+\: \widetilde{E}^\dagger \widetilde{\bf M}^2_E \widetilde{E}\:
+\: M^2_{H_u} H^\dagger_u H_u\: +\: M^2_{H_d} H^\dagger_d H_d\:
+\: \Big(B\mu\, H_u H_d\ +\ {\rm h.c.}\Big)\nonumber\\
&&+\: \Big(  \widetilde{U}^\dagger {\bf a}_u \widetilde{Q} H_u  \:
+\: \widetilde{D}^\dagger {\bf a}_d H_d \widetilde{Q}  \: +\: 
\widetilde{E}^\dagger {\bf a}_e H_d \widetilde{L} \ +\ {\rm h.c.}\Big)\;. 
\end{eqnarray}
Here $M_{1,2,3}$ are the soft SUSY-breaking masses associated with the
U(1)$_Y$, SU(2)$_L$ and SU(3)$_c$ gauginos, respectively. In addition,
$M^2_{H_{u,d}}$ and  $B\mu$ are the  soft masses related to  the Higgs
doublets $H_{u,d}$ and their bilinear mixing. Finally, $\widetilde{\bf
M}^2_{Q,L,D,U,E}$ are  the $3\times  3$ soft mass-squared  matrices of
squarks  and sleptons,  and  ${\bf a}_{u,d,e}$  are the  corresponding
$3\times  3$ soft  Yukawa  mass matrices~\footnote{Alternatively,  the
soft  Yukawa mass  matrices ${\bf  a}_{u,d,e}$ may  be defined  by the
relation:  $({\bf a}_{u,d,e})_{ij}  =  ({\bf h}_{u,d,e})_{ij}\,  ({\bf
A}_{u,d,e})_{ij}$, where  the parameters $({\bf  A}_{u,d,e})_{ij}$ are
generically of order $M_{\rm  SUSY}$ in gravity-mediated SUSY breaking
models.   In our paper,  both definitions  for the  soft SUSY-breaking
Yukawa couplings will be used, where convenient.}.  Hence, in addition
to the  $\mu$ term, the  unconstrained CP-violating MSSM  contains 109
real mass parameters.

One  frequently considers the  constrained MSSM  (CMSSM), which  has a
common gaugino mass $m_{1/2}$, a common soft SUSY-breaking scalar mass
$m_0$ and a common soft  trilinear Yukawa coupling $A$ for all squarks
and sleptons at  the GUT scale. The number  of independent mass scales
is greatly reduced since, even  allowing for maximal CP violation, the
free  parameters are  just $m_{1/2}$,  $\mu$, $m_0$,  $A$  and $B\mu$,
where all but $m_0$ are complex variables. The phase ${\rm arg}\, \mu$
may be removed by means  of a global Peccei--Quinn (PQ) symmetry under
which $H_u$ and $H_d$ have the same charges.  Imposing the two CP-even
tadpole  conditions on  the Higgs  potential, one  may replace  $\mu =
|\mu|$  and ${\rm  Re}\,(B\mu)$ by  the $Z$-boson  mass $M_Z$  and the
ratio  $\tan\beta  =  v_u/v_d$  of  the VEVs  of  the  Higgs  doublets
$H_{u,d}$,  in  the phase  convention  where  $v_{u,d}$  are real  and
positive. Linked to this, there  is one extra CP-odd tadpole condition
which  can  be used  to  eliminate  ${\rm  Im}\,(B\mu)$ in  favour  of
maintaining the same phase convention  for the VEVs, order by order in
perturbation theory~\cite{APLB}.  Thus, a convenient set of input mass
parameters of the constrained CP-violating MSSM is
\begin{equation}
  \label{cCPVMSSM}
\tan\beta (m_t)\,,\quad m_{1/2} (M_{\rm GUT})\,,\quad 
m_0 (M_{\rm GUT})\,,\quad A (M_{\rm GUT})\; ,
\end{equation}
where the relative sign of $\mu$ can always be absorbed into the phase
definition  of the  complex parameters  $m_{1/2}$ and  $A$.   Thus, in
addition to $\tan \beta$, this CP-violating CMSSM has just 5 real mass
parameters, two  more than  in its CP-conserving  counter-part, namely
the CP-odd parameters: ${\rm Im}\, m_{1/2}$ and ${\rm Im}\, A$.

How can the general notion of  MFV can be extended to this constrained
CP-violating  MSSM? In  such a  constrained model,  the  physical FCNC
observables remain independent of details of the Yukawa texture chosen
at the GUT scale. They depend only on the CKM mixing matrix ${\bf V}$,
the  fermion  masses,  $\tan\beta$  and  the 5  real  mass  parameters
mentioned above.  If  the CKM matrix ${\bf V}$ were  equal to the unit
matrix   {\bf   1},   the   FCNC   observables   would   vanish,   but
flavour-conserving,  CP-violating  effects  would  still  be  present,
associated with  ${\rm Im}\, m_{1/2}$  and ${\rm Im}\,  A$.  Moreover,
these parameters  also contribute to CP-violating  FCNC observables in
the presence  of non-trivial CKM mixing. Most  noticeably, ${\rm Im}\,
m_{1/2}$ and  ${\rm Im}\, A$  cannot generically mimic the  effects of
the usual CKM phase $\delta$.

We now consider how the above notion of MFV can be further extended
within the more general CP-violating MSSM.  To address this question,
we first notice that under the unitary flavour rotations of the quark
and lepton superfields,
\begin{equation}
  \label{SUPERrot}
\widehat{Q}'\ =\ {\bf U}_Q\, \widehat{Q}\, ,\quad  
\widehat{L}'\ =\ {\bf U}_L\, \widehat{L}\, ,\quad
\widehat{U}'^C\ =\ {\bf U}^*_U\, \widehat{U}^C\, ,\quad  
\widehat{D}'^C\ =\ {\bf U}^*_D\, \widehat{D}^C\, ,\quad
\widehat{E}'^C\ =\ {\bf U}^*_E\, \widehat{E}^C\; ,  
\end{equation}
the complete MSSM Lagrangian  of the theory remains invariant provided
the model parameters are redefined as follows:
\begin{eqnarray}
  \label{RGrot}
{\bf h}_{u,d} &\to & {\bf U}_{U,D}^\dagger\, {\bf h}_{u,d}\, 
{\bf  U}_Q\,,\qquad {\bf h}_e\, \ \to\, \ 
{\bf U}_E^\dagger\, {\bf h}_e\, {\bf  U}_L\,,\nonumber\\
\widetilde{\bf M}^2_{Q,L,U,D,E}  &\to & {\bf  U}_{Q,L,U,D,E}^\dagger\, 
\widetilde{\bf  M}^2_{Q,L,U,D,E}\, {\bf  U}_{Q,L,U,D,E}\,,\nonumber\\
{\bf a}_{u,d} &\to & {\bf U}_{U,D}^\dagger\, {\bf a}_{u,d}\, {\bf
  U}_Q\,,\qquad
{\bf a}_e \,\ \to \,\ {\bf U}_E^\dagger\, {\bf a}_e\, {\bf U}_L\; . 
\end{eqnarray}
The  remaining mass  scales, $\mu$,  $M_{1,2,3}$,  $M^2_{H_{u,d}}$ and
$B\mu$,    do    not    transform    under   the    unitary    flavour
rotations~(\ref{SUPERrot}). In fact, it  is apparent that the one-loop
RGEs  presented  in  Appendix~\ref{sec:RGE}  are invariant  under  the
redefinitions in~(\ref{RGrot}),  provided the  unitary flavour
matrices ${\bf U}_{Q,L,U,D,E}$  are taken to be independent  of the RG
scale.     The   effective    Lagrangian   formalism  we  describe   in
Section~\ref{sec:EPF} respects manifestly the property of flavour
covariance under the unitary transformations~(\ref{SUPERrot}).

It  is apparent  from~(\ref{RGrot}) that  the maximal
set of flavour-singlet mass scales includes:
\begin{equation}
  \label{MCPMFV}
M_{1,2,3}\,,\quad M^2_{H_{u,d}}\,,\qquad 
\widetilde{\bf  M}^2_{Q,L,U,D,E}\ =\ \widetilde{M}^2_{Q,L,U,D,E}\,{\bf
  1}_3\,,\qquad
{\bf A}_{u,d,e}\ =\ A_{u,d,e}\, {\bf 1}_3\; ,
\end{equation}
where the mass parameters $\mu$ and $B\mu$ can be eliminated by virtue
of a  global PQ  symmetry and by  the CP-even and  CP-odd minimization
conditions  on the  Higgs  potential. The scenario (\ref{MCPMFV}) has a total
of 19 mass parameters  that respect the general  MFV property,
6 of  which are CP-odd, namely  ${\rm Im}\, M_{1,2,3}$  and ${\rm
Im}\,  A_{u,d,e}$.   

{\it We  term this scenario  the maximally CP-violating  and minimally
flavour-violating (MCPMFV) variant of the  MSSM, or in short, the MSSM
with MCPMFV.}

It  is worth  noting that,  in  addition to  the flavour-singlet  mass
scales mentioned above, there may exist flavour {\em non-singlet} mass
scales in the  MSSM.  For example, one could  impose an unconventional
boundary   condition   on   the   left-handed   squark   mass   matrix
$\widetilde{\bf M}^2_Q$, such that
\begin{equation}
  \label{MQMX}
\widetilde{\bf M}^2_Q (M_X)\ =\ \widetilde{M}^2_Q\, {\bf 1}_3\: +\:
\widetilde{m}^2_1\, ({\bf h}^\dagger_d {\bf h}_d)\: +\:
\widetilde{m}^2_2\, ({\bf h}^\dagger_u {\bf h}_u)\: +\: 
\widetilde{m}^2_3\, ({\bf h}^\dagger_d {\bf h}_d)\, 
({\bf h}^\dagger_u {\bf h}_u)\: +\ \dots,
\end{equation}
where $M_X$  could be  $M_{\rm GUT}$ or  some other  scale. Evidently,
there are in principle a  considerable number of extra mass parameters
$\widetilde{m}^2_n$  that  could also  be  present in  $\widetilde{\bf
M}^2_Q    (M_X)$,    beyond    the    flavour-singlet    mass    scale
$\widetilde{M}^2_Q$.   In fact,  these additional  flavour non-singlet
mass  parameters $\widetilde{m}^2_n$ can  be as  many as  9 (including
$\widetilde{M}^2_Q$),  as  determined  by  the dimensionality  of  the
$3\times  3$  hermitian  matrix  $\widetilde{\bf M}^2_Q  (M_X)$.   The
generalized boundary  condition~(\ref{MQMX}) on $\widetilde{\bf M}^2_Q
(M_X)$ is in agreement with the  notion of MFV for solving the flavour
problem  by   suppressing  the  GIM-breaking   effects,  provided  the
hierarchy  $\widetilde{m}^2_n \ll  \widetilde{M}^2_Q$ is  assumed.  In
particular,    if    these    flavour-non-singlet   mass    parameters
$\widetilde{m}^2_n$ are induced by RG running, they may be generically
much   smaller   than   $\widetilde{M}^2_Q$.    In  this   case,   the
$\widetilde{m}^2_n$ will  not all be independent of  each other, e.g.,
in  our  MCPMFV scenario,  the  RG-induced flavour-non-singlet  scales
$\widetilde{m}^2_n$  would be  functionals of  the  19 flavour-singlet
mass parameters  stated in~(\ref{MCPMFV}).  In  general, a non-singlet
mass parameter  could either  be introduced by  hand or induced  by RG
running of  a theory  beyond the MSSM  with more  flavour-singlet mass
scales~\cite{Antusch:2007re}.       However,     since     introducing
$\widetilde{m}^2_n  \ll  \widetilde{M}^2_Q$  by  hand  has  no  strong
theoretical   motivation,  we   focus  our   attention  here   on  the
flavour-singlet MSSM framework embodied by the MCPMFV.

Before calculating FCNC observables in  the MSSM with MCPMFV, we first
develop in  the next section  an effective Lagrangian approach  to the
computation of Higgs-mediated effects, which play an important role in
our analysis.

\setcounter{equation}{0} 
\section{Effective Lagrangian Formalism}
  \label{sec:EPF}

Here  we present a  manifestly flavour-covariant  effective Lagrangian
formalism.   This  formalism enables  one  to  show the  flavour-basis
independence  of  FCNC   observables  in  general  soft  SUSY-breaking
scenarios of the MSSM.  It will also be used in Section~\ref{sec:FCNC}
to calculate FCNC processes in the MSSM with MCPMFV.

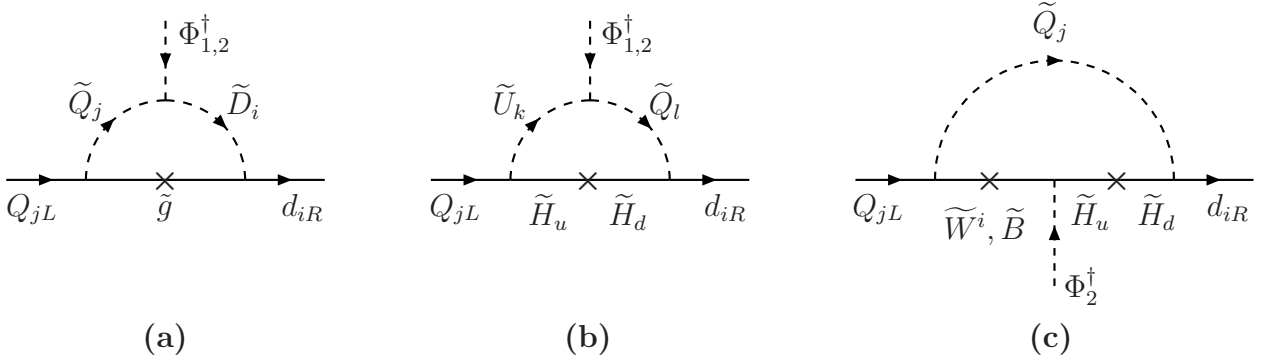
\begin{figure}

\begin{center}
\begin{picture}(450,150)(0,0)
\SetWidth{0.8}
 
\ArrowLine(0,70)(30,70)\Line(30,70)(60,70)
\Line(60,70)(90,70)\ArrowLine(90,70)(120,70)
\DashArrowArcn(60,70)(30,180,90){3}\DashArrowArcn(60,70)(30,90,0){3}
\DashArrowLine(60,130)(60,100){3}
\Text(60,70)[]{\boldmath $\times$}
\Text(0,65)[lt]{$Q_{jL}$}\Text(120,65)[rt]{$d_{iR}$}
\Text(60,65)[t]{$\tilde{g}$}\Text(65,125)[l]{$\Phi^\dagger_{1,2}$}
\Text(37,100)[r]{$\widetilde{Q}_{j}$}
\Text(83,100)[l]{$\widetilde{D}_{i}$}

\Text(60,10)[]{\bf (a)}

\ArrowLine(160,70)(190,70)\Line(190,70)(220,70)
\Line(220,70)(250,70)\ArrowLine(250,70)(280,70)
\DashArrowArcn(220,70)(30,180,90){3}\DashArrowArcn(220,70)(30,90,0){3}
\DashArrowLine(220,130)(220,100){3}
\Text(220,70)[]{\boldmath $\times$}
\Text(160,65)[lt]{$Q_{jL}$}\Text(280,65)[rt]{$d_{iR}$}
\Text(205,65)[t]{$\widetilde{H}_{u}$}
\Text(235,65)[t]{$\widetilde{H}_{d}$}
\Text(225,125)[l]{$\Phi^\dagger_{1,2}$}
\Text(197,100)[r]{$\widetilde{U}_k$}
\Text(243,100)[l]{$\widetilde{Q}_l$}

\Text(220,10)[]{\bf (b)}

\ArrowLine(320,70)(350,70)\Line(350,70)(372,70)
\Line(372,70)(395,70)\Text(372,70)[]{\boldmath  $\times$}
\Line(395,70)(420,70)\Line(420,70)(440,70)
\Text(420,70)[]{\boldmath $\times$}\ArrowLine(440,70)(470,70)
\DashArrowArcn(395,70)(45,180,0){3}\DashArrowLine(395,30)(395,70){3}
\Text(320,65)[lt]{$Q_{jL}$}\Text(470,65)[rt]{$d_{iR}$}
\Text(370,60)[t]{$\widetilde{W}^i,\widetilde{B}$}
\Text(410,65)[t]{$\widetilde{H}_{u}$}
\Text(435,65)[t]{$\widetilde{H}_{d}$}
\Text(395,130)[]{$\widetilde{Q}_j$}
\Text(400,30)[l]{$\Phi^\dagger_2$}

\Text(395,10)[]{\bf (c)}

\end{picture}
\end{center}
\caption{\it Gauge- and flavour-invariant one-loop  self-energy graphs
for down-type quarks in the single-Higgs insertion
approximation, with $H_u \equiv \Phi_2$ and $H_d \equiv i\tau_2
\Phi^*_1$.}\label{fig:self}
\end{figure}

To make  contact between our notation  and that used  elsewhere in the
literature~\cite{CEPW},  we redefine the  Higgs doublets  $H_{u,d}$ as
$H_u \equiv \Phi_2$ and  $H_d \equiv i\tau_2 \Phi^*_1$, 
where $\tau_{1,2,3}$ are the usual Pauli matrices.
We start  our discussion by
considering    the   effective    Lagrangian   that    describes   the
$\tan\beta$-enhanced  supersymmetric  contributions  to the  down-type
quark  self-energies as shown  in Fig.~\ref{fig:self}.   The effective
Lagrangian  can be  written in  gauge-symmetric  and flavour-covariant
form as follows:
\begin{equation}
  \label{Ldeff}
-\,{\cal L}^d_{\rm eff} [\Phi_1,\Phi_2] \ =\  \bar{d}^0_{iR}\,
\Big(\, {\bf h}_d\, \Phi^\dagger_1\: 
+\: \Delta {\bf h}_d [\Phi_1,\Phi_2]\,\Big)_{ij}\, Q^0_{jL}\ +\ {\rm h.c.},
\end{equation}
where   the  superscript   `0'  indicates   weak--eigenstate  fields.
In~(\ref{Ldeff}), the  first term denotes  the tree-level contribution
and  $\Delta  {\bf   h}_d$  is  a  $3\times  3$   matrix  which  is  a
Coleman--Weinberg--type~\cite{CW}    effective   functional    of   the
background  Higgs  doublets  $\Phi_{1,2}$.  We note  that  the  one-loop
effective functional  $\Delta {\bf h}_d [\Phi_1,\Phi_2]$  has the same
gauge   and   flavour   transformation   properties  as   ${\bf   h}_d
\Phi^\dagger_1$.   Its  analytic  and  flavour-covariant form  may  be
calculated via
\begin{eqnarray}
  \label{Dhd}
(\Delta {\bf h}_d)_{ij} &=& \int \frac{d^n k}{(2\pi)^n i}\ \Bigg[\,
P_L\, \frac{2\, g_s^2\, C_F\, M^*_3 }{k^2 - |M^2_3|}\  
\Bigg(\, \frac{1}{k^2 {\bf 1}_{12} - \widetilde{\bf M}^2}\,
\Bigg)_{\widetilde{D}_i \widetilde{Q}^\dagger_j} 
\\
&&\hspace{-5mm}+\, P_L\, \Bigg(\, \frac{1}{\not\! k {\bf 1}_8 - 
{\bf M}_C P_L - {\bf M}^\dagger_C P_R}\, \Bigg)_{\widetilde{H}_u
\widetilde{H}_d} P_L\ ({\bf h}_d)_{il}\, 
 \Bigg(\, \frac{1}{k^2 {\bf 1}_{12} - \widetilde{\bf M}^2}\,
\Bigg)_{\widetilde{Q}_l \widetilde{U}^\dagger_k} ({\bf h}_u)_{kj}
\nonumber\\
&&\hspace{-5mm}+\, P_L
\, \Bigg(\, \frac{1}{\not\! k {\bf 1}_8 - 
{\bf M}_C P_L - {\bf M}^\dagger_C P_R}\,\Bigg)_{\widetilde{H}_d\widetilde{B}} 
P_L\ ({\bf h}_d)_{il}\, 
 \Bigg(\, \frac{1}{k^2 {\bf 1}_{12} - \widetilde{\bf M}^2}\,
\Bigg)_{\widetilde{Q}_l \widetilde{Q}^\dagger_j}\,\left(\sqrt{2} g^\prime\right)
\nonumber\\
&&\hspace{-5mm}+\sum_{k=1}^3\, P_L
\, \Bigg(\, \frac{1}{\not\! k {\bf 1}_8 -
{\bf M}_C P_L - {\bf M}^\dagger_C P_R}\,\Bigg)_{\widetilde{H}_d \widetilde{W}^k}
P_L\ ({\bf h}_d)_{il}\,
 \Bigg(\, \frac{1}{k^2 {\bf 1}_{12} - \widetilde{\bf M}^2}\,
\Bigg)_{\widetilde{Q}_l \widetilde{Q}^\dagger_j}\,
\left(\frac{g\tau_k}{\sqrt{2}}\right) \Bigg]\; ,
\nonumber
\end{eqnarray}
where   $n   =    4   -   2\epsilon$   is   the    usual   number   of
analytically--continued dimensions in dimensional regularization (DR),
${\bf 1}_N$  stands for the  $N\times N$-dimensional unit  matrix,
$P_{L(R)}  = \frac{1}{2}\,  [ 1  -(+)\  \gamma_5 ]$  are the  standard
chirality--projection  operators, and $C_F=4/3$ is the quadratic Casimir invariant of
QCD in the fundamental representation.   The  $8\times  8$-  and  $12\times
12$-dimensional  matrices   ${\bf  M}_C$  and   $\widetilde{\bf  M}^2$
describe  the  squark and  chargino-neutralino  mass  spectrum in  the
background of non-vanishing Higgs doublets $\Phi_{1,2}$.

It   proves  convenient   to  express   the   $8\times  8$-dimensional
chargino-neutralino  mass  matrix  ${\bf  M}_C$  in  the  Weyl  basis
$(\widetilde{B},\       \widetilde{W}^{1,2,3},\      \widetilde{H}_u,\
\widetilde{H}_d   )$,   where   $\widetilde{H}_{u,d}$  are   SU(2)$_L$
doublets: $\widetilde{H}_u  = (\tilde{h}^+_u ,  \tilde{h}^0_u )^T$ and
$\widetilde{H}_d = (\tilde{h}^0_d ,  \tilde{h}^-_d )^T$.  In this weak
basis, the Higgs-field-dependent chargino-neutralino mass matrix ${\bf
M}_C [\Phi_1, \Phi_2]$ reads:
\begin{equation}
  \label{MC}
{\bf M}_C [\Phi_1, \Phi_2]\ =\ \left(\! \begin{array}{cccc}
M_1 & 0   & -\frac{1}{\sqrt{2}}\, g'\Phi^\dagger_2 & 
                    \frac{1}{\sqrt{2}}\, g'\Phi^T_1\, (i\tau_2 )  \\
0   & M_2\,{\bf 1}_3 & \frac{1}{\sqrt{2}}\, g \Phi^\dagger_2\,\tau_i & 
                    -\frac{1}{\sqrt{2}}\, g \Phi^T_1\, (i\tau_2 )\,\tau_i \\ 
-\,\frac{1}{\sqrt{2}}\, g' \Phi^*_2 & 
\frac{1}{\sqrt{2}}\, g \tau_i^T\,\Phi^*_2  & {\bf 0}_2 & \mu\, ( i\tau_2) \\ 
-\, \frac{1}{\sqrt{2}}\, (i\tau_2 )\, g'\Phi_1  & 
    \frac{1}{\sqrt{2}}\, g \tau^T_i\,(i\tau_2) \Phi_1 & 
             -\mu\, (i\tau_2) & {\bf 0}_2 \end{array}\! \right)\; ,
\end{equation}
where $g$  and $g'$  are the SU(2)$_L$  and U(1)$_Y$  gauge couplings,
respectively. Correspondingly, in  the presence of non-vanishing Higgs
doublets  $\Phi_{1,2}$,  the  $12\times  12$-dimensional  squark  mass
matrix $\widetilde{\bf M}^2 [\Phi_1, \Phi_2]$ is given by
\begin{equation}
  \label{Msquark}
\widetilde{\bf M}^2 [\Phi_1, \Phi_2]\ =\ \left( \begin{array}{ccc}
(\widetilde{\bf M}^2)_{\widetilde{Q}^\dagger \widetilde{Q}} & 
(\widetilde{\bf M}^2)_{\widetilde{Q}^\dagger \widetilde{U}} &
(\widetilde{\bf M}^2)_{\widetilde{Q}^\dagger \widetilde{D}} \\
(\widetilde{\bf M}^2)_{\widetilde{U}^\dagger \widetilde{Q}} & 
(\widetilde{\bf M}^2)_{\widetilde{U}^\dagger \widetilde{U}} &
(\widetilde{\bf M}^2)_{\widetilde{U}^\dagger \widetilde{D}} \\
(\widetilde{\bf M}^2)_{\widetilde{D}^\dagger \widetilde{Q}} & 
(\widetilde{\bf M}^2)_{\widetilde{D}^\dagger \widetilde{U}} &
(\widetilde{\bf M}^2)_{\widetilde{D}^\dagger \widetilde{D}}
\end{array}\right)_{ij}\, ,
\end{equation}
with
\begin{eqnarray}
  \label{MSelem}
(\widetilde{\bf M}^2)_{\widetilde{Q}^\dagger_i \widetilde{Q}_j} &=&
(\widetilde{\bf M}^2_Q)_{ij}\, {\bf 1}_2\ +\ 
({\bf h}^\dagger_d {\bf h}_d)_{ij}\, 
\Phi_1\Phi^\dagger_1\ +\ ({\bf h}^\dagger_u {\bf h}_u)_{ij}\, 
\Big( \Phi^\dagger_2 \Phi_2\, {\bf 1}_2\, -\,
\Phi_2\Phi^\dagger_2 \Big)\nonumber\\
&& -\ \frac{1}{2}\, \delta_{ij}\, g^2\, \Big( \Phi_1 \Phi^\dagger_1\, -\, 
\Phi_2 \Phi^\dagger_2\Big)\: + \: \delta_{ij}\,
\bigg(\, \frac{1}{4}\, g^2\, -\, \frac{1}{12}\, g'^2\,\bigg)\,
\Big( \Phi^\dagger_1 \Phi_1\, -\, \Phi^\dagger_2\Phi_2 \Big)\, 
{\bf 1}_2\, ,\nonumber\\ 
(\widetilde{\bf M}^2)_{\widetilde{U}^\dagger_i \widetilde{Q}_j} &=& 
(\widetilde{\bf M}^2)^\dagger_{\widetilde{Q}^\dagger_j
  \widetilde{U}_i}\ =\ -\,({\bf a}_u)_{ij}\,
\Phi^T_2 i\tau_2\ +\ ({\bf h}_u)_{ij}\, \mu^*\, \Phi^T_1 i\tau_2\,
,\nonumber\\ 
(\widetilde{\bf M}^2)_{\widetilde{D}^\dagger_i 
\widetilde{Q}_j} &=& (\widetilde{\bf  M}^2)^\dagger_{\widetilde{Q}^\dagger_j
 \widetilde{D}_i}\ =\ ({\bf a}_d)_{ij}\,
\Phi^\dagger_1\ -\ ({\bf h}_d)_{ij}\, \mu^*\, \Phi^\dagger_2\, ,\nonumber\\
(\widetilde{\bf M}^2)_{\widetilde{U}^\dagger_i \widetilde{U}_j} &=&
(\widetilde{\bf M}^2_U)_{ij}\ +\ ({\bf h}_u {\bf h}^\dagger_u)_{ij}\,
\Phi^\dagger_2\Phi_2\ +\ \frac{1}{3}\, \delta_{ij}\,
g'^2\, \Big( \Phi^\dagger_1 \Phi_1\, -\, \Phi^\dagger_2 \Phi_2
\Big)\,,\nonumber\\ 
(\widetilde{\bf M}^2)_{\widetilde{D}^\dagger_i \widetilde{D}_j} &=& 
(\widetilde{\bf M}^2_D)_{ij}\ +\ ({\bf h}_d {\bf
  h}^\dagger_d)_{ij}\,\Phi^\dagger_1\Phi_1\ -\ \frac{1}{6}\, \delta_{ij}\,
g'^2\, \Big( \Phi^\dagger_1 \Phi_1\, -\, 
\Phi^\dagger_2 \Phi_2 \Big)\,,\nonumber\\
(\widetilde{\bf M}^2)_{\widetilde{U}^\dagger_i \widetilde{D}_j} &=&
(\widetilde{\bf M}^2)^\dagger_{\widetilde{D}^\dagger_j \widetilde{U}_i}\
=\ ({\bf h}_u {\bf h}_d^\dagger)_{ij}\, \Phi^T_1 i\tau_2\Phi_2\; ,
\end{eqnarray} 
where $\delta_{ij}$ is the usual Kronecker symbol.

The form of the derived  effective Lagrangian depends, to some extent,
on the choice  of renormalization scheme. As usual,  one may adopt the
$\overline{\mbox{MS}}$    or    $\overline{\mbox{DR}}$   schemes    of
renormalization.   In  general,   the  different  schemes  affect  the
holomorphic part of  the Lagrangian at the one-loop  level.  Thanks to
the non-renormalization  theorems of SUSY, the  Yukawa couplings ${\bf
h}_{u,d}$   are  not   renormalized,   and  the   wave  functions   of
$\Phi_{1,2}$, $Q_{iL}$,  $u_{iR}$ and $d_{iR}$  remove the ultraviolet
(UV) divergences  of the one-loop corrections to  the Yukawa couplings
$\bar{d}_{iR}   \Phi^\dagger_1   Q_{jL}$   and  $\bar{u}_{iR}   \Phi_2
Q_{jL}$.  The left-over UV-finite  terms are  not $\tan\beta$-enhanced
and  can be absorbed  into the  definition of  ${\bf h}_{u,d}$,  up to
higher-order scheme-dependent corrections.   Although the latter could
be consistently included  in our gauge-symmetric and flavour-covariant
formalism, we  ignore these small UV-finite holomorphic  terms as they
are  higher-order effects  beyond  the one-loop  approximation of  our
interest.

By analogy, the gauge-  and flavour-covariant effective Lagrangian for
the up-type quark self-energies may be written down as follows:
\begin{equation}
  \label{Lueff}
-\,{\cal L}^u_{\rm eff} [\Phi_1,\Phi_2] \ =\  \bar{u}^0_{iR}\,
\Big(\, {\bf h}_u\, \Phi^T_2\,(-i\tau_2)\: 
+\: \Delta {\bf h}_u [\Phi_1,\Phi_2]\,\Big)_{ij}\, Q^0_{jL}\ +\ {\rm
  h.c.}, 
\end{equation}
where  $\Delta  {\bf  h}_u  [\Phi_1,\Phi_2]$ may  be  calculated  from
Feynman diagrams  analogous to Fig.~\ref{fig:self}. As  opposed to the
down-type quark self-energy case,  these radiative corrections are not
enhanced  for large values  of $\tan\beta$  and so  are ignored  in our
numerical analysis  in Section~\ref{sec:num}. 

The weak quark chiral states, $u^0_{L,R}$ and $d^0_{L,R}$, are related
to their respective mass eigenstates, $u_{L,R}$ and $d_{L,R}$, through
the unitary transformations:
\begin{equation}
  \label{Utr}
u^0_L\ =\ {\bf U}^Q_L\, u_L\,,\quad 
d^0_L\ =\ {\bf U}^Q_L\, {\bf V}\, d_L\,,\quad
u^0_R\ =\ {\bf U}^u_R\, u_R\,,\quad
d^0_R\ =\ {\bf U}^d_R\, d_R\; , 
\end{equation}
where  ${\bf  U}^Q_L$,  ${\bf  U}^{u,d}_R$  are  $3\times  3$  unitary
matrices and  ${\bf V}$  is the CKM  mixing matrix. All  these unitary
matrices are determined by the simple mass renormalization conditions:
\begin{equation}
  \label{Mcond}
\Big<\,{\cal L}^d_{\rm eff}[\Phi_1, \Phi_2]\, \Big>\ =\ -\, \bar{d}_R\,
	\widehat{\bf M}_d\, d_L\ +\ {\rm h.c.},\qquad
\Big<\,{\cal L}^u_{\rm eff}[\Phi_1, \Phi_2]\, \Big>\ =\ -\, \bar{u}_R\,
	\widehat{\bf M}_u\, u_L\ +\ {\rm h.c.},
\end{equation}
where $\langle  \dots \rangle$ denotes  the value when the Higgs
doublets  $\Phi_{1,2}$ acquire their VEVs, and  $\widehat{\bf M}_{u,d}$  are  the physical
diagonal  mass matrices  for the  up-  and down-type  quarks.  Imposing  the
conditions~(\ref{Mcond}) yields~\cite{DP}
\begin{equation}
  \label{hcond}
{\bf U}^{d\, \dagger}_R\, {\bf h}_d\, {\bf U}^Q_L\ =\ 
\frac{\sqrt{2}}{v_1}\, \widehat{\bf M}_d\, {\bf V}^\dagger\, 
                                            {\bf R}^{-1}_d\,,\qquad
{\bf U}^{u\, \dagger}_R\, {\bf h}_u\, {\bf U}^Q_L\ =\ 
\frac{\sqrt{2}}{v_2}\, \widehat{\bf M}_u\, {\bf R}^{-1}_u\; ,
\end{equation} 
where
\begin{eqnarray}
  \label{Rud}
{\bf R}_d &=& {\bf 1}\ +\ \frac{\sqrt{2}}{v_1}\, {\bf U}^{Q\,
  \dagger}_L\,  
\Big<\, {\bf h}^{-1}_d\, \Delta {\bf h}_d [\Phi_1, \Phi_2]\, \Big>\, 
{\bf U}^Q_L\,,\nonumber\\
{\bf R}_u &=& {\bf 1}\ +\ \frac{\sqrt{2}}{v_2}\, {\bf U}^{Q\,
  \dagger}_L\,  
\Big<\, {\bf h}^{-1}_u\, \Delta {\bf h}_u [\Phi_1, \Phi_2]\, \Big>\, 
{\bf U}^Q_L\; .
\end{eqnarray}
In~(\ref{Rud}) and  in the following,  the symbol ${\bf 1}$  without a
subscript will always denote the  $3\times 3$ unit matrix.  We observe
that the unitary matrices ${\bf  U}^Q_L$, ${\bf U}^{u,d}_R$ can all be
set  to ${\bf  1}$  by  virtue of  the  flavour transformations  given
in~(\ref{SUPERrot}).   The   Yukawa  couplings  ${\bf   h}_{u,d}$  are
determined  by  the  physical  mass conditions~(\ref{hcond}).   It  is
important  to  remark  here~\cite{DP}  that these  conditions  form  a
coupled  system   of  non-linear  equations  with   respect  to  ${\bf
h}_{u,d}$,  since the  Yukawa  couplings also  enter  the right  sides
of~(\ref{hcond})    through    the    expressions   ${\bf    R}_{d,u}$
in~(\ref{Rud}). In  addition, one should notice that  the physical CKM
mixing  matrix  ${\bf V}$  remains  unitary  throughout our  effective
Lagrangian  approach.  As  we will  see below  and more  explicitly in
Appendix~\ref{sec:WIs},  the  unitarity of  ${\bf  V}$ throughout  the
renormalization  process is  a  crucial property  for maintaining  the
gauge symmetries  through the Ward  identities (WIs) in  our effective
Lagrangian formalism.   

We now  consider the effective FCNC Lagrangian  related to Higgs
interactions to down-type quarks. {}From~(\ref{Ldeff}), we find that 
\begin{eqnarray}
  \label{LdHeff}
-\, {\cal L}^{d,H}_{\rm eff} &=& \bar{d}_R\, \frac{{\bf
 h}_d}{\sqrt{2}}\; \Bigg[\, \phi_1\, \Big( {\bf 1} + {\bf \Delta}^{\phi_1}_d
 \Big)\: -\: i a_1\, \Big( {\bf 1} + {\bf \Delta}^{a_1}_d \Big)\: +\:
\phi_2\,{\bf \Delta}^{\phi_2}_d\: -\: i a_2 {\bf \Delta}^{a_2}_d\, \Bigg]\, 
{\bf V}\, d_L\nonumber\\
&&+\, \bar{d}_R\, {\bf h}_d\, \Big[\, \phi^-_1\, 
                 \Big( {\bf 1} + {\bf \Delta}^{\phi^-_1}_d \Big)\: 
                 +\: \phi^-_2 {\bf \Delta}^{\phi^-_2}_d\,\Big]\,u_L\ 
                                                        +\ {\rm h.c.}, 
\end{eqnarray}
where the individual components of the Higgs doublets $\Phi_{1,2}$ are
given by 
\begin{equation}
\Phi_{1,2}\ =\ \left(\! \begin{array}{c}
\phi^+_{1,2}\\ 
\frac{1}{\sqrt{2}}\, \Big(\, v_{1,2}\: +\: \phi_{1,2}\: +\: i a_{1,2}\,\Big)
\end{array}\! \right)\; .
\end{equation}
Moreover,  the  $3\times  3$ matrices  ${\bf  \Delta}^{\phi_{1,2}}_d$,
${\bf  \Delta}^{a_{1,2}}_d$   and  ${\bf  \Delta}^{\phi^\pm_{1,2}}_d$  are
given by
\begin{equation}
  \label{Deltad}
{\bf \Delta}^{\phi_{1,2}}_d \ =\ \sqrt{2}\;
\Big<\, \frac{\delta}{\delta \phi_{1,2}}\, 
{\bf \Delta }_d\, \Big>\; ,\quad
{\bf \Delta}^{a_{1,2}}_d \ =\ i\,\sqrt{2}\;
\Big<\, \frac{\delta}{\delta a_{1,2}}\, 
{\bf \Delta}_d\, \Big>\; ,\quad
{\bf \Delta}^{\phi^\pm_{1,2}}_d \ =\ 
\Big<\, \frac{\delta}{\delta \phi^\pm_{1,2}}\, 
{\bf \Delta}_d\, \Big>\; ,
\end{equation}
where  we have used  the short-hand  notation, ${\bf  \Delta}_d \equiv
{\bf h}^{-1}_d\,  \Delta {\bf  h}_d [\Phi_1, \Phi_2]$,  and suppressed
the vanishing  iso-doublet components on the  LHS's of (\ref{Deltad}).
In the  CP-violating MSSM,  the weak-state Higgs  fields $\phi_{1,2}$,
$a_{1,2}$ and $\phi^-_{1,2}$ are  related to the neutral CP-mixed mass
eigenstates $H_{1,2,3}$~\cite{PW,CEPW}, the  charged Higgs boson $H^-$
and the would-be Goldstone bosons $G^0$ and $G^-$, associated with the
$Z$ and $W^-$ bosons, through:
\begin{eqnarray}
  \label{Hmix}
\phi_1 \!&=&\! O_{1i}\, H_i\,,\qquad\qquad\qquad\qquad\ 
\phi_2 \ =\ O_{2i}\, H_i\, ,\nonumber\\
a_1 \!&=&\! c_\beta\, G^0\: -\: s_\beta\, O_{3i}\, H_i\,,\qquad\quad\
a_2 \ =\ s_\beta\, G^0\: +\: c_\beta\, O_{3i}\, H_i\,,\nonumber\\
\phi^-_1 \!&=&\! c_\beta\, G^-\: -\: s_\beta\, H^-\,,\qquad\qquad  
\phi^-_2 \ =\ s_\beta\, G^-\: +\: c_\beta\, H^-\; ,
\end{eqnarray}
where $s_\beta  \equiv \sin\beta$, $c_\beta \equiv  \cos\beta$ and $O$
is an orthogonal $3\times 3$ Higgs-boson-mixing matrix.

One may now exploit the properties of gauge- and flavour-covariance of
the effective  functional ${\bf \Delta}_d [\Phi_1,  \Phi_2]$ to obtain
useful relations  in the large-$\tan\beta$  limit. Specifically, ${\bf
\Delta}_d [\Phi_1, \Phi_2]$ should have the form:
\begin{equation}
  \label{Ddform}
{\bf \Delta}_d [\Phi_1,  \Phi_2]\ =\ \Phi^\dagger_1\, {\bf f}_1\:
+\: \Phi^\dagger_2\, {\bf f}_2\;, 
\end{equation}
where ${\bf f}_{1,2} \Big( \Phi^\dagger_1\Phi_1, \Phi^\dagger_2\Phi_2,
\Phi^\dagger_1   \Phi_2,  \Phi^\dagger_2\Phi_1\Big)$   are  calculable
$3\times   3$-dimensional  functionals   which   transform  as   ${\bf
h}^\dagger_d  {\bf h}_d$ or  ${\bf h}^\dagger_u  {\bf h}_u$  under the
flavour rotations~(\ref{SUPERrot}).  Given the form~(\ref{Ddform}). it
is then not  difficult to show that in  the infinite-$\tan\beta$ limit
($v_1 \to 0$),
\begin{equation}
  \label{Ddrel}
\lim_{v_1 \to 0} i\,\sqrt{2}\; \Big<\, \frac{\delta}{\delta a_2}\, {\bf
  \Delta}_d\,\Big>\ =\ \frac{\sqrt{2}}{v_2}\, \big<\, {\bf
  \Delta}_d\,\big>\,,\qquad
\lim_{v_1 \to 0} \Big<\, \frac{\delta}{\delta \phi^-_2}\, {\bf
  \Delta}_d\,\Big>\ =\ \frac{\sqrt{2}}{v_2}\, \big<\, {\bf
  \Delta}_d\,\big>\, .
\end{equation}
Very similar  relations may be  derived for the up-type  quark sector,
but  in   the  limit  of   vanishing  $\tan\beta$.   As  we   show  in
Appendix~\ref{sec:WIs}, Ward identities  (WIs) involving the $W^-$ and
$Z$-boson couplings to  quarks give rise to the  following {\em exact}
relations:
\begin{equation}
  \label{DdG}
{\bf \Delta}^{G^0}_d \ \equiv\ i\,\sqrt{2}\; 
\Big<\, \frac{\delta}{\delta G^0}\, {\bf
  \Delta}_d\,\Big>\ =\ \frac{\sqrt{2}}{v}\, \big<\, {\bf
  \Delta}_d\,\big>\,,\qquad
{\bf \Delta}^{G^-}_d \ \equiv\ \Big<\, \frac{\delta}{\delta G^-}\, {\bf
  \Delta}_d\,\Big>\ =\ \frac{\sqrt{2}}{v}\, \big<\, {\bf
  \Delta}_d\,\big>\,,
\end{equation}
where $v = \sqrt{v^2_1 + v^2_2}$ is  the VEV of the Higgs boson in the
SM.  Relations very analogous to those stated in~(\ref{DdG}) hold true
for  the up-type  sector as  well, i.e.~${\bf  \Delta}^{G^0}_u  = {\bf
\Delta}^{G^+}_u = -  \sqrt{2}\,\big<\, {\bf \Delta}_u\,\big>/v$, where
the extra  minus sign comes from  the opposite isospin  of the up-type
quarks with respect to the down-type quarks.

For  our phenomenological analysis  in Section~\ref{sec:FCNC},  we may
conveniently  express  the  general  flavour-changing  (FC)  effective
Lagrangian  for the  interactions  of the  neutral  and charged  Higgs
fields to the up- and down-type quarks $u,\ d$ in the following form:
\begin{eqnarray}
  \label{LeffFCNC}
{\cal L}_{\rm  FC} &=& -\,  \frac{g}{2 M_W}\ \Bigg[\,  H_i\; \bar{d}\,
\Big(\, \widehat{\bf  M}_d\, {\bf g}^L_{H_i\bar{d}d}\,  P_L\: +\: {\bf
g}^R_{H_i\bar{d}d}\,  \widehat{\bf M}_d\,  P_R\, \Big)\,  d\  +\ G^0\;
\bar{d}\,  \widehat{\bf M}_d\, i\gamma_5\,  d \nonumber\\  
&&+\, H_i\; \bar{u}\, \Big(\, \widehat{\bf  M}_u\, 
{\bf g}^L_{H_i\bar{u}u}\, P_L\: +\: {\bf g}^R_{H_i\bar{u}u}\, 
\widehat{\bf  M}_u\, P_R\, \Big)\, u\ -\
G^0\, \bar{u}\, \widehat{\bf  M}_u\, i\gamma_5\, u\; \Bigg]\nonumber\\
&&-\,  \frac{g}{\sqrt{2}   M_W}\  \Bigg[\,  H^-\;   \bar{d}\,  \Big(\,
\widehat{\bf   M}_d\,  {\bf   g}^L_{H^-\bar{d}u}\,   P_L\:  +\:   {\bf
g}^R_{H^-\bar{d}u}\,  \widehat{\bf M}_u\,  P_R\,  \Big)\, u\nonumber\\
&&+\ G^-\; \bar{d}\,  \Big(\, \widehat{\bf M}_d\,{\bf V}^\dagger P_L\:
-\:  {\bf V}^\dagger\,\widehat{\bf  M}_u\,  P_R\, \Big)\,  u\ +\  {\rm
H.c.}\;\Bigg]\;,
\end{eqnarray}
where the Higgs couplings in the flavour basis ${\bf U}^Q_L = {\bf U}^u_R
= {\bf U}^d_R = {\bf 1}$ are given by 
\begin{eqnarray}
  \label{gLd}
{\bf g}^L_{H_i\bar{d}d} &=& \frac{O_{1i}}{c_\beta}\;
{\bf V}^\dagger\, {\bf R}^{-1}_d\, \Big( {\bf 1} + {\bf
  \Delta}^{\phi_1}_d \Big)\, {\bf V}\ +\ 
\frac{O_{2i}}{c_\beta}\; {\bf V}^\dagger\, {\bf R}^{-1}_d\, 
{\bf \Delta}^{\phi_2}_d\, {\bf V}\nonumber\\
&&+\, iO_{3i}\, t_\beta\, {\bf V}^\dagger\, {\bf R}^{-1}_d\,
\Big( {\bf 1} + {\bf \Delta}^{a_1}_d - \frac{1}{t_\beta}\, 
{\bf \Delta}^{a_2}_d \Big)\, {\bf V}\; ,\\[3mm]
  \label{gRd}
{\bf g}^R_{H_i\bar{d}d} &=& ({\bf g}^L_{H_i\bar{d}d})^\dagger\; ,\\[3mm]
  \label{gLu}
{\bf g}^L_{H_i\bar{u}u} &=& \frac{O_{1i}}{s_\beta}\;
{\bf R}^{-1}_u\, {\bf
  \Delta}^{\phi_1}_u\ +\ 
\frac{O_{2i}}{s_\beta}\; {\bf R}^{-1}_u\, 
\Big( {\bf 1} + {\bf \Delta}^{\phi_2}_u \Big)\nonumber\\
&&+\, iO_{3i}\, t^{-1}_\beta\, {\bf R}^{-1}_u\,
\Big( {\bf 1} - {\bf \Delta}^{a_2}_u + t_\beta\, 
{\bf \Delta}^{a_1}_u \Big)\; ,\\[3mm]
  \label{gRu}
{\bf g}^R_{H_i\bar{u}u} &=& ({\bf g}^L_{H_i\bar{u} u})^\dagger\; ,\\[3mm]
  \label{gLud}
{\bf g}^L_{H^-\bar{d} u} &=& 
-\, t_\beta\, {\bf V}^\dagger\, {\bf R}^{-1}_d\, \Big(
  {\bf 1} + {\bf \Delta}^{\phi^-_1}_d \Big)\  +\ 
{\bf V}^\dagger\, {\bf R}^{-1}_d\, {\bf \Delta}^{\phi^-_2}_d \; ,\\[3mm]
  \label{gRud}
{\bf g}^R_{H^-\bar{d} u} &=& -\, t^{-1}_\beta\, {\bf V}^\dagger\, 
 \Big(  {\bf 1} - ({\bf
  \Delta}^{\phi^+_2}_u )^\dagger \Big)\, ({\bf R}^{-1}_u)^\dagger\ -\
{\bf V}^\dagger\, ({\bf \Delta}^{\phi^+_1}_u )^\dagger\, 
({\bf R}^{-1}_u )^\dagger\; ,
\end{eqnarray}
and $t_\beta \equiv \tan \beta$.
We  note  that  the  Higgs-boson vertex-correction  matrices  for  the
up-type      quarks,     ${\bf      \Delta}_u^{\phi_{1,2}}$,     ${\bf
\Delta}_u^{a_{1,2}}$   and   ${\bf  \Delta}_u^{\phi^\pm_{1,2}}$,   are
defined as in~(\ref{Deltad}).

The above general form of the effective Lagrangian ${\cal L}_{\rm FC}$
extends  the one derived  in~\cite{DP} in  several aspects.  First, it
consistently includes  all higher-order terms of  the form $(t_\beta\,
m_b  \mu/M^2_{\rm  SUSY})^{n\ge 1}$,  which  can  become important  in
scenarios with  large bottom-squark mixing~\cite{CGNW}.   Secondly, it
does not suffer from the  limitation that the soft SUSY-breaking scale
should be much higher than the electroweak scale $M_Z$.  Specifically,
SM    electroweak    corrections    may    be    included    in    the
Coleman--Weinberg-type   effective  functionals   ${\bf  \Delta}_{d,u}
[\Phi_1,  \Phi_2]$, provided  the  theory is  quantized in  non-linear
gauges~\cite{APLB98} that preserve  the Higgs-boson low-energy theorem
(HLET)~\cite{EGN}.  Finally,  the effective Lagrangian  ${\cal L}_{\rm
FC}$ implements properly  all the gauge symmetries through  the WIs as
discussed in Appendix~\ref{sec:WIs}.

The general FC effective Lagrangian~(\ref{LeffFCNC}) takes on the form
presented  in~\cite{DP} in  the  single-Higgs-insertion approximation.
In  this case,  the $\tan\beta$-enhanced  threshold  corrections ${\bf
\Delta}^{a_2}_d$,        ${\bf        \Delta}^{\phi_2}_d$,       ${\bf
\Delta}^{\phi^\pm_2}_d$  and  $\langle  {\bf  \Delta}_d  \rangle$  are
inter-related as follows:
\begin{equation}
  \label{Drel}
\frac{\sqrt{2}}{v_2}\, \big<\, {\bf \Delta}_d\,\big>\ =\ {\bf
  \Delta}^{a_2}_d\ =\ {\bf \Delta}^{\phi_2}_d\ =\ {\bf
  \Delta}^{\phi^-_2}_d\ =\ \Big({\bf \Delta}^{\phi^+_2}_d\Big)^\dagger\; ,
\end{equation} 
where $\big< {\bf \Delta}_d\big>$ is given in the MSSM with MCPMFV by
\begin{eqnarray}
  \label{Dds}
\frac{\sqrt{2}}{v_2}\,\Big< {\bf \Delta}_d \Big> \!&=&\!  {\bf 1}\;
\frac{2\alpha_3}{3\pi}\ \mu^* M^*_3\; I \Big( \widetilde{M}^2_Q\, ,\, 
\widetilde{M}^2_D\, ,\, |M_3|^2 \Big)\ +\
\frac{{\bf h}^\dagger_u {\bf h}_u}{16\,\pi^2}\ \mu^* A_u\;
I\Big(\widetilde{M}^2_Q\,,\, \widetilde{M}^2_U\, ,\, |\mu
|^2\Big)\nonumber\\
&& +\ \dots, 
\end{eqnarray}
and $I(x,y,z)$ is the one-loop function:
\begin{equation}
  \label{Ixyz}
I(x,y,z)\ =\ \frac{xy\,\ln (x/y)\: +\: yz\,\ln (y/z)\: +\:
             xz\, \ln (z/x)}{(x-y)\,(y-z)\,(x-z)}\ .
\end{equation}
The ellipses in~(\ref{Dds}) denote the small contributions coming from
the  Feynman diagram  in  Fig.~\ref{fig:self}(c), which  has the  same
flavour     structure    as     the    gluino-mediated     graph    in
Fig.~\ref{fig:self}(a), i.e., this  contribution is flavour-singlet in
the single-Higgs-insertion approximation.  We remark, finally, that in
writing down~(\ref{Dds}) we have not considered the RG-running effects
on the squark mass matrices  between $M_{\rm GUT}$ and $M_{\rm SUSY}$.
These  effects  are important,  and  are  taken  into account  in  our
numerical analysis in Section~\ref{sec:num}.

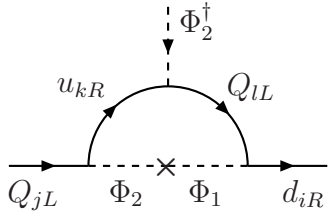
\begin{figure}

\begin{center}
\begin{picture}(450,100)(0,50)
\SetWidth{0.8}

\ArrowLine(160,70)(190,70)\DashLine(190,70)(220,70){3}
\DashLine(220,70)(250,70){3}\ArrowLine(250,70)(280,70)
\ArrowArcn(220,70)(30,180,90)\ArrowArcn(220,70)(30,90,0)
\DashArrowLine(220,130)(220,100){3}
\Text(220,70)[]{\boldmath $\times$}
\Text(160,65)[lt]{$Q_{jL}$}\Text(280,65)[rt]{$d_{iR}$}
\Text(205,65)[t]{$\Phi_2$}
\Text(235,65)[t]{$\Phi_1$}
\Text(225,125)[l]{$\Phi^\dagger_2$}
\Text(197,100)[r]{$u_{kR}$}
\Text(243,100)[l]{$Q_{lL}$}
 
\end{picture}
\end{center}
\caption{\it Two-Higgs-doublet
model (2HDM) contribution to the one-loop  self-energy graphs
for down-type quarks in the single-Higgs-insertion
approximation.}\label{fig:2HDM}
\end{figure}

In addition to graphs  involving SUSY particles, the two-Higgs-doublet
model (2HDM) sector  of the MSSM may also  contribute significantly to
the one-loop self-energy graphs of the down quarks.  This contribution
is  shown in  Fig.~\ref{fig:2HDM} and  is formally  enhanced  at large
$\tan\beta$,  since  it  is  proportional  to  ${\bf  h}_d$.   In  the
single-Higgs-insertion approximation,  the 2HDM contribution  is given
by
\begin{equation}
  \label{2HDMapprox}
\frac{\sqrt{2}}{v_2}\, \Big<\, {\bf \Delta}^{\rm 2HDM}_d\,\Big>\ =\
\frac{{\bf h}^\dagger_u {\bf h}_u}{16\,\pi^2}\
\frac{B^*\mu^*}{M^2_{H_d}\: -\: M^2_{H_u}}\ \ln\Bigg|\frac{M^2_{H_d} + 
|\mu|^2}{M^2_{H_u} + |\mu |^2}\Bigg|\ .
\end{equation}
This  contribution turns  out to  be  subleading with  respect to  the
Feynman diagram~\ref{fig:self}(b) and  exhibits a very similar flavour
structure.   Beyond   the  single-Higgs-insertion  approximation,  the
effective functional $\Delta  {\bf h}^{\rm 2HDM}_d [\Phi_1,\Phi_2]$ is
calculated as
\begin{eqnarray}
  \label{2HDM}
(\Delta{\bf h}^{\rm  2HDM}_d)_{ij} &=& 
\int \frac{d^n k}{(2\pi)^n i}\ ({\bf h}_d)_{il}\,  
P_L\, \Bigg(\, \frac{1}{\not\! k {\bf 1}_6 - 
{\bf M}_q P_L - {\bf M}^\dagger_q P_R}\, \Bigg)_{Q_l \bar{u}_k} P_L\  
({\bf h}_u)_{kj}\nonumber\\
&&\times\, 
\Bigg(\, \frac{1}{k^2 {\bf 1}_4 - {\bf M}_H^2}\,
\Bigg)_{\Phi_1 \Phi^\dagger_2}\; , 
\end{eqnarray}
where  ${\bf M}_q [\Phi_1,\Phi_2]$  and ${\bf  M}_H^2 [\Phi_1,\Phi_2]$
are the $6\times 6$- and $4\times 4$-dimensional quark and Higgs-boson
mass matrices in the background of non-zero $\Phi_{1,2}$. The $6\times
6$-dimensional quark mass matrix is given by
\begin{equation}
  \label{Mquark}
{\bf M}_q [\Phi_1,\Phi_2]\ =\ 
\left(\! \begin{array}{c}  
({\bf M}_q)_{\bar{u}_iQ_j} \\
({\bf M}_q)_{\bar{d}_iQ_j} 
\end{array}\!\right)
\ =\ 
\left(\! \begin{array}{c} 
({\bf h}_u)_{ij}\, \Phi^T_2 (-i\tau_2) \\
({\bf h}_d)_{ij}\, \Phi^\dagger_1
\end{array}\!\right)\; .
\end{equation}
The Higgs-boson  background mass matrix  ${\bf M}_H^2 [\Phi_1,\Phi_2]$
receives   appreciable   radiative   corrections   beyond   the   tree
level~\cite{ERZ,APLB,PW,CEPW}.   At  the   tree  level,  the  $4\times
4$-dimensional  matrix ${\bf  M}_H^2 [\Phi_1,\Phi_2]$  is given  in the
weak basis $(\Phi_1, \Phi_2)$ by
\begin{equation}
  \label{MH2}
{\bf  M}_H^2 [\Phi_1,\Phi_2]\ =\ 
\left(\! \begin{array}{cc}  
({\bf M}^2_H)_{\Phi^\dagger_1\Phi_1} & ({\bf M}^2_H)_{\Phi^\dagger_1\Phi_2} \\
({\bf M}^2_H)_{\Phi^\dagger_2\Phi_1} & 
({\bf M}^2_H)_{\Phi^\dagger_2\Phi_2}
\end{array}\!\right)\;,
\end{equation}
where
\begin{eqnarray}
  \label{Mphi12}
({\bf M}^2_H)_{\Phi^\dagger_1\Phi_1} &=& \Bigg(\, M^2_{H_d}\: +\:
  |\mu|^2 \: +\: 
\frac{g^2+g'^2}{2} \Phi^\dagger_1\Phi_1\: +\: \frac{g^2-g'^2}{4} 
\Phi^\dagger_2\Phi_2\,\Bigg)\, {\bf 1}_2\:
+\: \frac{g^2}{2} \Phi_2\Phi^\dagger_2\; ,\nonumber\\
({\bf M}^2_H)_{\Phi^\dagger_2\Phi_2} &=&\Bigg(\, M^2_{H_u}\: +\: |\mu|^2 \: +\:
\frac{g^2+g'^2}{2} \Phi^\dagger_2\Phi_2\: 
+\: \frac{g^2-g'^2}{4} \Phi^\dagger_1\Phi_1\,\Bigg)\,{\bf 1}_2\:
+\: \frac{g^2}{2} \Phi_1\Phi^\dagger_1\; ,\nonumber\\
({\bf M}^2_H)_{\Phi^\dagger_1\Phi_2} &=& 
({\bf M}^2_H)^\dagger_{\Phi^\dagger_2\Phi_1}\ =\
\Bigg( -B\mu\:  +\: 
\frac{g^2}{2} \Phi^\dagger_2\Phi_1\, \Bigg)\, {\bf 1}_2\: 
+\: \frac{g^2-g'^2}{4} \Phi_1\Phi^\dagger_2\; .
\end{eqnarray}

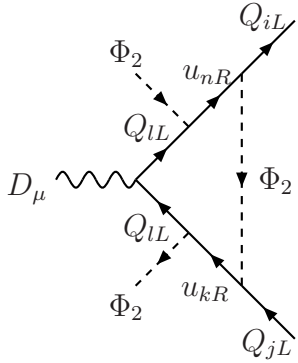
\begin{figure}

\begin{center}
\begin{picture}(300,180)(0,20)
\SetWidth{0.8}

\Photon(100,100)(130,100){3}{3}\ArrowLine(130,100)(150,120)
\ArrowLine(150,120)(170,140)\ArrowLine(170,140)(190,160)
\ArrowLine(150,80)(130,100)\ArrowLine(170,60)(150,80)
\ArrowLine(190,40)(170,60)\DashArrowLine(170,140)(170,60){3}
\DashArrowLine(130,140)(150,120){3}\DashArrowLine(150,80)(130,60){3}
\Text(177,100)[l]{$\Phi_2$}\Text(127,143)[b]{$\Phi_2$}
\Text(127,57)[t]{$\Phi_2$}
\Text(97,97)[r]{$D_\mu$}
\Text(188,162)[r]{$Q_{iL}$}\Text(190,38)[r]{$Q_{jL}$}
\Text(166,58)[r]{$u_{kR}$}\Text(167,140)[r]{$u_{nR}$}
\Text(144,120)[r]{$Q_{lL}$}\Text(144,82)[r]{$Q_{lL}$}

\end{picture}
\end{center}
\caption{\it   Dominant  gauge-  and   flavour-invariant  contribution
leading to a modification  of the tree-level Goldstone-boson couplings
to quarks.}\label{fig:gold}
\end{figure}

In  the  one-loop  effective  Lagrangian  ${\cal  L}_{\rm  FC}$  given
in~(\ref{LeffFCNC}), the  couplings of the Goldstone  bosons $G^0$ and
$G^\pm$ to  quarks retain their  tree-level form.  This result  is not
accidental, but a consequence  of the Goldstone theorem, which applies
when  the momenta  of  the external  particles  are all  set to  zero.
However, the tree-level form  of the Goldstone couplings gets modified
when momentum-dependent (derivative)  terms are considered. To leading
order  in a  derivative  expansion,  one would  have  to consider  the
effective Lagrangian
\begin{eqnarray}
  \label{LDslash}
{\cal L}_{\not\! D} &=& i\, \bar{Q}_L\, \bigg[\; {\bf Z}_Q\, \not\!\!
  D\ +\ 
{\bf A}^{(i,j)}_Q\, \Big(\,\Phi^\dagger_i\, (\not\!\! D \Phi_j)\: -\:
(\not\!\! D \Phi_j)^\dagger\, \Phi_i \Big)\nonumber\\
&&+\; {\bf B}^{(i,j)}_Q\, \Big(\,\Phi_i\, (\not\!\! D
  \Phi_j)^\dagger\: -\: 
(\not\!\! D \Phi_j)\, \Phi^\dagger_i \Big)\,\bigg]\, Q_L\ +\ \dots,
\end{eqnarray}
where the  dots denote  analogous terms for  the right-handed  up- and
down-type quarks $u_R$  and $d_R$.  The first term  depending on ${\bf
Z}_Q$  is a  functional  of $\Phi_{1,2}$  for  the left-handed  quarks
$Q_L$.    Such    a    term    is   not    $\tan\beta$-enhanced    and
renormalization-scheme dependent. As  mentioned above, these terms can
be neglected to a  good approximation. The effective functionals ${\bf
A}^{(i,j)}_Q [\Phi_1,\Phi_2]$  and ${\bf B}^{(i,j)}_Q [\Phi_1,\Phi_2]$
are UV finite and include  large Yukawa-coupling effects due to 
$h_t$.\footnote{These effects have first been identified and studied
 in~\cite{Pepe} within the Standard Model.}
In particular,  this is  the case for  the effective  functionals with
$i=j=2$.  One  typical graph  of such a  contribution is  displayed in
Fig.~\ref{fig:gold}.    Because   of   gauge   invariance,   analogous
contributions  will be  present  in the  one-loop  $Z$- and  $W$-boson
couplings. All these effects are  not enhanced by $\tan\beta$, and can
be consistently neglected without spoiling the gauge symmetries of the
effective Lagrangian ${\cal L}_{\rm FC}$.

In the  next two sections,  we present analytic and  numerical results
related   to   FCNC  $B$-meson   observables,   using  the   effective
Lagrangian~(\ref{LeffFCNC})      and      including      the      2HDM
contribution~(\ref{2HDM}).

\setcounter{equation}{0} 
\section{FCNC {\boldmath $B$}-Meson Observables}
  \label{sec:FCNC}

In this section,  our interest will be in  FCNC $B$-meson observables,
such  as the  $B^0_{d,s}$-$\bar{B}^0_{d,s}$  mass differences  $\Delta
M_{B_{d,s}}$, and  the decays $B_{s,d} \to \mu^+\mu^-$,  $B_u \to \tau
\nu$ and $B \to X_s \gamma$.

\subsection{$\Delta  M_{B_{d,s}}$}

Our discussion  and conventions here follow  closely~\cite{DP}. In the
approximation  of   equal  $B$-meson   lifetimes,  the  SM   and  SUSY
contributions to  $\Delta M_{B_{d,s}}$  may be written  separately, as
follows:
\begin{eqnarray} 
\Delta M_{B_q}\ =\ 2\, 
| \bra{\bar{B}_q^0}\, H_{\rm eff}^{\Delta B=2}\, \ket{B^0_q}_{\rm SM}\ + \
\bra{\bar{B}_q^0}\, H_{\rm eff}^{\Delta B=2}\, \ket{B^0_q}_{\rm SUSY}|\;,
\end{eqnarray}
where  $q \equiv  d,s$  and  $H_{\rm eff}^{\Delta  B=2}$  is the  effective
$\Delta B=2$ Hamiltonian.  Neglecting the subdominant SM contribution,
the SUSY  contributions to the  $\Delta B=2$ transition  amplitudes are
given by
\begin{eqnarray}
  \label{Bsusy}
\bra{\bar{B}^0_d}\,H_{\rm eff}^{\Delta B=2}\,
\ket{B^0_d}_{\rm SUSY} \!\!&=&\!\!  1711 {\rm ~ps}^{-1} 
\Bigg(\,\frac{\hat{B}_{B_d}^{1/2}\, F_{B_d}}{230 {\rm ~MeV}}\, \Bigg)^2\,
\Bigg(\,\frac{\eta_B}{0.55}\,\Bigg) \nonumber\\[3mm] 
&&\hspace{-2.7cm}
\times\, \Big[\, 0.88\, \Big(\,
C_2^{\rm  LR\, (DP)}\: +\: C_2^{\rm  LR\, (2HDM)}\, \Big)\ -\
0.52\, \Big(\, C_1^{\rm  SLL\,(DP)}\: +\: 
C_1^{\rm  SRR\,(DP)}\, \Big)\, \Big]\;, \nonumber \\[3mm]
\bra{\bar{B}^0_s}\, H_{\rm eff}^{\Delta B=2}\, \ket{B^0_s}_{\rm SUSY} 
\!\!&=&\!\! 
2310 {\rm ~ps}^{-1} \Bigg(\,
\frac{\hat{B}_{B_s}^{1/2}\, F_{B_s}}{265 {\rm ~MeV}}\, \Bigg)^2
\Bigg(\,\frac{\eta_B}{0.55}\,\Bigg) \nonumber\\[3mm] 
&&\hspace{-2.7cm}
\times\, \Big[\, 0.88\, \Big(\,
C_2^{\rm  LR\, (DP)}\: +\: C_2^{\rm  LR\, (2HDM)}\, \Big)\ -\
0.52\, \Big(\, C_1^{\rm  SLL\,(DP)}\: +\: 
C_1^{\rm  SRR\,(DP)}\, \Big)\, \Big]\; ,\qquad\quad
\end{eqnarray}
where  DP stands  for  the Higgs-mediated  double-penguin effect.   In
addition,  we  have used  the  next-to-leading  order  QCD factors
determined in~\cite{Jager,Ciuchini,Misiak,Becirevic,Chankowski}, along
with their hadronic matrix elements at the scale $\mu=4.2$~GeV:
\begin{eqnarray}
\bar{P}_1^{\rm LR}\ =\ -0.58 \;,\qquad
\bar{P}_2^{\rm LR}\ =\ 0.88 \;,\qquad
\bar{P}_1^{\rm SLL}\ =\ -0.52 \;,\qquad
\bar{P}_2^{\rm SLL}\ =\ -1.1 \;.
\end{eqnarray}
The Wilson coefficients occurring in~(\ref{Bsusy}) are given by
\begin{eqnarray}
  \label{dpkaon}
C_1^{\rm SLL\,(DP)} &=& -\,\frac{16\pi^2 m_b^2}{\sqrt{2}\, G_F M_W^2}\,
\sum_{i=1}^3\, \frac{{\bf g}_{H_i\bar{b}q}^L\, {\bf g}_{H_i\bar{b}q}^L }{M_{H_i}^2} \;, 
\nonumber \\[3mm]
C_1^{\rm SRR\,(DP)} &=& -\,\frac{16\pi^2 m_q^2}{\sqrt{2}\, G_F M_W^2}\,
\sum_{i=1}^3\, \frac{{\bf g}_{H_i\bar{b}q}^R\, {\bf g}_{H_i\bar{b}q}^R }{M_{H_i}^2} \;,  
\nonumber \\[3mm]
C_2^{\rm LR\,(DP)} &=& -\,\frac{32\pi^2 m_b m_q}{\sqrt{2}\, G_F M_W^2}\,
\sum_{i=1}^3\, \frac{{\bf g}_{H_i\bar{b}q}^L\, {\bf g}_{H_i\bar{b}q}^R }{M_{H_i}^2} \;,
\end{eqnarray}
where the $\tan^2\beta$-enhanced couplings ${\bf g}_{H_i\bar{s}d}^{L,R}$ may
be obtained from~(\ref{LeffFCNC}).   Hence, the DP Wilson coefficients
in~(\ref{dpkaon}) have a  $\tan^4\beta$ dependence and, although 
two-loop suppressed, they become  significant for large values of
$\tan\beta \stackrel{>}{{}_\sim} 40$.

There are  two   relevant  one-loop
contributions       to       $\bra{\bar{B}^0}H_{\rm       eff}^{\Delta
B=2}\ket{B^0}_{\rm  SUSY}$ at large  $\tan\beta$: (i)  the $t$-$H^\pm$
box  contribution  to $C_2^{\rm  LR}$  of the
2HDM  type,   and  (ii)  the  one-loop   chargino-stop  box  diagram
contributing to  $C_1^{\rm  SLL}$.  To a  good approximation, 
$C_2^{\rm LR\,(2HDM)}$ may be given by~\cite{Chankowski}
\begin{eqnarray}
  \label{2HDMC2}
C_2^{\rm LR\,(2HDM)} \approx -\,\frac{2 m_b m_q}{M_W^2}\, 
(V_{tb}^*V_{tq})^2\, \tan^2\beta \;. 
\end{eqnarray}
In the kinematic region $M_{H^\pm}\approx m_t$, the above contribution
can amount to as much as  10\% of the DP effects mentioned above. This
estimate  is   obtained  by  noticing  that   the  light-quark  masses
in~(\ref{dpkaon}) and (\ref{2HDMC2}) are  running and are evaluated at
the top-quark mass  scale, i.e., $m_s(m_t) \simeq 90$  MeV, $m_d (m_t)
\simeq  4$ MeV~\cite{quarkmass}.   The second  contribution~(ii) turns
out   to   be   non-negligible   only   for  small   values   of   the
$\mu$-parameter~\cite{Chankowski},      i.e.,    for      $|\mu     |
\stackrel{<}{{}_\sim} 200$~GeV.

\subsection{$\bar{B}^0_{d,s} \to  \mu^+ \mu^-$}

The leptonic decays of  neutral $B$ mesons, $\bar{B}^0_{d,s} \to \mu^+
\mu^-$,      are      enhanced       at      large      values      of
$\tan\beta$~\cite{FCNC,FCNCCP,Frank,ADKT,Ambrosio,Buras,DP,Bphases,LR,Carena}.
Neglecting  contributions  proportional to  the  lighter quark  masses
$m_{d,s}$, the  relevant effective  Hamiltonian for $\Delta  B=1$ FCNC
transitions is given by
\begin{eqnarray}
  \label{DB1}
H_{\rm eff}^{\Delta B=1}\ =\ -\,2\,\sqrt{2}\, G_F\, V_{tb}V_{tq}^*\,
\Big(\, C_S\, {\cal O}_S\ +\ C_P\, {\cal O}_P\ +\ C_{10}\, {\cal O}_{10}
\Big)\;,
\end{eqnarray}
where 
\begin{eqnarray}
{\cal O}_S &=& \frac{e^2}{16\pi^2}\, m_b\, 
(\bar{q} P_R b)\, (\bar{\mu}\mu) \;, \nonumber \\[2mm]
{\cal O}_P &=& \frac{e^2}{16\pi^2}\, m_b\, (\bar{q} P_R b)\, 
(\bar{\mu}\gamma_5 \mu) \;,\nonumber \\[2mm]
{\cal O}_{10} &=& \frac{e^2}{16\pi^2}\,  (\bar{q}\gamma^\mu P_L b)\,
 (\bar{\mu}\gamma_\mu \gamma_5 \mu) \;.
\end{eqnarray}
Using  the resummed  FCNC  effective Lagrangian~(\ref{LeffFCNC}),  the
Wilson coefficients $C_S$  and $C_P$ in the region  of large values of
$\tan\beta$ are given by
\begin{eqnarray}
  \label{CSCP}
C_S &=& \frac{2 \pi m_\mu}{\alpha_{\rm em}}\, 
\frac{1}{V_{tb} V_{tq}^*}\, \sum_{i=1}^3\, \frac{{\bf g}_{H_i\bar{q}b}^R\, 
g_{H_i\bar{\mu}\mu}^S}{M_{H_i}^2} \ ,\nonumber\\[3mm]
C_P &=& i\, \frac{2 \pi m_\mu}{\alpha_{\rm em}}\, \frac{1}{V_{tb} V_{tq}^*}\,
\sum_{i=1}^3\, \frac{{\bf g}_{H_i\bar{q}b}^R\, 
g_{H_i\bar{\mu}\mu}^P}{M_{H_i}^2} \ ,
\end{eqnarray}
where  $C_{10}=-4.221$  denotes   the  leading  SM  contribution.   In
addition,  the  reduced scalar  and  pseudoscalar  Higgs couplings  to
charged  leptons   $g_{H_i\bar{\mu}\mu}^{S,P}$  in~(\ref{CSCP})  are
given by
\begin{equation}
  \label{eq:HIMM}
g_{H_i \bar{\mu} \mu}^S\ =\ \frac{O_{1i}}{\cos\beta}\ , \qquad
g_{H_i \bar{\mu} \mu}^P\ =\ -\,\tan\beta\, O_{3i} \; .
\end{equation}
Here  we neglect the  non-holomorphic vertex  effects on  the leptonic
sector since they are unobservably small.

Taking  into  consideration  the  aforementioned  approximations,  the
branching  ratio  for $\bar{B}^0_{d,s}  \to  \mu^+\mu^-$  is found  to
be~\cite{Frank}
\begin{eqnarray}
  \label{Bll}
B(\bar{B}^0_q \to \mu^+ \mu^-) &=& \\
&&\hspace{-2cm}
\frac{G_F^2 \alpha_{\rm em}^2}{16\pi^3}\, M_{B_q} \tau_{B_q}\, 
|V_{tb}V_{tq}^*|^2\, \sqrt{1-\frac{4 m_\mu^2}{M_{B_q}^2}}\
\Bigg[\, \Bigg(\,1-\frac{4 m_\mu^2}{M_{B_q}^2} \Bigg)\, |F^q_S|^2\ +\
|F_P^q\: +\: 2 m_\mu F_A^q|^2\, \Bigg] \;,\nonumber
\end{eqnarray}
where $q  = d,s$ and $\tau_{B_q}$  is the total lifetime  of the $B_q$
meson. Moreover, the form factors $F^q_{S,P,A}$ are given by
\begin{eqnarray}
  \label{FSP}
F_{S,P}^q\ =\ -\,\frac{i}{2}\, M_{B_q}^2 F_{B_q}\, 
\frac{m_b}{m_b+m_q}\, C_{S,P}\;,\qquad 
F_A^q\ =\ -\,\frac{i}{2}\,F_{B_q}\, C_{10} \;.
\end{eqnarray}
Although the Wilson coefficient $C_{10}$ is subdominant for $\tan\beta
\stackrel{>}{{}_\sim}  40$,  its  effect  has  been  included  in  our
numerical estimates.

\subsection{$B_u \to \tau \nu$}

There is  an important tree-level charged-Higgs  boson contribution to
$B_u  \to  \tau \nu$  decay~\cite{Akeroyd:2003zr,Itoh:2004ye}.   It  is not  helicity
suppressed  and  interferes  destructively  with the  SM  contribution
~\cite{Hou:1992sy}.  The ratio of the  branching ratio to the SM value
is given by
\begin{equation}
R_{B\tau\nu}\ =\
\frac{B(B^-\to \tau^- \bar{\nu})}{B^{\rm SM}(B^-\to \tau^-
  \bar{\nu})}\ =\
\left|1+\tan\beta\,\frac{({\bf
      g}^{L\,\dagger}_{H^-\bar{d}u})_{13}}{{\bf V}_{13}}\, 
\left(\frac{M_{B^\pm}}{M_{H^\pm}}\right)^2\right|^2\,,
\end{equation}
where  ${\bf g}^L_{H^-\bar{d}u}=-\tan\beta\,{\bf  V}^\dagger$  at tree
level  [cf.~(\ref{gLud})], leading to  the negative  interference with
the SM contribution.


\subsection{$B \to X_s \gamma$}

The relevant effective Hamiltonian for $B \to X_s \gamma$ is given by
\begin{equation}
H_{\rm eff}^{b\to s\gamma} = -\frac{4G_F}{\sqrt{2}}\,V_{tb}V^*_{ts}\,\left\{
\sum_{i=2,7,8}C_i(\mu_b){\cal O}_i(\mu_b)+
C^\prime_7(\mu_b){\cal O}^\prime_7(\mu_b)+
C^\prime_8(\mu_b){\cal O}^\prime_8(\mu_b)
\right\} ,
\end{equation}
with
\begin{eqnarray}
{\cal O}_2&=& \bar{s}_L \gamma_\mu c_L\, \bar{c}_L\gamma^\mu b_L\,,
\nonumber \\
{\cal O}_7&=& \frac{e\,m_b}{16\,\pi^2}\,\bar{s}_L \sigma_{\mu\nu} F^{\mu\nu} b_R\,;
\ \ \ \ \
{\cal O}^\prime_7= \frac{e\,m_b}{16\,\pi^2}\,\bar{s}_R \sigma_{\mu\nu} F^{\mu\nu} b_L\,,
\nonumber \\
{\cal O}_8&=& \frac{g_s\,m_b}{16\,\pi^2}\,\bar{s}_L \sigma_{\mu\nu} F^{\mu\nu} b_R\,;
\ \ \ \ \
{\cal O}^\prime_8= \frac{g_s\,m_b}{16\,\pi^2}\,\bar{s}_R \sigma_{\mu\nu} F^{\mu\nu} b_L\,.
\end{eqnarray}
We closely follow the calculations of Refs.~\cite{KN} for 
the branching ratio $B(B\rightarrow X_s \, \gamma)$ 
and the direct CP asymmetry in the decay. 
For the running $c$ quark mass, we use $m_c(m_c^{\rm pole})$ to
capture a part of NNLO corrections \cite{b2sg-nnlo}.
We refer to, for example, Appendix B of Ref.~\cite{microomegas} for the detailed
expression of the branching ratio in terms of the Wilson coefficients which we
are gong to present below.

The LO charged-Higgs contribution is given by
\begin{eqnarray}
C_{7,8}^{(0)\,H^\pm}(M_W)=\frac{1}{3}\,
\frac{({\bf g}^{R\,\dagger}_{H^-\bar{d}u})_{33}}{{\bf V}_{33}}\,
\frac{({\bf g}^R_{H^-\bar{d}u})_{23}}{{\bf V}^\dagger_{23}}\,
F^{(1)}_{7,8}(y)+
\frac{({\bf g}^{L\,\dagger}_{H^-\bar{d}u})_{33}}{{\bf V}_{33}}\,
\frac{({\bf g}^R_{H^-\bar{d}u})_{23}}{{\bf V}^\dagger_{23}}\,
F^{(2)}_{7,8}(y) ,
\end{eqnarray}
where $y  \equiv \overline{m}^2_t(M_W)/M_{H^\pm}^2$, the  ratio of the
top-quark running mass  at the scale $M_W$ to  the charged Higgs-boson
pole   mass.   In  the   numerical  analysis,   we  include   the  NLO
contribution.  Note  that ${\bf g}^R_{H^-\bar{d}u}=-t^{-1}_\beta\,{\bf
V}^\dagger$ and ${\bf g}^L_{H^-\bar{d}u}=-t_\beta\,{\bf V}^\dagger$ at
tree  level, see  Eqs.~(\ref{gLud}) and  (\ref{gRud}).   The functions
$F^{(1),(2)}_{7,8}$ can be found in Ref.\cite{b2sg-ch,microomegas}.

The chargino contributions are
\begin{eqnarray}
C_{7,8}^{\,\chi^\pm}(\mu_{\rm SUSY}) &=&
\sum_{i=1,2}\left\{ 
\frac{2}{3}\,\frac{M_W^2}{\widetilde{m}_q^2}\,
\left|(C_R)_{i1}\right|^2\,F_{7,8}^{(1)}(x_{\widetilde{q}\chi^-_i})
\right.  \nonumber \\
&& \hspace{1.0cm}
-\,\frac{({\bf V}^\dagger {\bf R}_d^{-1})^\dagger_{13}\,{\bf V}^\dagger_{21}+
({\bf V}^\dagger {\bf R}_d^{-1})^\dagger_{23}\,{\bf V}^\dagger_{22}}
{c_\beta\,{\bf V}_{33}\,{\bf V}^\dagger_{23}}\,
\frac{(C_L)_{i2}\,(C_R)_{i1}^*\,M_W}{\sqrt{2}\,m_{\chi_i^-}}
F_{7,8}^{(3)}(x_{\widetilde{q}\chi^-_i}) 
\nonumber \\
&&
-\frac{2}{3}\sum_{j=1,2}\left|(C_R)_{i1}(U^{\widetilde{t}}_{1j})^*
-\frac{({\bf \widehat{M}}_u{\bf R}_u^{-1})_{33}}{\sqrt{2}\,s_\beta\,M_W}
(C_R)_{i2}(U^{\widetilde{t}}_{2j})^*\right|^2
\frac{M_W^2}{m_{\widetilde{t}_j}^2}
\,F_{7,8}^{(1)}(x_{\widetilde{t}_j\chi^-_i}) 
\nonumber \\
&&
+\frac{({\bf V}^\dagger {\bf R}_d^{-1})^\dagger_{33}}{c_\beta\,{\bf V}_{33}}
\sum_{j=1,2}
\left( -\frac{(C_L)_{i2}\,(C_R)_{i1}^*\,M_W}{\sqrt{2}\,m_{\chi_i^-}}
\left|U^{\widetilde{t}}_{1j}\right|^2\,
\right.
\nonumber \\
&& \hspace{1.0cm}\left.\left.
+(U^{\widetilde{t}}_{1j})^*\,U^{\widetilde{t}}_{2j}\,
\frac{(C_L)_{i2}\,(C_R)_{i2}^*\,({\bf \widehat{M}}_u{\bf
R}_u^{-1})^\dagger_{33}}{2\,s_\beta\,m_{\chi_i^-}}\,\right)
F_{7,8}^{(3)}(x_{\widetilde{t}_j\chi^-_i})
\right\}\,,
\label{eq:c78_chargino}
\end{eqnarray}
where  $x_{ij}\equiv m_i^2/m_j^2$.   We  refer to~\cite{b2sg-chargino}
for  the  functions  $F^{(3)}_{7,8}$  and to~\cite{cpsuperh}  for  the
chargino mixing matrices $C_{L,R}$ and the stop mixing matrix
$U^{\widetilde{t}}$.

Finally, the gluino contributions to the Wilson coefficients $C_{7,8}$
are given by
\begin{eqnarray}
C_7^{\widetilde{g}}(\mu_{\rm SUSY})&=&
-\,\frac{8\pi\alpha_s}{9\sqrt{2}G_F|M_3|^2\lambda_t}\,\sum_{i=1}^6\,x_i\,
(G^d_L)_{i2}^{*} \nonumber \\ &&
\hspace{3.0cm}
\times \left[(G^d_L)_{i3}f_2(x_i)+(G^d_R)_{i3}\frac{M_3}{m_b}f_4(x_i)\right]\,,
\nonumber \\ 
C_8^{\widetilde{g}}(\mu_{\rm SUSY})&=&
-\,\frac{\pi\alpha_s}{\sqrt{2}G_F|M_3|^2\lambda_t}\,\sum_{i=1}^6\,x_i\, (G^d_L)_{i2}^{*}
\left\{ (G^d_L)_{i3}\left[3f_1(x_i)+\frac{1}{3}f_2(x_i)\right] \right.
\nonumber \\ && \left.
\hspace{3.0cm}
+(G^d_R)_{i3}\frac{M_3}{m_b}\left[3f_3(x_i)+\frac{1}{3}f_4(x_i)\right]\right\}\,,
\end{eqnarray}
where    $\lambda_t\equiv    {\bf    V}_{33}   {\bf    V}^\dagger_{23}
=V_{tb}V_{ts}^*$  and $x_i\equiv  |M_3|^2/m^2_{\widetilde{d}_i}$.  The
loop     functions    $f_{1,2,3,4}(x_i)$     may    be     found    in
Ref.~\cite{b2sg-gluino}.   The  Wilson  coefficients  for  the  primed
operators  ${\cal O}_{7,8}^\prime$  can  be obtained  by the  exchange
$L\leftrightarrow R$ and $M_3 \rightarrow M_3^*$:
\begin{eqnarray}
C_7^{\prime\,\widetilde{g}}(\mu_{\rm SUSY})&=&
-\,\frac{8\pi\alpha_s}{9\sqrt{2}G_F|M_3|^2\lambda_t}\,\sum_{i=1}^6\,x_i\,
(G^d_R)_{i2}^{*}\nonumber \\ && 
\hspace{3.0cm}
\times \left[(G^d_R)_{i3}f_2(x_i)+(G^d_L)_{i3}\frac{M_3^*}{m_b}f_4(x_i)\right]\,,
\nonumber \\
C_8^{\prime\,\widetilde{g}}(\mu_{\rm SUSY})&=&
-\,\frac{\pi\alpha_s}{\sqrt{2}G_F|M_3|^2\lambda_t}\,\sum_{i=1}^6\,x_i\, (G^d_R)_{i2}^{*}
\left\{ (G^d_R)_{i3}\left[3f_1(x_i)+\frac{1}{3}f_2(x_i)\right] \right.
\nonumber \\ && \left.
\hspace{3.0cm}
+\ (G^d_L)_{i3}\frac{M_3^*}{m_b}\left[3f_3(x_i)+\frac{1}{3}f_4(x_i) 
\right]\right\}\,. 
\end{eqnarray}
In  the   above,  $C_{7,8}^{(\prime)\,\widetilde{g}}$,  the  down-type
squark-gluino-quark  couplings  $G_{L,R}^d$  are defined  through  the
interaction Lagrangian (suppressing the colour indices)
\begin{eqnarray}
{\cal L}_{\widetilde{d}\,\widetilde{g}\,d}&=&-\sqrt{2}\,g_s\,\Big\{\,
\widetilde{d}_i^{\,*}\,t^a\,\overline{\widetilde{g}^a}\left[
(G^d_L)_{i\alpha}\,P_L\,+\,(G^d_R)_{i\alpha}\,P_R \right]\,d_\alpha 
\nonumber \\
&& \hspace{1.5cm}\,+\, 
\overline{d_\alpha}\left[ (G^d_L)^*_{i\alpha}\,P_R\,+\,
(G^d_R)^*_{i\alpha}\,P_L \right]\,\widetilde{g}^a\,t^a\,\widetilde{d}_i
\,\Big\}\,,
\end{eqnarray}
where $t^a$  are the  usual Gell-Mann matrices,  $i=1,2,\dots,6$ label
the mass eigenstates of down-type squarks, and $\alpha=1,2,3$ the mass
eigenstates  of  down-type quarks.  The  couplings  are  given by  the
down-type squark mixing matrix as
\begin{equation}
(G^d_L)_{i\alpha}= 
\left(U^{\widetilde{d}\,\dagger}\right)_{i\,\alpha}\,, 
\ \ \ \
(G^d_R)_{i\,\alpha}= 
-\left(U^{\widetilde{d}\,\dagger}\right)_{i\,\alpha+3}\,.
\end{equation}
The $6\times 6$ unitary matrix $U^{\widetilde{d}}$ diagonalizes the
down-type squark mass matrix as
\begin{equation}
U^{\widetilde{d}\,\dagger}\,{\bf M}^2_{\widetilde{d}}\,U^{\widetilde{d}}\ =\ {\rm
diag}(m^2_{\widetilde{d}_1}\, ,\,m^2_{\widetilde{d}_2}\, ,\dots\, ,
\, m^2_{\widetilde{d}_6}) ,
\end{equation}
where $\widetilde{d}_1$ is the lightest and $\widetilde{d}_6$ the heaviest.
In the super-CKM basis, in which the down squarks are aligned with the down
quarks and ${\bf U}^Q_L = {\bf U}^u_R = {\bf U}^d_R = {\bf 1}$,
the $6\times 6$ down-type squark mass matrix ${\bf
  M}^2_{\widetilde{d}}$ takes on the form 
\begin{equation}
{\bf M}^2_{\widetilde{d}}=\left(
\begin{array}{cc}
{\bf V}^\dagger \, \widetilde{\bf M}^2_{LL}\,{\bf V} & 
{\bf V}^\dagger \, \widetilde{\bf M}^2_{LR} \\
\widetilde{\bf M}^2_{RL}\,{\bf V} & 
\widetilde{\bf M}^2_{RR}
\end{array}\right) ,
\end{equation}
where the $3\times 3$ submatrices are given by
\begin{eqnarray}
\widetilde{\bf M}^2_{LL} &=& {\bf \widetilde{M}}^2_Q 
\,+\,\frac{v_1^2}{2}\, ({\bf h}_d^\dagger {\bf h}_d)
\,+\,c_{2\beta}\,M_Z^2\,\left(-\frac{1}{2}+\frac{1}{3}s_W^2\right)\,{\bf 1}\, ,
\nonumber \\
\widetilde{\bf M}^2_{LR} &=& \frac{1}{\sqrt{2}}\,{\bf a}_d^\dagger\,v_1 
\,-\,\frac{1}{\sqrt{2}}\,{\bf h}_d^\dagger\,\mu\, v_2\, ,
\nonumber \\
\widetilde{\bf M}^2_{RL} &=& \frac{1}{\sqrt{2}}\,{\bf a}_d\,v_1
\,-\,\frac{1}{\sqrt{2}}\,{\bf h}_d\,\mu^*\, v_2\, ,
\nonumber \\
\widetilde{\bf M}^2_{RR} &=&{\bf \widetilde{M}}^2_D
\,+\,\frac{v_1^2}{2}\, ({\bf h}_d {\bf h}_d^\dagger)
\,+\,c_{2\beta}\,M_Z^2\,\left(-\frac{1}{3}s_W^2\right)\,{\bf 1}\, ,
\end{eqnarray}
with     ${\bf     h}_d=\frac{\sqrt{2}}{v_1}\,{\bf\widehat{M}}_d\,{\bf
V^\dagger}\,{\bf R}_d^{-1}$.   As a byproduct of  the chosen super-CKM
basis,   we  observe   the  absence   of  flavour   mixing   in  ${\bf
M}_{\widetilde{d}}^2$, for all ${\bf h}_d$-dependent terms, when ${\bf
R}_d \propto {\bf 1}$.

\setcounter{equation}{0} 
\section{Numerical Examples}
  \label{sec:num}

For our numerical estimates  of FCNC observables at large $\tan\beta$,
we take the GUT  scale to be the same as in  the usual CMSSM with MFV,
and  a  dedicated program  has  been  developed  to calculate  the  RG
evolution  from the  GUT scale  to the  low-energy SUSY  scale  in the
MCPMFV framework  of the  MSSM.  For the  Higgs mass spectrum  and the
mixing matrix  $O_{\alpha i}$  at the $M_{\rm  SUSY}$ scale,  the code
{\tt CPsuperH}  \cite{cpsuperh} has been used.  In  the calculation of
the   flavour-changing   effective   couplings,   only   the   leading
contributions   have   been   kept   in   the   single-Higgs-insertion
approximation, neglecting the EW corrections and the generically small
flavour-off-diagonal elements of the squark mass matrices.

In order  to study  the effects of  CP-violating phases in  the MCPMFV
framework,  we consider  a  CP-violating variant  of  a typical  CMSSM
scenario:
\begin{eqnarray}
&&\left|M_{1,2,3}\right|=250~~{\rm GeV}\,, \nonumber \\
&&M^2_{H_u}=M^2_{H_d}=\widetilde{M}^2_Q=\widetilde{M}^2_U=\widetilde{M}^2_D
=\widetilde{M}^2_L=\widetilde{M}^2_E=(100~~{\rm GeV})^2\,, \nonumber \\
&&\left|A_u\right|=\left|A_d\right|=\left|A_e\right|=100~~{\rm GeV}\,,
\label{eq:cpsps1a}
\end{eqnarray}
at the GUT scale with $\tan\beta\,(M_{\rm SUSY})=10$, which corresponds to
$\tan\beta\,(m_t^{\rm pole})\simeq10.2$. As for the CP-violating phases,
we adopt the convention that $\Phi_\mu=0^\circ$,
and we vary the following three phases:
\begin{equation}
\Phi_{12}\equiv\Phi_1=\Phi_2\,; \ \ \ \Phi_{3}\,; \ \ \
\Phi_{A}^{\rm GUT}\equiv\Phi_{A_u}=\Phi_{A_u}=\Phi_{A_e} ,
\end{equation}
where,  for simplicity, common  phases $\Phi_{12}$  and $\Phi_{A}^{\rm
GUT}$  are  taken  for  the  phases of  $M_{1,2}  (M_{\rm  GUT})$  and
$A_{u,d,e} (M_{\rm  GUT})$, respectively.  We note that  the phases of
the gaugino mass parameters,  $\Phi_{1,2,3}$, and the $\mu$ parameter,
$\Phi_\mu$, are  unchanged by the  RG evolution, whilst the  phases of
the elements  of the matrix  ${\bf A}_{u,d,e}$ could  be significantly
different at low scales from the  values given at the GUT scale.  This
scenario      becomes     the     SPS1a      point~\cite{SPS}     when
$\Phi_{1,2,3}=0^\circ$  and  $\Phi^{\rm  GUT}_A=180^\circ$.   We  have
found  that $M_{\rm SUSY}$  varies between  530 GeV  and 540  GeV, and
$M_{\rm GUT}/10^{16}~{\rm  GeV}$ between 1.825 and  1.838 depending on
the values of the CP-violating phases.

We  do  not  consider  in  this section  the  electric  dipole  moment
constraints~\cite{EDMs} on the MCPMFV  parameter space of the MSSM.  A
systematic implementation of these constraints and their impact on the
FCNC observables will be given in a forthcoming communication.

\subsection{Phases and Masses}

We first consider the (3,3) elements $A_{f_3}\equiv({\bf
a}_f)_{33}/({\bf h}_f)_{33}$ at $M_{\rm SUSY}$ with $f=u,d,e$ and
$f_3=t,b,\tau$. We find that the complex quantity $A_{f_3}$ can be
written in terms of the complex $A_f$ and $M_j$ at the GUT scale as:
\begin{equation}
A_{f_3}(M_{\rm SUSY})\approx C^{A_f}_{f_3}\,A_f(M_{\rm GUT})-
C^{M_i}_{f_3}\,M_i(M_{\rm GUT}) ,
\label{eq:amsusy}
\end{equation}
where  the real coefficients  $C^{A_f}_{f_3}$ and  $C^{M_i}_{f_3}$ are
functions  of  the Yukawa  and  gauge  couplings.  This expression  is
similar to that found in Ref.~\cite{Goto}.
In    general,    $C^{A_{u,d}}_{t,b}$    are   much    smaller    than
$C^{M_{3}}_{t,b}$.    Indeed,    they    are   even    smaller    than
$C^{M_{1,2}}_{t,b}$ with  $C^{A_u}_{t} < C^{A_d}_{b}$.   For $A_\tau$,
$C^{A_e}_{\tau}$  is not  so much  smaller  than $C^{M_{1,2}}_{\tau}$,
whilst $C^{M_{3}}_{\tau}$  is negligible.  This is because  the strong
coupling  amplifies the  influence  of $M_3$, while  the large  Yukawa
couplings suppress those of  the $A$ terms via renormalization effects
\cite{Goto}.
For  the parameter  set  (\ref{eq:cpsps1a}) with  $\tan\beta =10$,  we
observe   that    the   phases   $\Phi_{A_{t}}(M_{\rm    SUSY})$   and
$\Phi_{A_{b}}(M_{\rm  SUSY})$  are  largely  determined  by  $\Phi_3$,
whereas the phase $\Phi_{A_{\tau}}(M_{\rm  SUSY})$ is more affected by
$\Phi_{1,2}$ than by $\Phi^{\rm GUT}_A$.
This  situation becomes  different for  larger values  of $\tan\beta$,
i.e.~we find  that $C^{M_3}_\tau$ becomes  significant and $C^{A_d}_b$
decreases when $\tan\beta$ increases.

\begin{figure}[ht]
\hspace{ 0.0cm}
\centerline{\epsfig{figure=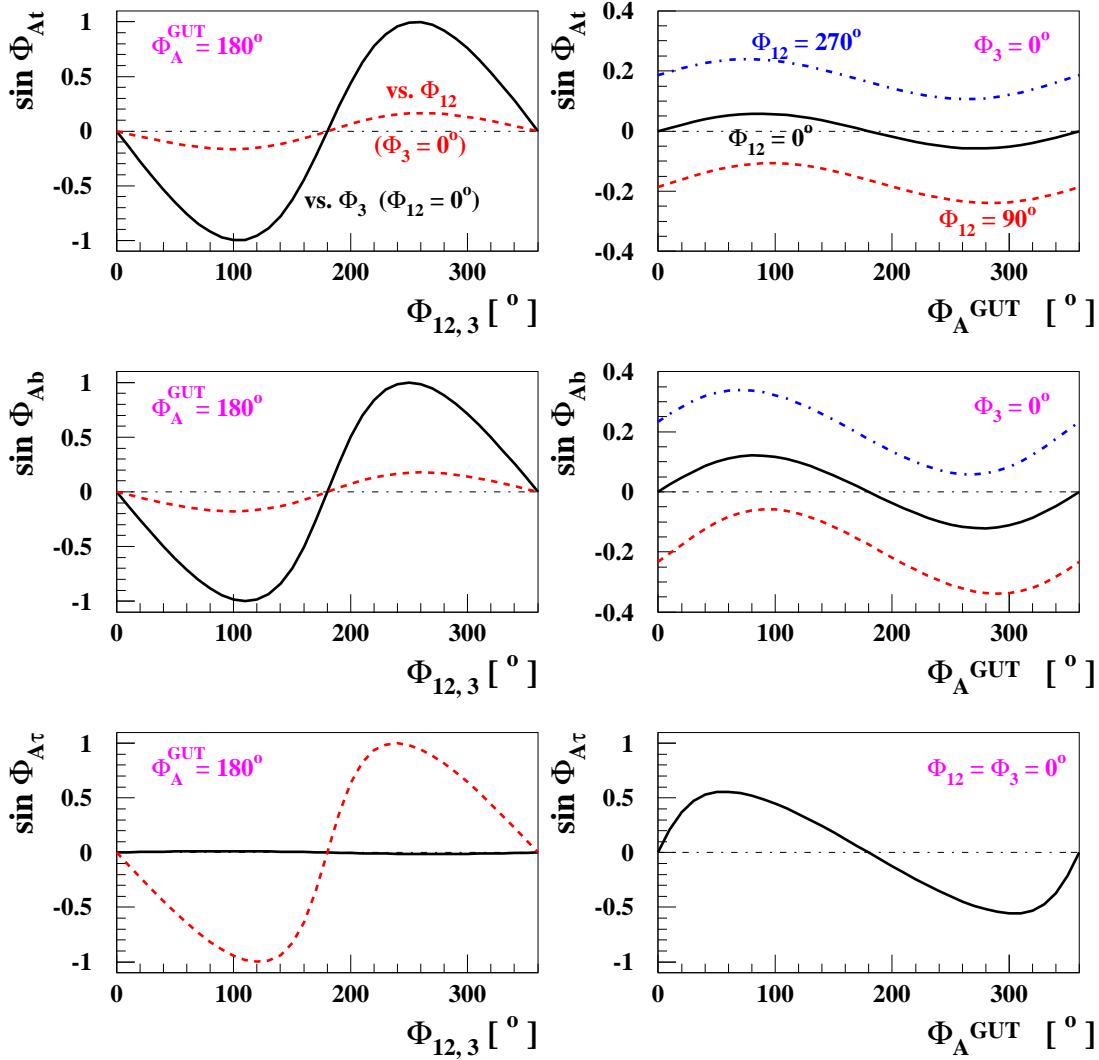,height=16cm,width=16cm}}
\vspace{-0.9cm}
\caption{\it  In the left frames, taking $\Phi^{\rm GUT}_A=180^\circ$,
$\sin\Phi_{A_{t}}$ (upper),
$\sin\Phi_{A_{b}}$ (middle), and
$\sin\Phi_{A_{\tau}}$ (lower) are shown
as functions of $\Phi_{3}$ taking $\Phi_{12}=0^\circ$ (solid lines)
and $\Phi_{12}$ taking $\Phi_3=0^\circ$ (dashed lines). 
In the right frames they are shown as functions of
$\Phi^{\rm GUT}_A$ taking  $\Phi_3=0^\circ$ or $\Phi_{12}=0^\circ$.
For $\sin\Phi_{A_{t}}$ and $\sin\Phi_{A_{b}}$, 
three cases are shown: $\Phi_{12}=270^\circ$ (blue dash-dotted lines),
$0^\circ$ (black solid lines), and
$90^\circ$ (red dashed lines).
For $\sin\Phi_{A_{\tau}}$, we set $\Phi_3=0^\circ$ as well.
The parameters are taken as in Eq.~(\ref{eq:cpsps1a}) with $\tan\beta(M_{\rm
SUSY})=10$.
}
\label{fig:phia_msusy}
\end{figure}

In     Fig.~\ref{fig:phia_msusy}    we     show    $\sin\Phi_{A_{t}}$,
$\sin\Phi_{A_{b}}$,  and $\sin\Phi_{A_{\tau}}$  for the  parameter set
(\ref{eq:cpsps1a})  with $\tan\beta(M_{\rm  SUSY})=10$.   In the  left
frames, we observe that  $\Phi_{A_{t,b}}$ and $\Phi_{A_{\tau}}$ can be
fully  generated from  $\Phi_3$ and  $\Phi_{1,2}$,  respectively, even
when   $A_{u,d,e}$   at   the   GUT   scale   are   real,   $\Phi^{\rm
GUT}_A=180^\circ$.  Whilst  the  dependence  of  $\Phi_{A_{\tau}}$  on
$\Phi_{3}$  is negligible (solid  line in  the left-lower  frame), the
dependences  of  $\Phi_{A_{t,b}}$  on  $\Phi_{1,2}$  can  be  sizeable
(dashed lines in the left-upper and left-middle frames).
In    the   right    frames,   the    cases    with   $\Phi_3=0^\circ$
($\Phi_{A_{t,b}}$)  and  $\Phi_{12}=0^\circ$  ($\Phi_{A_{\tau}}$)  are
considered, showing  how large the  $A$-term phases may become  at the
$M_{\rm SUSY}$ scale for real  $M_3$ and/or real $M_1$ and $M_2$. When
the   gaugino   masses   are   all  real,   $|\sin\Phi_{A_{t}}|$   and
$|\sin\Phi_{A_{b}}|$ turn  out to be $0.06$  and $0.12$, respectively,
whereas $|\sin\Phi_{A_{\tau}}|$  can be  as large as  $0.55$. Somewhat
larger  CP-violating  phases   are  possible  for  $\Phi_{A_{t}}$  and
$\Phi_{A_{b}}$ when $M_1$ and $M_2$ are pure imaginary (see dashed and
dash-dotted  lines  in  the  right-upper and  right-middle  frames  of
Fig.~\ref{fig:phia_msusy}). Finally,  there are no  visible effects of
$\Phi_3$ on $\Phi_{A_{\tau}}$.

\begin{figure}[ht]
\hspace{ 0.0cm}
\centerline{\epsfig{figure=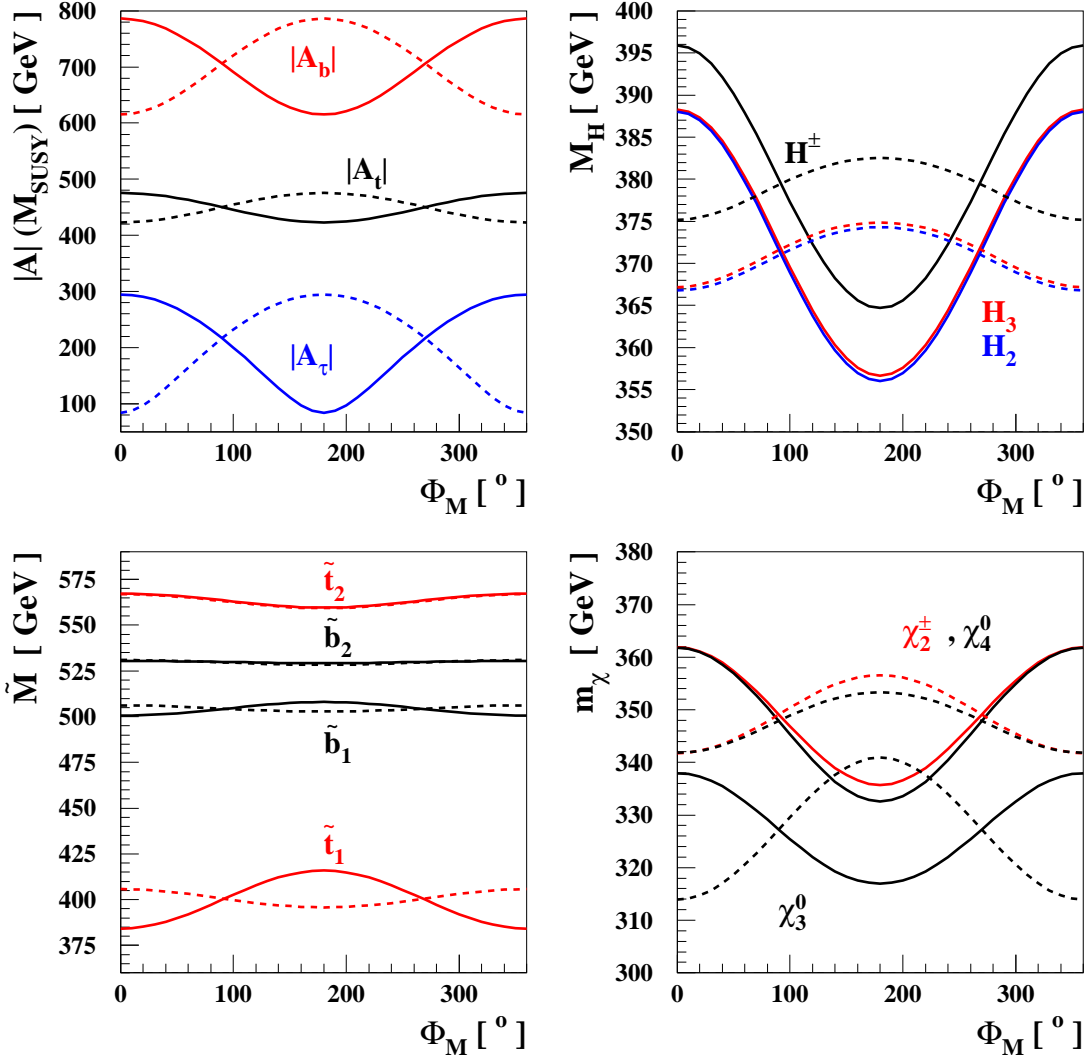,height=16cm,width=16cm}}
\vspace{-0.9cm}
\caption{\it The absolute values of $A_{t,b,\tau}$ (upper-left) and
the masses of the heavy Higgs bosons (upper right), sbottoms and stops
(lower left), and charginos and neutralinos (lower right) as functions
of a common phase $\Phi_M\equiv\Phi_1=\Phi_2=\Phi_3$.  The solid lines
are for $\Phi_A^{\rm GUT}=180^\circ$ and the dashed lines for
$\Phi_A^{\rm GUT}=0^\circ$.  The parameters are listed in
Eq.~(\ref{eq:cpsps1a}).  }
\label{fig:masses}
\end{figure}

We now discuss the effects of CP-violating phases on the masses of Higgs
bosons, third-generation squarks and heavy neutralinos and chargino.
In the upper-left frame of Fig.~\ref{fig:masses}, we show the absolute
values   of   $A_{t,b,\tau}$   as   functions  of   a   common   phase
$\Phi_M\equiv\Phi_1=\Phi_2=\Phi_3$  for  two  values  of  $\Phi_A^{\rm
GUT}$: $0^\circ$ (dashed lines) and $180^\circ$ (solid lines). In this
case,  one can show  the absolute  values squared  depend only  on the
difference $\Phi_A^{\rm GUT}-\Phi_M$:
\begin{equation}
|A_f|^2\approx\alpha_f -\beta_f\cos(\Phi_A^{\rm GUT}-\Phi_M) ,
\end{equation}
using Eq.~(\ref{eq:amsusy}), with $\alpha_f\,,\beta_f > 0$. 
From   Fig.~\ref{fig:masses},  we   observe  that   there   is  strong
correlation between  $|A_{t,b,\tau}|$ and the  particle mass spectrum.
This correlation  is due to  the phase-dependent terms  ${\rm Tr}({\bf
a}^\dagger_u{\bf a}_u)$ and  ${\rm Tr}({\bf a}^\dagger_d{\bf a}_d)$ in
$dM_{H_u,H_d}^2/dt$  and $d\widetilde{\bf M}^2_{Q,U,D}/dt$.   The fact
that   $|M_{H_u}^2|$  decreases   (increases)   when  ${\rm   Tr}({\bf
a}^\dagger_u{\bf  a}_u)$  decreases  (increases) explains  the  CP-odd
phase dependence  of heavier Higgs-boson  masses, as can be  seen from
the upper-right frame  of Fig.~\ref{fig:masses}.  The same correlation
is observed for the heavy  chargino and neutralinos in the lower-right
frame of Fig.~\ref{fig:masses}, since a decreased (increased) value of
$|M_{H_u}^2|$ leads  to smaller (larger)  values of $|\mu|$.   We find
that the  variations in the masses  of the lightest  Higgs boson $H_1$
and the lightest neutralino $\tilde{\chi}^0_{1}$ amount to 2 GeV and 3
GeV, respectively.
The   CP-odd    phase   dependences   of    $\widetilde{\bf   M}^2_Q$,
$\widetilde{\bf  M}^2_U$,  and $\widetilde{\bf  M}^2_D$  at the  scale
$M_{\rm SUSY}$ can be understood similarly.  Here the (3,3) components
of  the   mass  matrices  decrease  (increase)   when  ${\rm  Tr}({\bf
a}^\dagger_u{\bf a}_u)$  increases (decreases).  For  the chosen value
of    $\tan\beta(M_{\rm   SUSY})=10$,    the   (3,3)    component   of
$\widetilde{\bf    M}^2_U$   shows    the   largest    effect,   since
$d\widetilde{\bf     M}^2_U/dt$     contains     $2\,{\rm     Tr}({\bf
a}^\dagger_u{\bf  a}_u)$ compared  to ${\rm  Tr}({\bf a}^\dagger_u{\bf
a}_u)+{\rm   Tr}({\bf  a}^\dagger_d{\bf  a}_d)$   in  $d\widetilde{\bf
M}^2_Q/dt$   and   $2\,{\rm   Tr}({\bf  a}^\dagger_d{\bf   a}_d)$   in
$d\widetilde{\bf M}^2_D/dt$.   Furthermore, we note  that $\tilde{t}_1
\sim  \tilde{t}_R$  and  $\tilde{b}_1  \sim \tilde{b}_L$.  From  these
observations,  one   can  understand  the   qualitative  CP-odd  phase
dependence of the stop and  sbottom masses, as shown in the lower-left
frame of Fig.~\ref{fig:masses}.

\subsection{Effects on $\Delta M_{B_s}$ and $\Delta M_{B_d}$}

\begin{figure}[ht]
\hspace{ 0.0cm}
\centerline{\epsfig{figure=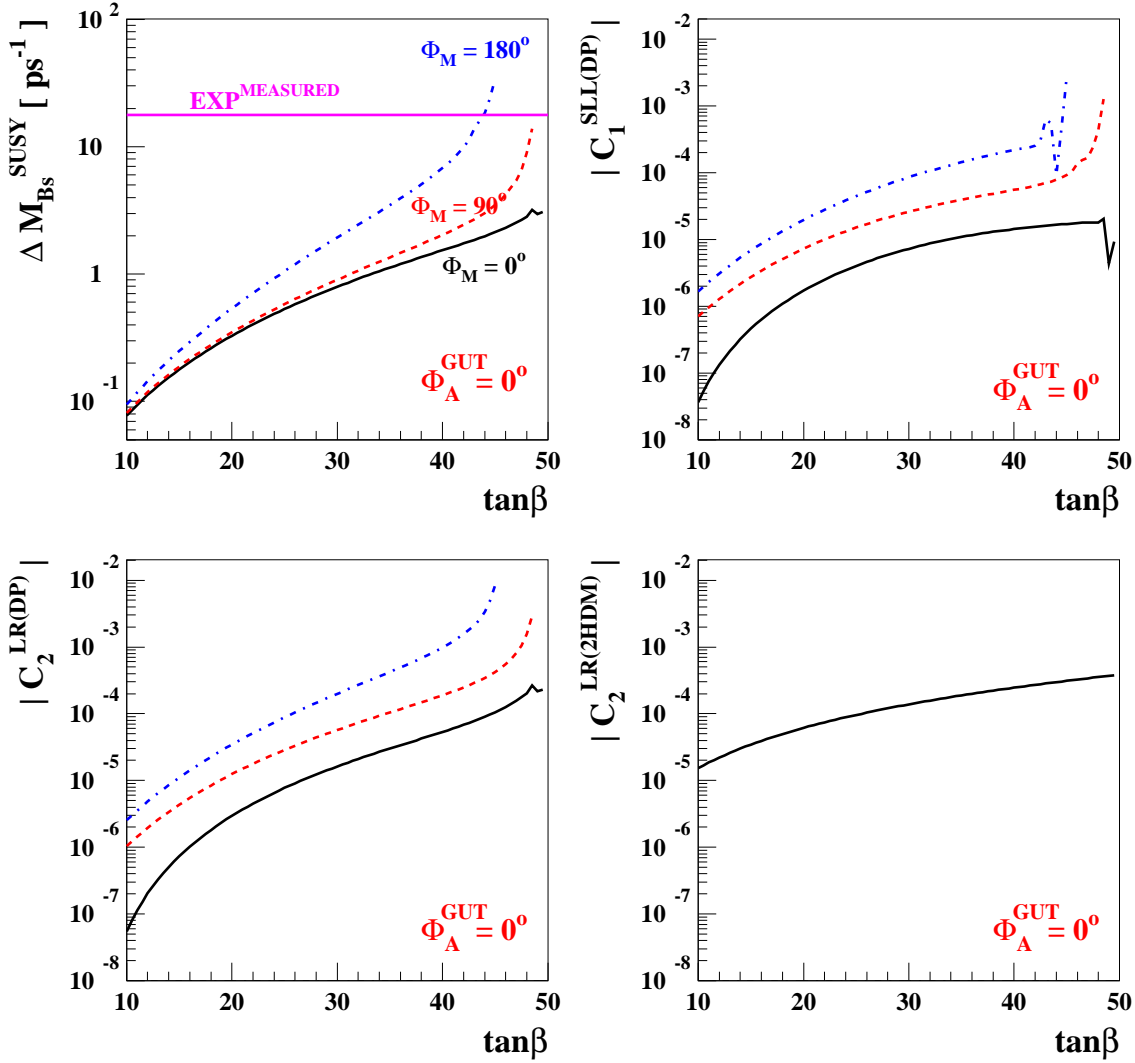,height=16cm,width=16cm}}
\vspace{-0.9cm}
\caption{\it  The SUSY contribution  to $\Delta  M_{B_s}$ in  units of
${\rm ps}^{-1}$  (upper-left) and the relevant couplings  in the other
three  frames, as  functions of  $\tan\beta(M_{\rm SUSY})$,  for three
values of the common phase: $\Phi_M=0^\circ$ (solid lines), $90^\circ$
(dashed   lines),  and  $180^\circ$   (dash-dotted  lines).    We  fix
$\Phi_A^{\rm  GUT}=0^\circ$  and  the   parameters  are  taken  as  in
Eq.~(\ref{eq:cpsps1a}),     except     that     here     we     choose
$\widetilde{M}_{L,E}=200$ GeV so as to avoid a very light or tachyonic
$\tilde{\tau}_1$  state  for  large  $\tan\beta$.  In  the  upper-left
frame, we show the currently measured value as the horizontal line.}
\label{fig:dmbs}
\end{figure}
\begin{figure}[ht]
\hspace{ 0.0cm}
\centerline{\epsfig{figure=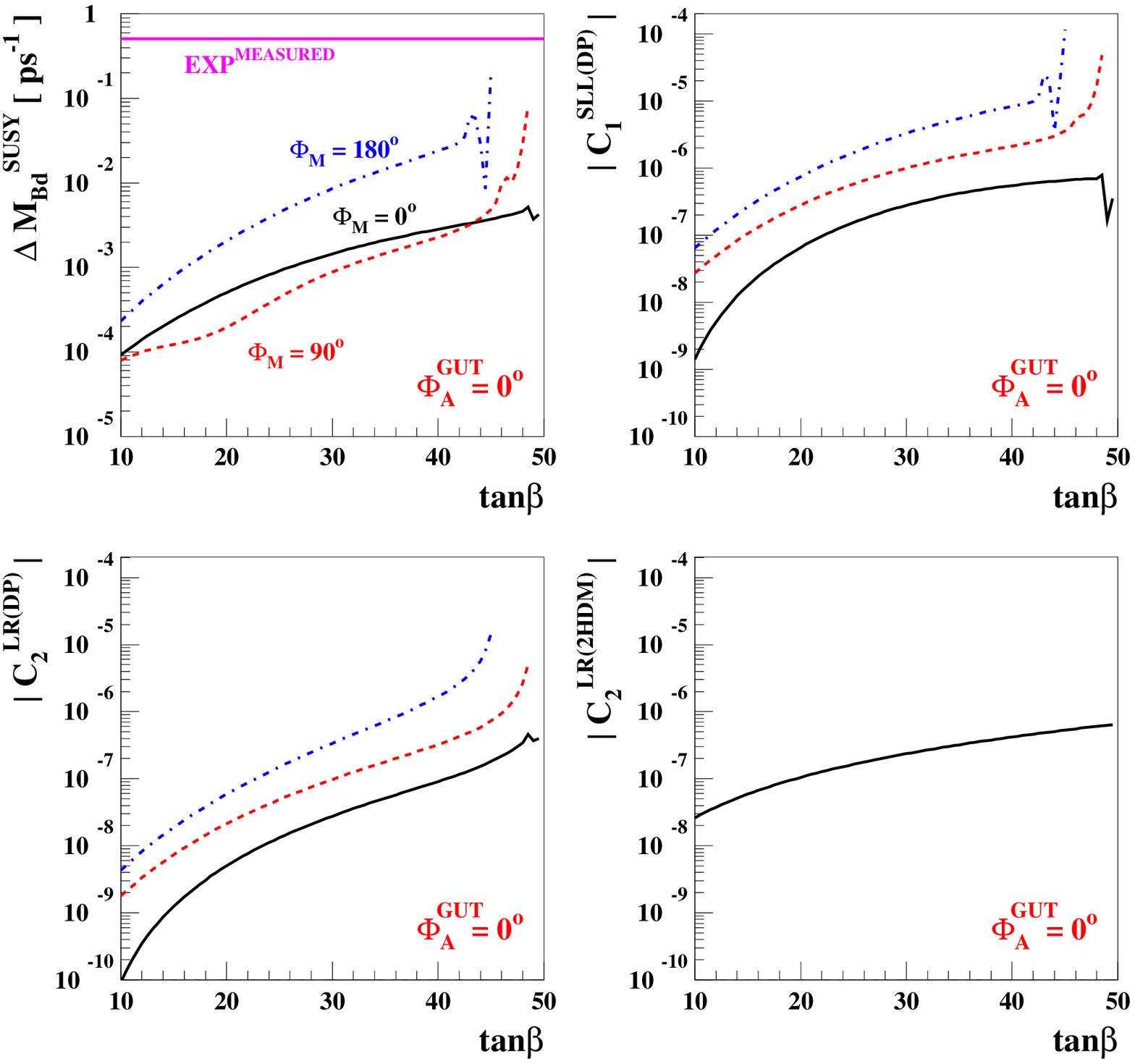,height=16cm,width=16cm}}
\vspace{-0.9cm}
\caption{\it The SUSY contribution to $\Delta M_{B_d}$ in units of
${\rm ps}^{-1}$ (upper-left), and the relevant couplings in the other
three frames. The line conventions and the parameters are the same as
in Fig.~\ref{fig:dmbs}. }
\label{fig:dmbd}
\end{figure}

In the upper-left frame of Fig.~\ref{fig:dmbs}, we show the SUSY
contribution to $\Delta M_{B_s}$ in units of ${\rm ps}^{-1}$ as a
function of $\tan\beta(M_{\rm SUSY})$
for three values of the common phase, namely
$\Phi_M=0^\circ$ (solid line), $90^\circ$ (dashed line), and
$180^\circ$ (dash-dotted line). 
The horizontal line is for the measured
value: $\Delta M_{B_s}^{\rm EXP}=17.77\pm 0.10~({\rm stat.})\pm 0.07~({\rm
syst.})~{\rm ps}^{-1}$~\cite{CDFD0}. We observe that the SUSY
contribution can be larger than the current observed value for
$\Phi_M=180^\circ$ when $\tan\beta$ is large. Indeed, for $\Phi_M =
180^\circ~(90^\circ)$, we find $\tan \beta < 44~(48)$, whereas there is no
restriction on $\tan \beta$ for $\Phi_M = 0^\circ$.

The   SUSY   contribution  $C_1^{\rm   SRR(DP)}$   is  suppressed   by
$m_s^2/m_b^2$    with   respect    to    $C_1^{\rm   SLL(DP)}$    [see
Eq.~(\ref{dpkaon})].   The  $|C_2^{\rm   LR(DP)}|$  is  comparable  to
$|C_1^{\rm   SLL(DP)}|$,  while   the  2HDM   contribution,  $C_2^{\rm
LR(2HDM)}$, becomes less important  as $\tan\beta$ increases.  The dip
of   the  coupling   $|C_1^{\rm   SLL(DP)}|$  for   $\Phi_M=180^\circ$
(upper-right frame)  at $\tan\beta\simeq 45$  is due to the  fact that
the   three  Higgs   bosons   become  degenerate   and  cancel   other
contributions.   Beyond this  point,  $M_{H_1}\sim M_{H_2}$  decreases
rapidly while $M_{H_3}\sim 110$ GeV remains nearly unchanged.

In  the upper-left  frame  of Fig.~\ref{fig:dmbd},  we  show the  SUSY
contribution  to $\Delta  M_{B_d}$ in  units of  ${\rm ps}^{-1}$  as a
function of $\tan\beta(M_{\rm SUSY})$, using the same line conventions
as  in Fig.~\ref{fig:dmbs}. The  horizontal line  is for  the measured
value:   $\Delta  M_{B_d}^{\rm   EXP}=0.507\pm   0.005~{\rm  ps}^{-1}$
~\cite{PDG}. We  observe that the SUSY contribution  is always smaller
than the measured value, although  it does exhibit a strong dependence
on the CP-violating phase  $\Phi_M$.  The dips at $\tan\beta\simeq 45$
$(\Phi_M=180^\circ)$  and   $\tan\beta\simeq  49$  $(\Phi_M=90^\circ)$
arise for the same reason as in the $\Delta M_{B_s}$ case.
The  dominant   contribution  comes  from   $C_1^{\rm  SLL(DP)}$,  and
$C_1^{\rm  SRR(DP)}$ is suppressed  by $m_d^2/m_b^2$.   The 
value of $|C_2^{\rm LR(DP)}|$ is smaller than  that of
$|C_1^{\rm SLL(DP)}|$.  Finally, as before,
the  2HDM contribution $C_2^{\rm  LR(2HDM)}$ becomes  less significant
for large values of $\tan\beta$.

\begin{figure}[ht]
\hspace{ 0.0cm}
\centerline{\epsfig{figure=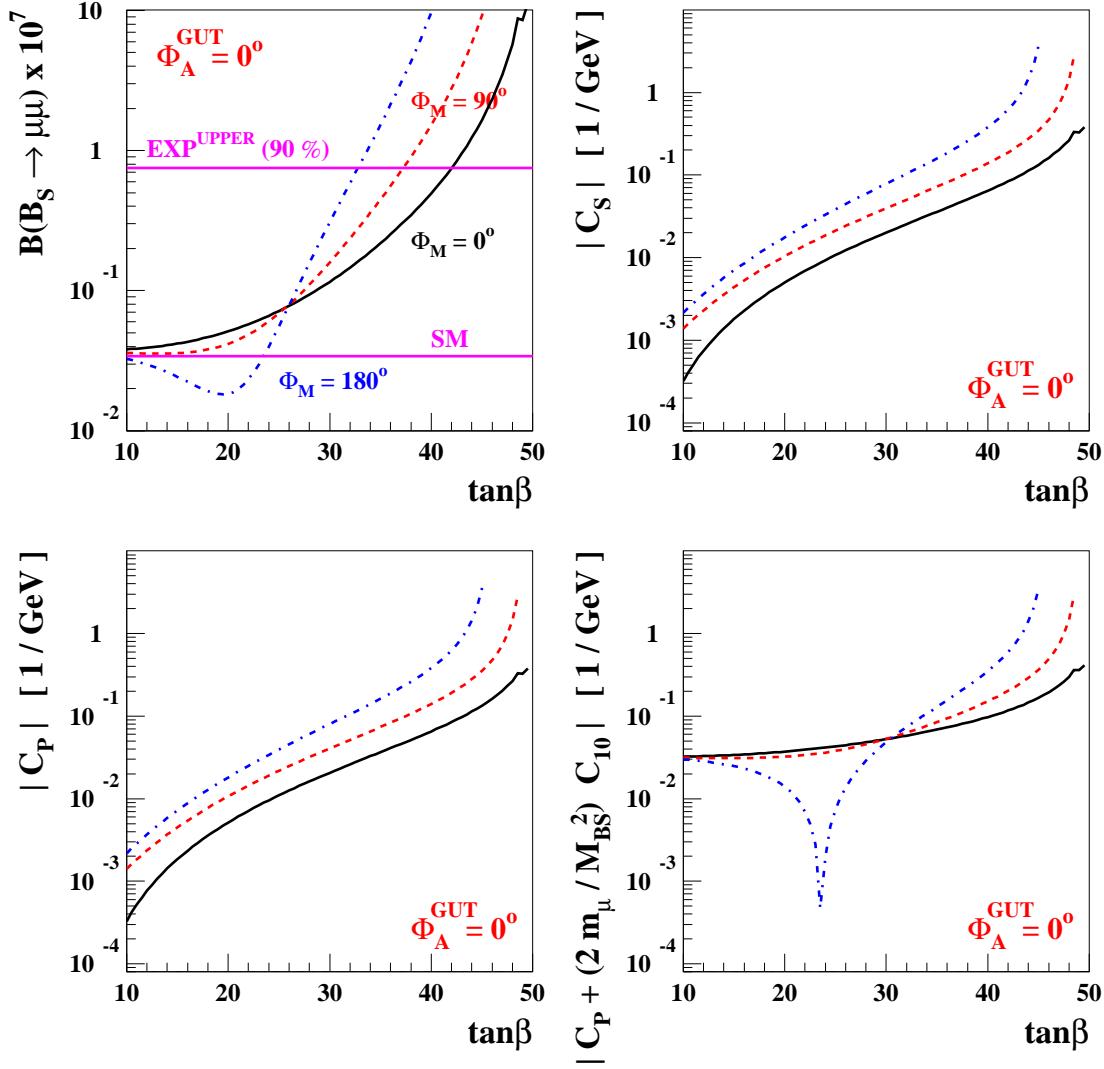,height=16cm,width=16cm}}
\vspace{-0.9cm}
\caption{\it The branching ratio $B(B_s\rightarrow \mu^+\mu^-)$ in the
upper-left frame and the relevant couplings in the other three frames,
in units of ${\rm GeV}^{-1}$ as functions of $\tan\beta(M_{\rm
SUSY})$.  The line conventions and the parameters chosen are the same
as in Fig.~\ref{fig:dmbs}, except that the two horizontal lines in the
upper-left frame are for the SM prediction and the current upper limit
at 90 \% C.L.}
\label{fig:bsmm}
\end{figure}

\subsection{Effects on $B_s\rightarrow \mu^+\mu^-$}
 
In the upper-left frame  of Fig.~\ref{fig:bsmm}, we show the branching
ratio    $B(B_s\rightarrow    \mu^+\mu^-)$    as   a    function    of
$\tan\beta(M_{\rm  SUSY})$  using  the  same line  conventions  as  in
Fig.~\ref{fig:dmbs}  for three  values of  the common  phase $\Phi_M$:
$\Phi_M=0^\circ$   (solid  line),   $90^\circ$   (dashed  line),   and
$180^\circ$  (dash-dotted  line).  The  two  horizontal  lines in  the
upper-left frame are for the SM prediction and the current upper limit
at 90  \% C.L., namely  $7.5 \times 10^{-8}$~\cite{CDFD0}.  We observe
that   the   branching  ratio   changes   substantially  as   $\Phi_M$
varies. Specifically, for  $\Phi_M = 180^\circ~(90^\circ)~0^\circ$, we
find that  the present  upper limit on  $B(B_s\rightarrow \mu^+\mu^-)$
imposes the upper limit $\tan \beta < 34~(38)~42$.

The phase  dependence of  the branching ratio  comes from that  of the
couplings $C_S$  and $C_P$ [see~(\ref{CSCP})], which are  shown in the
upper-right  and the  lower-left frames,  respectively.  We  find that
$|C_S|\simeq  |C_P|$,   since  $O_{1  1}\sim  O_{a  1}   \sim  0$  and
$M_{H_2}\sim M_{H_3}$ [cf.~(\ref{CSCP}) and~(\ref{eq:HIMM})].  We note
that,  for $\Phi_M=180^\circ$,  $B(B_s\rightarrow \mu^+\mu^-)$  can be
smaller  than  the SM  prediction  for  $\tan\beta\lsim  24$. This  is
because  the  Higgs-mediated contribution  $C_P$  cancels  the SM  one
$C_{10}$, as shown in the lower-right frame of Fig.~\ref{fig:bsmm}, in
which the factor $m_b/(m_b+m_s)$ [cf.~(\ref{FSP})] has been suppressed
in the label of the $y$-axis.

\subsection{Effects on $B_u \to \tau \nu$}

\begin{figure}[ht]
\hspace{ 0.0cm}
\centerline{\epsfig{figure=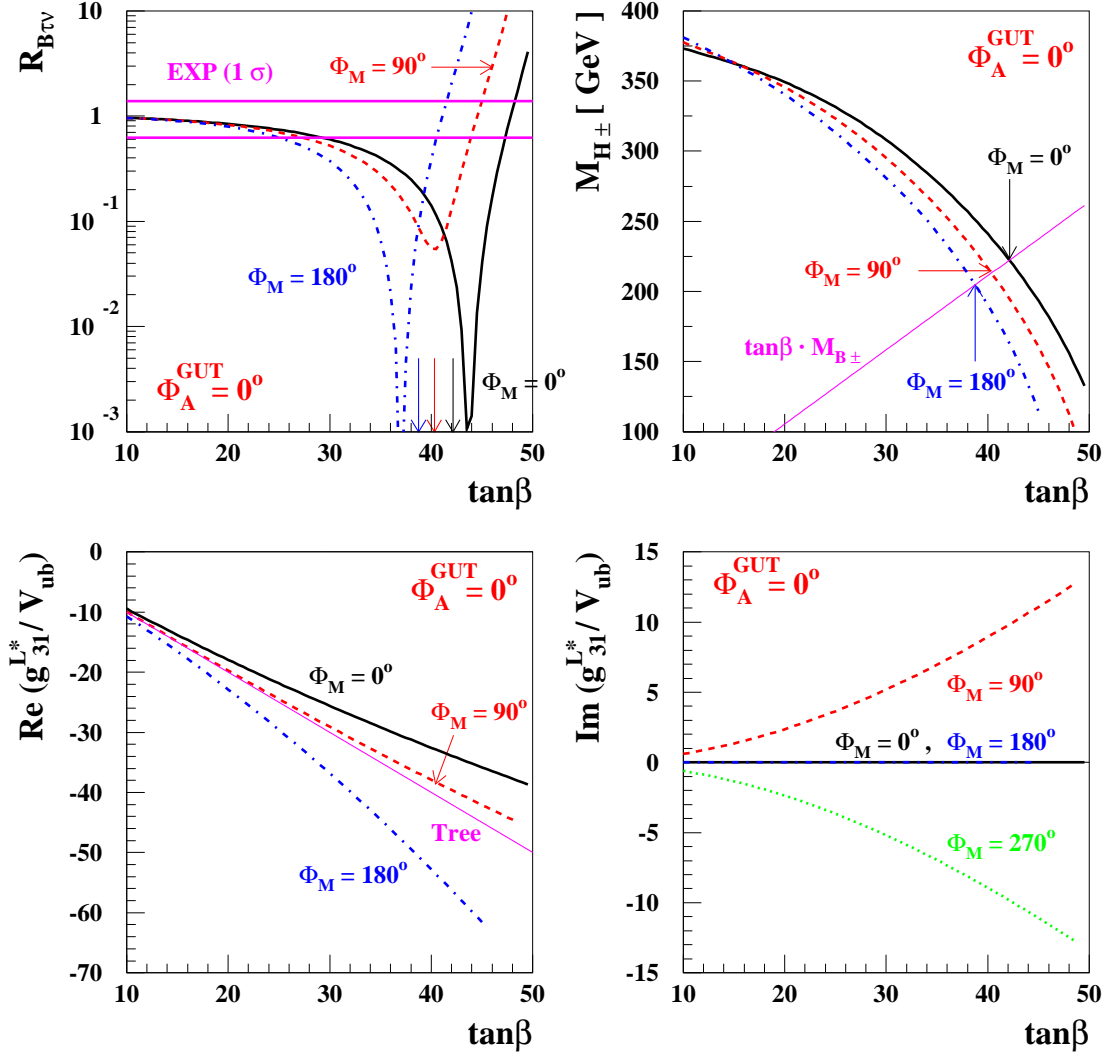,height=16cm,width=16cm}}
\vspace{-0.9cm}
\caption{\it The ratio  $R_{B\tau\nu}$ (upper-left), the charged-Higgs
boson  mass  in  GeV  (upper-right),  and the  real  (lower-left)  and
imaginary    (lower-right)    parts    of    the    coupling    $({\bf
g}^{L\,\dagger}_{H^-\bar{d}u})_{13}/{\bf         V}_{13}=        ({\bf
g}^{L\,*}_{H^-\bar{d}u})_{31}/V_{ub}$ as  functions of $\tan\beta$ for
three or four values of $\Phi_M$, taking $\Phi_A^{\rm GUT}=0^{\rm o}$.
The  experimentally  allowed 1-$\sigma$  region  is  bounded with  two
horizontal lines  in the upper-left  frame.  The straight line  with a
tag `Tree' in the lower-left frame shows the tree-level coupling.  The
parameters are the same as in Fig.~\ref{fig:dmbs}. }
\label{fig:rbtaunu}
\end{figure}

The recent BELLE and BABAR results for 
the branching ratio $B(B^-\to \tau^- \bar{\nu})$
are~\cite{Belle:Btaunu,Babar:Btaunu}
\begin{eqnarray}
B(B^-\to \tau^- \bar{\nu})^{\rm BELLE} &=&
\left(1.79^{+0.56}_{-0.49}~{\rm (stat)}^{+0.46}_{-0.51}~{\rm (syst)}\right)
\times 10^{-4} \,, \\
B(B^-\to \tau^- \bar{\nu})^{\rm BABAR} &=&
\left(1.2\pm 0.4~{\rm (stat)}\,\pm 0.3~{\rm (bkg~syst)}
\,\pm 0.2~{\rm (other~syst)}\right) \times 10^{-4}\,, \nonumber
\end{eqnarray}
which lead to $B(B^-\to \tau^- \bar{\nu})^{\rm EXP}=(1.4\pm 0.43)\times 10^{-4}$.
Combining the BELLE and BABAR results with the SM value
$B(B^-\to \tau^- \bar{\nu})^{\rm SM}=(1.41\pm 0.33)\times 10^{-4}$
obtained by the global fit
without using $B(B^-\to \tau^- \bar{\nu})$ as an input~\cite{utfit},
we have the following $1~\sigma$ range for
the ratio to the SM prediction~\footnote{This range is different from that used in
\cite{Isidori:2007jw} due to the new BABAR result~\cite{Babar:Btaunu}.
}:
\begin{equation}
R_{B\tau\nu}^{\rm EXP}=1.0\pm 0.38.
\label{eq:rbtaunu_exp}
\end{equation}
In the  upper-left frame  of Fig.~\ref{fig:rbtaunu}, we  show possible
values  of this  ratio  in the  MSSM  with MCPMFV,  together with  the
experimental  range given in  (\ref{eq:rbtaunu_exp}), as  functions of
$\tan\beta$  for  three  representative  values of  the  common  phase
$\Phi_M$ and for  $\Phi^{\rm GUT}_A=0$.  The three thin  arrows at the
bottom indicate  the positions  where the ratio  vanishes at  the tree
level without including  threshold corrections for $\Phi_M=180^\circ$,
$90^\circ$, and  $0^\circ$ (from left  to right).  Beyond  the minimum
point, the charged Higgs-boson contribution dominates over the SM one.
It  rapidly grows  as $\tan^4\beta$  initially and  then goes  over to
$\tan^2\beta$ due  to the  threshold corrections.  For  each displayed
value  of $\Phi_M$,  we find  two regions  of $\tan  \beta$  where the
experimental value  of $B(B^-\to \tau^- \bar{\nu})$  is obtained.  One
region   is   at   $\tan   \beta   <   25~(27)~29$   for   $\Phi_M   =
180^\circ~(90^\circ)~0^\circ$, and  corresponds to the  case where the
charged  Higgs-boson contribution is  a small  `correction' to  the SM
term.   The second  region is  at  $\tan \beta  \sim 41~(46)~48$,  for
$\Phi_M =  180^\circ~(90^\circ)~0^\circ$, and corresponds  to the case
where  the  charged Higgs-boson  contribution  dominates  over the  SM
term. We note that the locations of these second allowed regions would
not  be estimated  correctly  if the  threshold  corrections were  not
included.   These  regions  are  actually  excluded by  the  $B_s  \to
\mu^+\mu^-$ constraint discussed previously.

The tree-level vanishing points  are also indicated in the upper-right
frame  as  intersections   of  the  $M_{H^\pm}$  and  $\tan\beta\times
M_{B^\pm}$  lines.  We  observe  that the  resummed threshold  effects
enhance the  charged Higgs-boson contribution  when $\Phi_M=180^\circ$
and  suppress  it when  $\Phi_M=0^\circ$.  As  can  be seen  from  the
lower-left frame of Fig.~\ref{fig:rbtaunu}, for $\Phi_M=90^\circ$, the
$\tan\beta$-dependence of $R_{B\tau\nu}$ becomes rather similar to the
tree-level  one.  However, as  displayed in  the lower-right  frame of
Fig.~\ref{fig:rbtaunu}, there is a non-vanishing contribution from the
imaginary        part       of        the        coupling       $({\bf
g}^{L\,\dagger}_{H^-\bar{d}u})_{13}/{\bf V}_{13}$.

\subsection{Effects on $B \rightarrow X_s \gamma$}

\begin{figure}[ht]
\hspace{ 0.0cm}
\centerline{\epsfig{figure=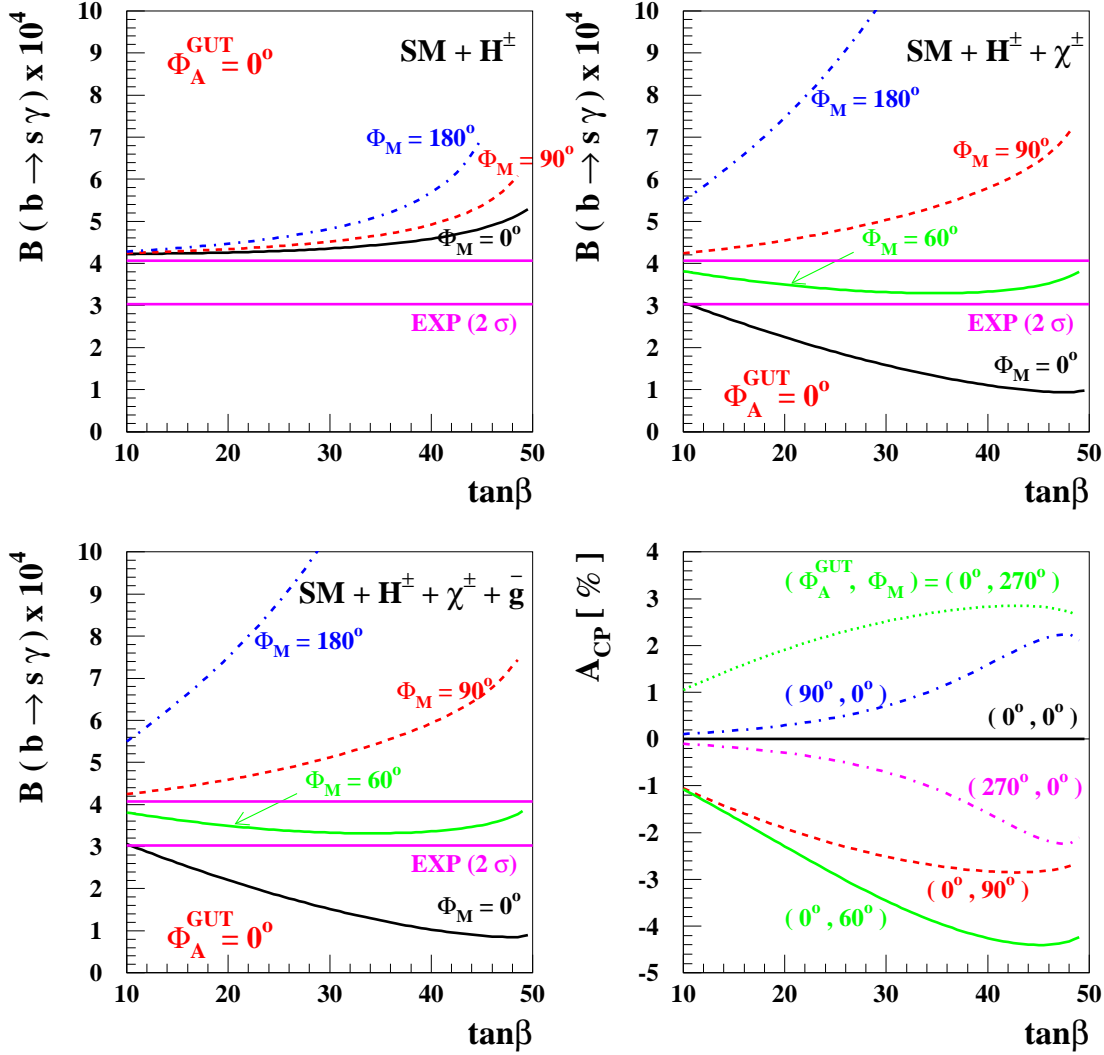,height=16cm,width=16cm}}
\vspace{-0.9cm}
\caption{\it The  branching ratio $B(B  \rightarrow X_s \gamma)$  as a
function  of $\tan\beta$  for  several values  of  the common  phase
$\Phi_M=\Phi_1=\Phi_2=\Phi_3$ and $\Phi_A^{\rm GUT}$.
The region allowed experimentally at the 2-$\sigma$ level is bounded
by  two   horizontal  lines.   In  the  upper-left   frame,  only  the
charged-Higgs  contribution is  added  to the  SM  prediction. In  the
upper-right  and   lower-left  frames,  the   SUSY  contributions  are
included.   The direct  CP  asymmetry ${\cal  A}^{\rm dir}_{\rm  CP}(B
\rightarrow X_s  \gamma)$ is also  shown in the lower-right  frame for
several combinations of $(\Phi_A^{\rm GUT}\,,\Phi_M)$.  The parameters
are the same as in Fig.~\ref{fig:dmbs}. }
\label{fig:bsg}
\end{figure}

\begin{figure}[ht]
\hspace{ 0.0cm}
\centerline{\epsfig{figure=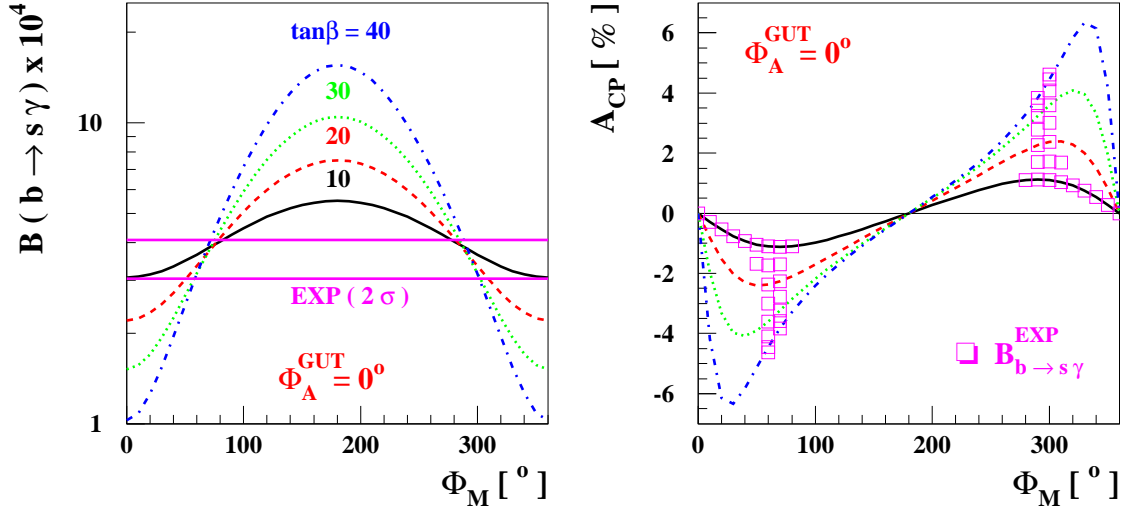,height=16cm,width=16cm}}
\vspace{-7.9cm}
\caption{\it The branching ratio  $B(B \rightarrow X_s \gamma)$ (left)
and the  CP asymmetry ${\cal  A}^{\rm dir}_{\rm CP}(B  \rightarrow X_s
\gamma)$  (right)  as  functions   of  $\Phi_M$  for  four  values  of
$\tan\beta$ taking  $\Phi_A^{\rm GUT}=0^{\rm o}$.   The region allowed
experimentally at the 2-$\sigma$  level is bounded by two horizontal
lines in the  left frame.  In the right  frame, points satisfying this
constraint are denoted  by open squares.  The parameters  are the same
as in Fig.~\ref{fig:dmbs}. }
\label{fig:bsgp}
\end{figure}

The current experimental bound on $B(B \to X_s \gamma)$ with a photon
energy cut of $E_\gamma > E_{\rm cut} = 1.6$ GeV is \cite{HFAG}
\begin{equation}
B(B \to X_s \gamma)^{\rm EXP} = (3.55 \pm 0.24^{+0.09}_{-0.10} \pm
0.03) \times 10^{-4}. 
\end{equation}
Our  estimate of the  SM prediction  based on  the NLO  calculation is
$3.35 \times 10^{-4}$, which is  about 1~$\sigma$ larger than the NNLO
result, $(3.15 \pm 0.23) \times 10^{-4}$ \cite{b2sg-nnlo}.
In Fig.~\ref{fig:bsg} we show the branching ratio $B(B \rightarrow X_s
\gamma)$ and  the direct CP  asymmetry ${\cal A}^{\rm  dir}_{\rm CP}(B
\rightarrow  X_s   \gamma)$  as  functions  of   $\tan\beta$.  In  the
upper-left  frame,  we include  only  the charged-Higgs  contribution,
which increases  the branching ratio.  The larger contribution  in the
high-$\tan\beta$  region  is  due  to  the  decrease  of  the  charged
Higgs-boson mass.
In   the  upper-right   frame  of   Fig.~\ref{fig:bsg},  we   add  the
contribution  from  the  chargino-mediated loops.   This  contribution
largely cancels  the charged-Higgs contribution,  when $\Phi_{M} \lsim
90^{\rm o}$.  Instead, if $\Phi_{M}$ is larger than $\sim 90^{\rm o}$,
the chargino  contribution interferes constructively with  the SM one,
resulting in  a rapid increase  of the branching ratio  as $\tan\beta$
grows.   This behaviour  can  be  understood from  the  fact that  the
dominant contribution to $C_{7,8}^{\chi^\pm}$ comes from the last term
of  Eq.~(\ref{eq:c78_chargino}), which is  proportional to  $\sim {\rm
e}^{i\,\Phi_{A_t}}/c_\beta$, and  the branching ratio  is proportional
to its real part, namely $\cos\Phi_{A_t}/c_\beta$.  We recall that the
phase $\Phi_{A_t}$ at  the low-energy scale can largely  be induced by
non-vanishing $\Phi_M$ even when  $\Phi_A^{\rm GUT}$ vanishes (see the
upper frames of Fig.~\ref{fig:phia_msusy}).
In the lower-left frame of Fig.~\ref{fig:bsg}, we show the full result
including  the contribution  of  the gluino-mediated  loops, which  is
non-vanishing  in the  presence  of flavour  mixing  in the  down-type
squark mass matrix. We find  that it is numerically negligible for the
parameters chosen.  In the same frame,  as well as  in the upper-right
one,  we show the  case of  the common  phase $\Phi_M=60^{\rm  o}$, in
which there  is a nearly  exact cancellation between the  chargino and
charged-Higgs contributions, and all the $\tan\beta$ region considered
is compatible with the current experimental bound. This observation is
also apparent in the left panel of Fig.~\ref{fig:bsgp}.
In the lower-right frame of  Fig.~\ref{fig:bsg}, we show the direct CP
asymmetry for several  combinations of ($\Phi^{\rm GUT}_A \,,\Phi_M)$,
finding   that  it   can  be   as  large   as  $\sim   -4$   \%,  when
$\Phi_M=60^\circ$.

To illustrate the strong dependences of the branching ratio and the CP
asymmetry on the  common phase $\Phi_M$, we show  them as functions of
$\Phi_M$ for  four values of $\tan\beta$  in Fig.~\ref{fig:bsgp}.  The
region allowed  experimentally at the  2-$\sigma$ level is  bounded by
two horizontal  lines in the left  frame.  In the  right frame, points
within this region are denoted with open squares.  We observe that the
branching   ratio   is  quite   insensitive   to  $\tan\beta$   around
$\Phi_M=60^{\rm o}$, whereas the CP  asymmetry can be as large as $\pm
5$ \% for points within  the current 2-$\sigma$ bound on the branching
ratio.  For comparison, we  note that the experimental range currently
allowed  is  $0.4  \pm  3.7$  \%~\cite{HFAG}, implying  that  the  new
contribution  in the  MSSM  with  MCPMFV could  be  comparable to  the
present experimental error, and  much larger than the SM contribution,
which is expected to be below  1\%. Finally, it is important to remark
that, in  the absence  of any cancellation  mechanism~\cite{EDMs}, EDM
constraints severely  restrict the  soft CP-odd phases  in constrained
models  of  low-scale  SUSY,  such  as the  constrained  MSSM.   In  a
forthcoming paper,  however, we will demonstrate in  detail, how these
constraints can be considerably relaxed in the MSSM with MCPMFV.

\setcounter{equation}{0} 
\section{Conclusions}
  \label{sec:concl}

In this paper we have formulated the maximally CP-violating version of
the MSSM with minimal flavour violation, the MSSM with MCPMFV, showing
that it has 19  parameters, including 6 additional CP-violating phases
beyond the CKM  phase in the SM. As preparation  for our discussion of
$B$-meson    observables,    we    have   developed    a    manifestly
flavour-covariant  effective  Lagrangian  formalism, including  a  new
class  of  dominant  subleading  contributions due  to  non-decoupling
effects of  the third-generation quarks. We  have presented analytical
results for a range  of different $B$-meson observables, including the
$B_s$  and $B_d$  mass  differences,  and the  decays  $B_s \to  \mu^+
\mu^-$, $B_u  \to \tau \nu$  and $b \to  s \gamma$. We  have presented
numerical  results  for  these  observables  in  one  specific  MCPMFV
scenario. This serves to demonstrate that the experimental constraints
on  $B$-meson mixings and  their decays  impose constraints,  e.g., on
$\tan \beta$, that  depend strongly on the CP-violating  phases in the
MCPMFV  model, most  notably  on  the soft  gluino-mass  phase in  the
specific example studied.

In summary, on the one hand,  our paper introduces a new class of MSSM
models  of   potential  phenomenological  interest   and  develops  an
appropriate  formalism  for  analyzing  them,  and on  the  other,  it
presents exploratory  numerical studies of the  constraints imposed by
experimental  limits on $B$-meson  observables. In  view of  the large
number of the theoretical parameters in the MSSM with MCPMFV, we leave
for future  work a more  complete exploration of its  parameter space,
including  the   correlation  with  other   experimental  constraints,
e.g.~those imposed by limits on electric dipole moments.

\subsection*{Acknowledgements}
We thank Robert N. Hodgkinson for pointing out typos in Eq.~(\ref{Dhd}).
The work of AP has been supported in part by the STFC research grant:
PP/D000157/1.  The  work of J.S.L. has  been supported in  part by the
Korea  Research Foundation and  the Korean  Federation of  Science and
Technology  Societies Grant  funded by  the Korea  Government (MOEHRD,
Basic Research Promotion Fund).

\newpage

\def\theequation{\Alph{section}.\arabic{equation}}
\begin{appendix}

\setcounter{equation}{0}
\section{Renormalization Group Equations}\label{sec:RGE}

Here  we list  all relevant  one-loop renormalization  group equations
(RGEs) for the  gauge and Yukawa couplings~\cite{GLT}, as  well as for
the    soft   SUSY-breaking   mass    parameters   of    the   general
MSSM~\cite{BBO,CPR}.   Defining the  RG evolution  parameter $t  = \ln
(Q^2/M^2_{\rm  GUT})$,  we  may   write  down  the  one-loop  RGEs  as
follows:\footnote{Our results are in agreement with~\cite{CPR}.}
\begin{eqnarray}
  \label{RGEs}
\frac{d   g_{1,2,3}}{dt}    \!& = &\!   \frac{1}{32\pi^2}\ \Bigg\{\,
\frac{33}{5}\, g^3_1\,,\ g^3_2\,,\ -3g^3_3\,\Bigg\}\; ,\\[3mm]
\frac{d   M_{1,2,3}}{dt}    \!& = &\!   \frac{1}{16\pi^2}\ \Bigg\{\,
\frac{33}{5}\, g^2_1 M_1\,,\ g^2_2 M_2\,,\ -3g^2_3
M_3\,\Bigg\}\; ,\\[3mm]
\frac{d {\bf h}_u}{dt}  \!& = &\! \frac{{\bf h}_u}{32\pi^2}\ \Bigg( -
\frac{13}{15}\, g^2_1\: -\: 3g^2_2\: -\: \frac{16}{3}\,
g^2_3\: +\: 3\,{\bf h}^\dagger_u {\bf h}_u\: +\:
{\bf h}^\dagger_d {\bf h}_d\: +\: 3\,{\rm Tr}\, ({\bf h}^\dagger_u
{\bf h}_u )\, \Bigg)\; ,\\[3mm]
\frac{d {\bf h}_d}{dt}  \!& = &\! \frac{{\bf h}_d}{32\pi^2}\ 
\Bigg( -
\frac{7}{15}\, g^2_1\: -\: 3g^2_2\: -\: \frac{16}{3}\,
g^2_3\: +\: 3\,{\bf h}^\dagger_d {\bf h}_d\: +\:
{\bf h}^\dagger_u {\bf h}_u\: +\: 3\,{\rm Tr}\, ({\bf h}^\dagger_d
{\bf h}_d )\nonumber\\
&& +\: {\rm Tr}\, ({\bf h}^\dagger_e {\bf h}_e)\, \Bigg)\; ,\\[3mm]
\frac{d {\bf h}_e}{dt}  \!& = &\! \frac{{\bf h}_e}{32\pi^2}\ \Bigg(
- \frac{9}{5}\, g^2_1\: -\: 3g^2_2\: +\: 3\,{\bf h}^\dagger_e {\bf h}_e\: +\:
3\,{\rm Tr}\, ({\bf h}^\dagger_d {\bf h}_d )\: +\:
{\rm Tr}\, ({\bf h}^\dagger_e {\bf h}_e)\, \Bigg)\; ,\\[3mm]
\frac{d {\bf a}_u}{dt} \!&=&\! \frac{1}{32\pi^2}\, 
\Bigg[\, \Bigg(\,\frac{26}{15}\, g^2_1 M_1\: +\: 
6 g^2_2 M_2\: +\: \frac{32}{3}\, g^2_3 M_3\,\Bigg)\, {\bf h}_u\:
-\:  \Bigg(\,\frac{13}{15}\, g^2_1\: +\: 
3 g^2_2\: +\: \frac{16}{3}\, g^2_3\,\Bigg)\, {\bf a}_u\nonumber\\
&&+\: 4\,{\bf h}_u {\bf h}^\dagger_u {\bf a}_u\:
+\: 5\,{\bf a}_u {\bf h}^\dagger_u {\bf h}_u\: +\:
6\, {\rm Tr}\, ({\bf h}^\dagger_u {\bf a}_u)\, {\bf h}_u\: +\:
3\, {\rm Tr}\, ({\bf h}^\dagger_u {\bf h}_u)\, {\bf a}_u\: +\:  
2\,{\bf h}_u {\bf h}^\dagger_d {\bf a}_d\nonumber\\
&&+\: {\bf a}_u {\bf h}^\dagger_d {\bf h}_d\, \Bigg]\; ,\\[3mm]
\frac{d {\bf a}_d}{dt} \!&=&\! \frac{1}{32\pi^2}\, 
\Bigg[\, \Bigg(\,\frac{14}{15}\, g^2_1 M_1\: +\: 
6 g^2_2 M_2\: +\: \frac{32}{3}\, g^2_3 M_3\,\Bigg)\, {\bf h}_d\:
-\:  \Bigg(\,\frac{7}{15}\, g^2_1\: +\: 
3 g^2_2\: +\: \frac{16}{3}\, g^2_3\,\Bigg)\, {\bf a}_d\nonumber\\
&&+\: 4\,{\bf h}_d {\bf h}^\dagger_d {\bf a}_d\:
+\: 5\,{\bf a}_d {\bf h}^\dagger_d {\bf h}_d\: +\:
6\, {\rm Tr}\, ({\bf h}^\dagger_d {\bf a}_d)\, {\bf h}_d\: +\:
3\, {\rm Tr}\, ({\bf h}^\dagger_d {\bf h}_d)\, {\bf a}_d\: +\:
2\,{\bf h}_d {\bf h}^\dagger_u {\bf a}_u\nonumber\\
&&+\: {\bf a}_d {\bf h}^\dagger_u {\bf h}_u\:
+\: 2\,{\rm Tr}\,({\bf h}^\dagger_e {\bf a}_e)\,{\bf h}_d\: +\:
{\rm Tr}\, ({\bf h}^\dagger_e {\bf h}_e)\, {\bf a}_d\, 
\Bigg]\; ,\\[3mm]
\frac{d {\bf a}_e}{dt} \!&=&\! \frac{1}{32\pi^2}\, 
\Bigg[\, \Big(\, 6 g^2_1 M_1\: +\: 
6 g^2_2 M_2\,\Big)\, {\bf h}_e\:
-\:  \Big(\, 3 g^2_1\: +\: 3 g^2_2\,\Big)\, {\bf a}_e\nonumber\\
&&+\: 4\,{\bf h}_e {\bf h}^\dagger_e {\bf a}_e\:
+\: 5\,{\bf a}_e {\bf h}^\dagger_e {\bf h}_e\: +\:
2\, {\rm Tr}\, ({\bf h}^\dagger_e {\bf a}_e)\, {\bf h}_e\: +\:
{\rm Tr}\, ({\bf h}^\dagger_e {\bf h}_e)\, {\bf a}_e\nonumber\\
&&+\: 6\,{\rm Tr}\,({\bf h}^\dagger_d {\bf a}_d)\,{\bf h}_e\: +\:
3\,{\rm Tr}\, ({\bf h}^\dagger_d {\bf h}_d)\, {\bf a}_e\, 
\Bigg]\; ,\\[3mm]
\frac{dB}{dt} \!&=&\! \frac{3}{16\pi^2}\, \Bigg(\,
\frac{1}{5}\, g^2_1 M_1\: +\: g^2_2 M_2\: +\: 
{\rm Tr}\, ({\bf h}^\dagger_u {\bf a}_u)\: +\: 
{\rm Tr}\, ({\bf h}^\dagger_d {\bf a}_d)\: +\: \frac{1}{3}\,
{\rm Tr}\, ({\bf h}^\dagger_e {\bf a}_e)\, \Bigg)\;,\\[3mm]
\frac{d\mu}{dt} \!&=&\! \frac{3\,\mu}{32\pi^2}\, \Bigg(
-\,\frac{1}{5}\, g^2_1\: -\: g^2_2\: +\: 
{\rm Tr}\, ({\bf h}^\dagger_u {\bf h}_u)\: +\: 
{\rm Tr}\, ({\bf h}^\dagger_d {\bf h}_d)\: +\: \frac{1}{3}\,
{\rm Tr}\, ({\bf h}^\dagger_e {\bf h}_e)\, \Bigg)\;,\\[3mm]
\frac{d M^2_{H_u}}{dt} \!&=&\! \frac{3}{16\pi^2}\, \Bigg(
-\,\frac{1}{5}\, g^2_1 |M_1|^2\: -\: g^2_2 |M_2|^2\: +\: 
{\rm Tr}\, ({\bf h}_u \widetilde{\bf M}^2_Q {\bf h}^\dagger_u)\: +\:
{\rm Tr}\, ({\bf h}^\dagger_u \widetilde{\bf M}^2_U {\bf h}_u)\nonumber\\
&& +\: M^2_{H_u}\, {\rm Tr}\, ({\bf h}^\dagger_u {\bf h}_u)\:
+\: {\rm Tr}\, ({\bf a}^\dagger_u {\bf a}_u)\: +\: \frac{1}{10}\,
g^2_1\, {\rm Tr}\, (Y {\bf M}^2)\, \Bigg)\; ,\\[3mm]
\frac{d M^2_{H_d}}{dt} \!&=&\! \frac{3}{16\pi^2}\, \Bigg(
-\,\frac{1}{5}\, g^2_1 |M_1|^2\: -\: g^2_2 |M_2|^2\: +\: 
{\rm Tr}\, ({\bf h}_d \widetilde{\bf M}^2_Q {\bf h}^\dagger_d)\: +\:
{\rm Tr}\, ({\bf h}^\dagger_d \widetilde{\bf M}^2_D {\bf h}_d)\nonumber\\
&& +\: M^2_{H_d}\, {\rm Tr}\, ({\bf h}^\dagger_d {\bf h}_d)\:
+\: {\rm Tr}\, ({\bf a}^\dagger_d {\bf a}_d)\: +\: \frac{1}{3}\,
{\rm Tr}\, ({\bf h}_e \widetilde{\bf M}^2_L {\bf h}^\dagger_e)\: 
+\: \frac{1}{3}\, 
{\rm Tr}\, ({\bf h}^\dagger_e \widetilde{\bf M}^2_E {\bf h}_e)
\nonumber\\
&& +\: \frac{1}{3}\, M^2_{H_d}\, {\rm Tr}\, ({\bf h}^\dagger_e {\bf h}_e)\:
+\: \frac{1}{3}\, {\rm Tr}\, ({\bf a}^\dagger_e {\bf a}_e)\: -\:
\frac{1}{10}\, g^2_1\, {\rm Tr}\, (Y {\bf M}^2)\, \Bigg)\;
,\\[3mm]
\frac{d \widetilde{\bf M}^2_Q}{dt} \! & =&\! \frac{1}{16\pi^2}\,
\Bigg[ -\Bigg(\, \frac{1}{15}\, g^2_1 |M_1|^2\: +\: 3 g^2_2 
|M_2|^2\: +\: \frac{16}{3}\, g^2_3 |M_3|^2\, \Bigg)\,{\bf 1}_3\: 
+\: \frac{1}{2}\, {\bf h}^\dagger_u {\bf h}_u
\widetilde{\bf M}^2_Q\nonumber\\
&& +\: \frac{1}{2}\, \widetilde{\bf M}^2_Q {\bf h}^\dagger_u 
{\bf h}_u\: +\: {\bf h}^\dagger_u \widetilde{\bf M}^2_U {\bf h}_u\: +\:
M^2_{H_u} {\bf h}^\dagger_u {\bf h}_u\: +\: {\bf a}^\dagger_u {\bf
  a}_u   \: +\: \frac{1}{2}\, {\bf h}^\dagger_d {\bf h}_d
\widetilde{\bf M}^2_Q\: +\: \frac{1}{2}\,
\widetilde{\bf M}^2_Q {\bf h}^\dagger_d	  {\bf h}_d\nonumber\\
&&+\: {\bf h}^\dagger_d \widetilde{\bf M}^2_D {\bf h}_d\: +\:
M^2_{H_d} {\bf h}^\dagger_d {\bf h}_d\: +\: {\bf a}^\dagger_d {\bf
  a}_d\: +\: \frac{1}{10}\,
g^2_1\, {\rm Tr}\, (Y {\bf M}^2)\, {\bf 1}_3\, \Bigg]\; ,\\[3mm]
\frac{d \widetilde{\bf M}^2_L}{dt} \! & =&\! \frac{1}{16\pi^2}\,
\Bigg[ -\Bigg(\,\frac{3}{5}\, g^2_1 |M_1|^2\: +\:
3\, g^2_2 |M_2|^2\,\Bigg)\, {\bf 1}_3\: 
+\: \frac{1}{2}\, {\bf h}^\dagger_e {\bf h}_e
\widetilde{\bf M}^2_L\: +\: \frac{1}{2}\,
\widetilde{\bf M}^2_L {\bf h}^\dagger_e	{\bf h}_e\nonumber\\
&&+\: {\bf h}^\dagger_e \widetilde{\bf M}^2_E {\bf h}_e\: +\:
M^2_{H_d} {\bf h}^\dagger_e {\bf h}_e\: +\: {\bf a}^\dagger_e {\bf
  a}_e\: -\: 
\frac{3}{10}\, g^2_1\, {\rm Tr}\, (Y {\bf M}^2)\, 
{\bf 1}_3\, \Bigg]\; ,\\[3mm]
\frac{d \widetilde{\bf M}^2_U}{dt} \! & =&\! \frac{1}{16\pi^2}\,
\Bigg[ -\Bigg(\, \frac{16}{15}\, g^2_1 |M_1|^2\: +\:
\frac{16}{3}\, g^2_3 |M_3|^2\, \Bigg)\, {\bf 1}_3\: 
+\: {\bf h}_u {\bf h}^\dagger_u
\widetilde{\bf M}^2_U\: +\: 
\widetilde{\bf M}^2_U {\bf h}_u	{\bf h}^\dagger_u\nonumber\\
&&+\: 2\, {\bf h}_u \widetilde{\bf M}^2_Q {\bf h}^\dagger_u\: +\:
2\, M^2_{H_u} {\bf h}_u {\bf h}^\dagger_u\: +\: 2\, {\bf a}_u {\bf a}^\dagger_u
\: -\: \frac{2}{5}\, g^2_1\, {\rm Tr}\, (Y {\bf M}^2)\, {\bf 1}_3\,
\Bigg]\; ,\\[3mm]
\frac{d \widetilde{\bf M}^2_D}{dt} \! & =&\! \frac{1}{16\pi^2}\,
\Bigg[ -\Bigg(\, \frac{4}{15}\, g^2_1 |M_1|^2\: +\:
\frac{16}{3}\, g^2_3 |M_3|^2\, \Bigg)\,{\bf 1}_3\: 
+\: {\bf h}_d {\bf h}^\dagger_d
\widetilde{\bf M}^2_D\: +\: 
\widetilde{\bf M}^2_D {\bf h}_d	{\bf h}^\dagger_d\nonumber\\
&&+\: 2\, {\bf h}_d \widetilde{\bf M}^2_Q {\bf h}^\dagger_d\: +\:
2\, M^2_{H_d} {\bf h}_d {\bf h}^\dagger_d\: +\: 2\, {\bf a}_d {\bf a}^\dagger_d\: 
+\: \frac{1}{5}\, g^2_1\, {\rm Tr}\, (Y {\bf M}^2)\, {\bf 1}_3\,
\Bigg]\; ,\\[3mm]
\frac{d \widetilde{\bf M}^2_E}{dt} \! & =&\! \frac{1}{16\pi^2}\,
\Bigg( -\, \frac{12}{5}\, g^2_1 |M_1|^2\, {\bf 1}_3\: 
+\: {\bf h}_e {\bf h}^\dagger_e
\widetilde{\bf M}^2_E\: +\: 
\widetilde{\bf M}^2_E {\bf h}_e	{\bf h}^\dagger_e\: +\: 
2\, {\bf h}_e \widetilde{\bf M}^2_L {\bf h}^\dagger_e\nonumber\\
&& +\:
2\, M^2_{H_d} {\bf h}_e {\bf h}^\dagger_e\: +\: 2\, {\bf a}_e {\bf
  a}^\dagger_e\: +\: 
\frac{3}{5}\, g^2_1\, 
{\rm Tr}\, (Y {\bf M}^2)\, {\bf 1}_3\,\Bigg)\; ,
\end{eqnarray}
where $g_1$ is the  GUT-normalized gauge coupling, which is related
to  the  U(1)$_Y$  gauge coupling  $g'$  of  the  SM through  $g_1  =
\sqrt{5/3}\, g'$. In addition, the expression
\begin{equation}
  \label{Mtilde}
{\rm Tr}\, (Y {\bf M}^2)\ =\ M^2_{H_u}\: -\: M^2_{H_d}\:
+\: {\rm Tr}\, \Big( \widetilde{\bf M}^2_Q\: -\: \widetilde{\bf M}^2_L\: 
-\: 2\,\widetilde{\bf M}^2_U\: +\: \widetilde{\bf M}^2_D\: +\: 
\widetilde{\bf M}^2_E \Big)
\end{equation}
is the Fayet--Iliopoulos $D$-term contribution to the one-loop RGEs.  It
can be shown that $d {\rm Tr}\, (Y {\bf M}^2)/dt \propto {\rm Tr}\, (Y
{\bf  M}^2)$,  i.e., the  expression  ${\rm  Tr}\, (Y  {\bf  M}^2)$  is
multiplicatively   renormalizable.   As  usual,   the  GUT   scale  is
determined by the boundary condition: $g_1 (M_{\rm GUT}) = g_2 (M_{\rm
GUT}) =  g_3 (M_{\rm  GUT})$.  We note, finally,  that the  one-loop RGEs
listed above  are invariant under the  unitary flavour transformations
given in~(\ref{RGrot}).

\setcounter{equation}{0}
\section{{\boldmath $Z$}- and {\boldmath $W^\pm$}-Boson 
Ward Identities}\label{sec:WIs}

In the absence of gauge quantum corrections, the $Z$- and $W^\pm$ boson
couplings   to    quarks   obey   the    following   tree-level
WIs~\cite{APRD}:
\begin{eqnarray}
  \label{ZWI}
\frac{q^\mu}{M_Z}\ i\Gamma^{Zff'}_\mu (q,p,p-q)\: +\:
\Gamma^{G^0ff'} (q,p,p-q) &=& \\
&&\hspace{-7cm} \frac{ig_w}{M_Z c_w}\; \bigg[\, 
\Big( T^{f'}_z\, P_L\: -\: 2Q_{f'} s^2_w\Big)\,
\Sigma_{ff'} (p)\ -\ 
\Big( T^f_z\, P_R\: -\: 2Q_f s^2_w\Big)\, \Sigma_{ff'} (p - q)\,
\bigg]\; ,\nonumber\\[3mm]
  \label{WWI}
\frac{q^\mu}{M_W}\ i\Gamma^{W^+ud}_\mu (q,p,p-q)\: +\:
i\Gamma^{G^+ud} (q,p,p-q) &=& \\
&&\hspace{-3cm} -\, \frac{ig_w}{M_W}\; \bigg[\, {\bf V}_{u'd}\, 
\Sigma_{uu'} (p)\, P_L\ -\ 
{\bf V}_{ud'}\, P_R\,\Sigma_{dd'} (p - q)\, \bigg]\; ,\nonumber
\end{eqnarray}
where $c_w = \sqrt{1 - s^2_w}$  is the cosine of the weak mixing angle
and  $T^{u(d)}_z   =  \frac{1}2\,  (-\frac{1}2)$  and   $Q_{u  (d)}  =
\frac{2}3\,  (-\frac{1}3)$ are  the weak  isospin and  electric charge
quantum  numbers  for the  $u$  and  $d$  quarks.  In~(\ref{ZWI})  and
(\ref{WWI}), $\Sigma_{ff'} (p)$ are quark self-energies describing the
fermionic  transition  $f'~\to~f$, with  $f=u,d$  and $f'=u',d'$.   In
addition,   $\Gamma^{Zff'}_\mu   (q,p,p-q)$  and   $\Gamma^{W^+ud}_\mu
(q,p,p-q)$ are  vertex functions that describe the  interaction of the
$Z$- and $W^+$-boson to quarks, respectively.  The momenta $q^\mu$ of
the  gauge bosons are defined as  flowing into  the vertex,  while the
momentum flow of  the quarks follows the fermion  arrow, where $p^\mu$
always denotes the outgoing momentum.

In general,  virtual strong and  electroweak gauge corrections  to the
$Z$-  and  $W^\pm$-boson vertices  usually  distort these  identities,
through terms  that depend on the gauge-fixing parameter, e.g, $\xi$.
One possible framework in which these  identities can be enforced is the
pinch technique~\cite{PT}, leading to  analytic results that
are independent  of $\xi$.  Recently, this approach  has been extended
to  super  Yang-Mills theories~\cite{Brodsky}.   We  ignore the  gauge
quantum corrections  in our phenomenological analysis,  since they are
rather small.

In  the  limit $q^\mu  \to  0$,  the  WIs~(\ref{ZWI}) and  (\ref{WWI})
simplify considerably.  Let us first consider the WI involving the $Z$
boson  and its  associate would-be  Goldstone boson  $G^0$.  Since the
vertex function $\Gamma^{Zff'}_\mu  (q,p,p-q)$ has no IR singularities
in  the limit  $q^\mu \to  0$, the  WI~(\ref{ZWI}) takes  on  the much
simpler form
\begin{equation}
  \label{ZWI0}
\Gamma^{G^0ff'} (0,p,p)\ =\ \frac{ig_w}{M_Z c_w}\; T^f_z\, \bigg[\,
\Sigma_{ff'} (p)\, P_L\ -\ P_R\, \Sigma_{ff'} (p)\, \bigg]\; .
\end{equation}
Decomposing the quark self-energies $\Sigma_{ff'} (p)$ with respect to
their spinorial structure,
\begin{equation}
  \label{Sigmap}
\Sigma_{ff'} (p)\ =\ \Sigma^L_{ff'} (p^2) \not\! p\, P_L\ +\
\Sigma^R (p^2) \not\! p\, P_R\ +\ \Sigma^D_{ff'} (p^2)\, P_L\ +\
\Sigma^{D *}_{f'f} (p^2)\, P_R\; .
\end{equation}
we may rewrite~(\ref{ZWI0}) as follows:
\begin{equation}
  \label{G0WI}
\Gamma^{G^0ff'} (0,p,p)\ =\ \frac{ig_w}{M_W}\; T^f_z\, \bigg[\,
\Sigma^D_{ff'} (p^2)\, P_L\ -\ P_R\, \Sigma^{D *}_{f'f} (p^2)\, \bigg]\; .
\end{equation}
Considering  the  proper normalizations  determined  by the  relations
given  in~(\ref{Deltad}),  it  is   possible  to  make  the  following
identifications in the effective potential limit $p^\mu \to 0$:
\begin{equation}
\Sigma^D_{ff'} (0)\ =\ {\bf U}^{f\dagger}_R\, {\bf h}_f\, \langle\, {\bf
  \Delta}_f\,\rangle\, {\bf U}^f_L\;, \qquad
P_L\,\Gamma^{G^0ff'}(0,0,0)\ =\ -\, \frac{i}{\sqrt{2}}\ 
{\bf U}^{f\dagger}_R\, {\bf h}_f\, {\bf \Delta}^{G^0}_f\, 
{\bf U}^f_L\, P_L\; ,
\end{equation}
where the unitary matrices ${\bf U}^{u,d}_{L,R}$ take care of the weak
to the mass basis transformations  as given in (\ref{Utr}), with ${\bf
U}^u_L  = {\bf  U}^Q_L$ and  ${\bf U}^d_L  = {\bf  U}^Q_L\,  {\bf V}$.
Then, the simplified WI~(\ref{G0WI}) implies that
\begin{equation}
  \label{G0}
{\bf \Delta}^{G^0}_f\ =\ -\; \frac{\sqrt{2}}{v}\ T^f_z \,
 \langle\, {\bf \Delta}_f\,\rangle\; ,
\end{equation}
which    is   the    relation    assumed   in    Section~\ref{sec:EPF}
[cf.~(\ref{DdG})].

We now turn our attention  to the WI involving the $W^+$ boson and
the  associated  would-be  Goldstone  boson  $G^+$.  In  the  effective
potential limit $q^\mu,p^\mu \to 0$, we obtain
\begin{equation}
  \label{WIGud}
i\Gamma^{G^+ud} (0,0,0) \ =\
-\, \frac{ig_w}{\sqrt{2}\, M_W}\; \bigg[\, {\bf V}_{u'd}\, 
\Sigma^D_{uu'} (0)\, P_L\ -\ 
{\bf V}_{ud'}\, \Sigma^{D*}_{d'd} (0)\, P_R\, \bigg]\; .
\end{equation}
Employing the  definitions~(\ref{Deltad}) and taking  the weak-to-mass
basis  rotations  of  the  quark  states into  account,  we  find  the
relations:
\begin{equation}
P_L\,\Gamma^{G^+ud}(0,0,0)\ =\ {\bf U}^{u\dagger}_R\,
{\bf h}_u\, {\bf \Delta}^{G^+}_u\, {\bf U}^d_L\, P_L \; ,\qquad
P_L\,\Gamma^{G^-du}(0,0,0)\ =\  {\bf U}^{d\dagger}_R\,
{\bf h}_d\, {\bf \Delta}^{G^-}_d\, {\bf U}^u_L\, P_L\; .
\end{equation}
{}From the simplified WI~(\ref{WIGud}) and its Hermitean conjugate, we
then derive that
\begin{equation}
  \label{Gplus}
{\bf \Delta}^{G^+}_u \ =\ -\, \frac{\sqrt{2}}{v}\, \big<\, {\bf
  \Delta}_u\,\big>\,,\qquad
{\bf \Delta}^{G^-}_d \ =\ \frac{\sqrt{2}}{v}\, \big<\, {\bf
  \Delta}_d\,\big>\,,
\end{equation}
which is in agreement with~(\ref{DdG}) and the discussion given below.
We note  that the  unitarity of  the radiatively-corrected  CKM matrix
${\bf  V}$ lies  at  the heart  of  deriving the  relations~(\ref{G0})
and~(\ref{Gplus}).

\setcounter{equation}{0}
\section{{\color{Red}CP}{\color{Blue}super}{\color{OliveGreen}H} Interface}
\end{appendix}

To solve the RGEs  given in Appendix~\ref{sec:RGE}, we have considered
the following input parameters:
\begin{itemize}
%
\item The gauge couplings at the scale $M_Z$:
\begin{equation}
\alpha_1(M_Z)=\frac{5}{3}\,\frac{g^{\prime\,2}(M_Z)}{4\pi} \ \ ; \ \
\alpha_2(M_Z)=\frac{g^{2}(M_Z)}{4\pi} \ \ ; \ \
\alpha_3(M_Z)\,,
\end{equation}
where $g(M_Z)=e(M_Z)/s_W$ and $g^\prime(M_Z)=e(M_Z)/c_W$ with
$\alpha_{\rm em}(M_Z)=e^2(M_Z)/4\pi$.

The evolutions  of $\alpha_{1,2}$ from  $M_Z$ to $m_t^{\rm  pole}$ are
determined by \cite{Barger:1992ac}
\begin{eqnarray}
\alpha_1^{-1}(m_t^{\rm pole})&=&
\alpha_1^{-1}(M_Z)^{-1}+\frac{53}{30\pi}\ln(M_Z/m_t^{\rm pole})\,, \nonumber \\
\alpha_2^{-1}(m_t^{\rm pole})&=&
\alpha_2^{-1}(M_Z)-\frac{11}{6\pi}\ln(M_Z/m_t^{\rm pole})\,.
\end{eqnarray}
On the other hand, $\alpha_{3}(m_t^{\rm pole})$ has been obtained by solving
the following equation iteratively
\cite{Marciano:1983pj}
\begin{eqnarray}
\alpha_3^{-1}(m_t^{\rm pole}) &=& \alpha_3^{-1}(M_Z)
-b_0\ln\left(\frac{m_t^{\rm pole}}{M_Z}\right)
-\frac{b_1}{b_0}\ln\left(\frac{\alpha_3(m_t^{\rm pole})}{\alpha_3(M_Z)}\right)
\nonumber \\
&&
-\left(\frac{b_2b_0-b_1^2}{b_0^2}\right)
\left[\alpha_3(m_t^{\rm pole})-\alpha_3(M_Z)\right]+{\cal O}(\alpha_3^2)
\end{eqnarray}
where $b_0=-(11-2\,N_F/3)/2\pi$,
$b_1=-(51-19\,N_F/3)/4\pi^2$, and
$b_2=-(2857-5033\,N_F/9+325\,N_F^2/27)/64\pi^3$ with $N_F=5$.
%
\item
The  masses of the  quarks and  the charged  leptons at  the top-quark
pole-mass scale~$m_t^{\rm pole}$. In particular, the top-quark running
mass  at  $m_t^{\rm pole}$  is  obtained  from: $m_t(m_t^{\rm  pole})=
m_t^{\rm  pole}/\left[1+4\alpha_3(m_t^{\rm  pole})/3\pi\right]$.   The
CKM  matrix  ${\bf V}$  is  assumed  to be  given  at  the same  scale
$m_t^{\rm pole}$.   Then, in general,  the complex $3\times  3$ Yukawa
matrices at $m_t^{\rm pole}$ are given by
\begin{equation}
{\bf h}_{u,e}(m_t^{\rm pole})=\frac{\sqrt{2}}{v}\,
\widehat{\bf M}_{u,e}(m_t^{\rm pole})\,,
\ \ \ \
{\bf h}_{d}(m_t^{\rm pole})=\frac{\sqrt{2}}{v}\,
\widehat{\bf M}_{d}(m_t^{\rm pole})\,{\bf V}^\dagger(m_t^{\rm pole})
\end{equation}
in the flavour basis ${\bf U}^Q_L = {\bf U}^u_R = {\bf U}^d_R = {\bf 1}_3$.
The diagonal quark and charged-lepton mass matrices are given by
\begin{eqnarray}
\widehat{\bf M}_{u}(m_t^{\rm pole})
&=&{\bf diag}\left[m_u(m_t^{\rm pole})\,,m_c(m_t^{\rm pole})\,,m_t(m_t^{\rm
pole})\right]\,,
\nonumber \\
\widehat{\bf M}_{d}(m_t^{\rm pole})
&=&{\bf diag}\left[m_d(m_t^{\rm pole})\,,m_s(m_t^{\rm pole})\,,m_b(m_t^{\rm
pole})\right]\,,
\nonumber \\
\widehat{\bf M}_{e}(m_t^{\rm pole})
&=&{\bf diag}\left[m_e(m_t^{\rm pole})\,,m_\mu(m_t^{\rm
pole})\,,m_\tau(m_t^{\rm pole})\right]\,.
\end{eqnarray}

Given $\alpha_{1, 2, 3}(m_t^{\rm pole})$ and ${\bf h}_{\,u, d,
e}(m_t^{\rm pole})$, the evolution from $m_t^{\rm pole}$ to the scale
$M_{\rm SUSY}$ have been obtained by solving the SM RGEs.
Here the SUSY scale $M_{\rm SUSY}$ has been determined by solving
\begin{equation}
Q^2\Big|_{Q=M_{\rm SUSY}} = {\rm max}[m_{\tilde{t}}^2(Q^2)\,,
  m_{\tilde{b}}^2(Q^2)] 
\end{equation}
iteratively,              where             $m_{\tilde{t}}^2\equiv{\rm
max}(m_{\tilde{Q}_3}^2+m_t^2,m_{\tilde{U}_3}^2+m_t^2)$              and
$m_{\tilde{b}}^2\equiv{\rm
max}(m_{\tilde{Q}_3}^2+m_b^2,m_{\tilde{D}_3}^2+m_b^2)$.             For
$m_{\tilde{Q}_3,\tilde{U}_3,\tilde{D}_3,\tilde{L}_3,\tilde{E}_3}^2
(Q^2)$, we have taken the  $(3,3)$ component of the corresponding mass
matrix as
\begin{equation}
m_{\tilde{Q}_3,\tilde{U}_3,\tilde{D}_3,\tilde{L}_3,\tilde{E}_3}^2 (Q^2)
=\left[\widetilde{\bf M}^2_{Q,U,D,L,E} (Q^2)\right]_{(3,3)}\,.
\end{equation}
At the scale $M_{\rm SUSY}$, the Yukawa matrices match as
\begin{eqnarray}
{\bf       h}_u(M_{\rm        SUSY}^+)       &=&{\bf       h}_u(M_{\rm
SUSY}^-)/\sin\beta(M_{\rm  SUSY})\,,  \nonumber   \\  {\bf  h}_{\,d  ,
e}(M_{\rm      SUSY}^+)&=&      {\bf      h}_{\,     d      ,e}(M_{\rm
SUSY}^-)/\cos\beta(M_{\rm SUSY})\,,
\end{eqnarray}
and, finally, the evolution from $M_{\rm SUSY}$ to $M_{\rm GUT}$
have been obtained by solving the MSSM RGEs.
\item The 19 flavour-singlet mass scales of the MSSM with MCPMFV, which
are parameterized as follows:
\begin{equation}
|M_{1,2,3}|\,{\rm e}^{i\,\Phi_{1,2,3}}\,, \ \ \ \
|A_{u,d,e}|\,{\rm e}^{i\,\Phi_{A_{u,d,e}}}\,, \ \ \ \
\widetilde{M}^2_{Q,U,D,L,E}\,, \ \ \ \
M^2_{H_{u,d}}\,.
\end{equation}
These are inputed at the GUT  scale $M_{\rm GUT}$, which is defined as
the scale  where the  couplings $g_1$ and  $g_2$ meet.  Any difference
between $g_3(M_{\rm GUT})$ and $g_1(M_{\rm GUT})$ may be attributed to
some unknown threshold effect at the GUT scale.
\end{itemize}

By solving the RGEs from the GUT scale $M_{\rm GUT}$ to the SUSY scale
$M_{\rm SUSY}$, we obtain:
\begin{itemize}
\item[]
\begin{itemize}
\item Three complex gaugino masses,
$|M_i|\,{\rm e}^{i\,\Phi_{i}}(Q=M_{\rm SUSY})$.
\item Three $3\times 3$ complex Yukawa coupling matrices,
${\bf h}_{u,d,e} (Q=M_{\rm SUSY})$.
\item Three $3\times 3$ complex ${\bf a}$-term matrices, 
${\bf a}_{u,d,e} (Q=M_{\rm SUSY})$.
\item The soft Higgs masses,  $M^2_{H_u\,,H_d}(Q=M_{\rm SUSY})$.
\item The complex $3\times 3$ sfermion mass matrices,  
$\widetilde{\bf M}^2_{Q,U,D,L,E}(Q=M_{\rm SUSY})$.
\end{itemize}
\end{itemize}

\medskip

The inputs for the code {\tt CPsuperH}  are:
\begin{eqnarray}
&&
\tan\beta(m_t^{\rm pole})\,, \ \ \
M^{\rm pole}_{H^\pm}\,, \ \ \
\mu(M_{\rm SUSY})\,, \ \ \
M_{1,2,3}(M_{\rm SUSY})\,, \ \ \
\nonumber \\
&&
m_{\tilde{Q}_3,\tilde{U}_3,\tilde{D}_3,\tilde{L}_3,\tilde{E}_3}
(M_{\rm SUSY})\,,  
\ \ \
A_t(M_{\rm SUSY})\,, \ \ \
A_b(M_{\rm SUSY})\,, \ \ \
A_\tau(M_{\rm SUSY})\,.
\end{eqnarray}
The ratio of the vacuum expectation values at $m_t^{\rm pole}$ is related to that at
$M_{\rm SUSY}$ by \cite{CEPW}
\begin{equation}
\tan\beta(m_t^{\rm pole})=\frac{\xi^-_2(m_t^{\rm pole})}{\xi^+_1(m_t^{\rm pole})}
\tan\beta(M_{\rm SUSY})
\end{equation}
with
\begin{equation}
\xi^{+(-)}_{1(2)}(m_t^{\rm pole})=1+\frac{3|h_{b(t)}|^2}{32\pi^2}
\ln\frac{M_{\rm SUSY}^2}{m_t^{{\rm pole}\,2}}\,.
\end{equation}
The gaugino mass parameters are  directly read from the results of the
RG running, the sfermion masses are given by
\begin{equation}
m_{\tilde{Q}_3,\tilde{U}_3,\tilde{D}_3,\tilde{L}_3,\tilde{E}_3} (M_{\rm SUSY})
=\left\{\left[
\widetilde{\bf M}^2_{Q,U,D,L,E} (M_{\rm SUSY})
\right]_{(3,3)}\right\}^{1/2}
\end{equation}
and the $A$ parameters, including their CP-violating phases, by
\begin{equation}
A_f(M_{\rm SUSY})=\frac
{\left[{\bf a}_f(M_{\rm SUSY})\right]_{(3,3)}}
{\left[{\bf h}_f(M_{\rm SUSY})\right]_{(3,3)}}\,.
\end{equation}

The $\mu$ parameter and charged Higgs-boson pole mass $M^{\rm pole}_{H^\pm}$ can be
obtained from $M^2_{H_u}(M_{\rm SUSY})$ and $M^2_{H_d}(M_{\rm SUSY})$ by 
imposing the two CP-even tadpole  conditions,
$T_{\phi_1}= T_{\phi_2}=0$ \cite{CEPW}. The tadpoles can be cast into the form
\begin{equation}
T_{\phi_1(\phi_2)}=v_{1(2)}\,\overline\mu_{1(2)}^2+
v_{2(1)}\,{\rm Re}\,\overline{m}_{12}^2+
v_{1(2)}\left[\overline\lambda_{1(2)}\,v_{1(2)}^2+
\frac{1}{2}(\overline\lambda_3+\overline\lambda_4)\,v_{2(1)}^2
\right]+v_{1(2)}X_{1(2)}
\end{equation}
where
\begin{equation}
X_{1(2)}\equiv
\frac{3}{8\pi^2}\, \left[
|h_{b(t)}|^2 m^2_{b(t)} \, \bigg( \ln\frac{m^2_{b(t)}}{m_t^{{\rm pole}\,2}}\ -\
1\,\bigg)\right]\,.
\end{equation}
The quantities $\overline\mu_{1,2}^2$ and $\overline\lambda_i$ are given by
\begin{eqnarray}
\overline\mu_{1,2}^2&=&-M_{H_d,H_u}^2-|\mu|^2 +\mu_{1,2}^{2(1)}(m_t^{\rm pole})\,,
\nonumber \\
\overline\lambda_i &=& \lambda_i + \lambda_i^{(1)}(m_t^{\rm pole}) 
+ \lambda_i^{(2)}(m_t^{\rm pole}) \,,
\end{eqnarray}
where
\begin{eqnarray}
\mu_{1}^{2(1)}(m_t^{\rm pole})&=&-\frac{3}{16\pi^2}\left[
|h_t|^2 |\mu|^2\ln\frac{M_{\tilde{t}}^2}{m_t^{{\rm pole}\,2}}+
|h_b|^2 |A_b|^2\ln\frac{M_{\tilde{b}}^2}{m_t^{{\rm pole}\,2}} \right]\,,
\nonumber \\
\mu_{2}^{2(1)}(m_t^{\rm pole})&=&-\frac{3}{16\pi^2}\left[
|h_t|^2 |A_t|^2\ln\frac{M_{\tilde{t}}^2}{m_t^{{\rm pole}\,2}}+
|h_b|^2 |\mu|^2\ln\frac{M_{\tilde{b}}^2}{m_t^{{\rm pole}\,2}} \right]\,.
\end{eqnarray}
The  couplings   $\lambda_i$,  $\lambda_i^{(1)}(m_t^{\rm  pole})$  and
$\lambda_i^{(2)}(m_t^{\rm  pole})$ may  be found  in Ref.~\cite{CEPW}.
The  squared   absolute  value   $|\mu|^2$  can  be   determined  from
$(T_{\phi_1}/v_2-T_{\phi_2}/v_1)=0$,  which does  not depend  on ${\rm
Re}\,\overline{m}_{12}^2$, since
\begin{equation}
|\mu|^2\ =\ \frac{(M_{H_d}^2-M_{H_u}^2 t_\beta^2)-(\overline\lambda_1
v_1^2-\overline\lambda_2 v_2^2 t_\beta^2)
+X_A
-(X_1-t_\beta^2 X_2)}
{(t_\beta^2-1)
+X_{tb} }
\end{equation}
with
\begin{eqnarray}
X_A&\equiv &
\frac{3}{16\pi^2}\left( |h_b|^2
|A_b|^2\ln\frac{M_{\tilde{b}}^2}{m_t^{{\rm pole}\,2}} - 
t_\beta^2|h_t|^2 |A_t|^2 \ln\frac{M_{\tilde{t}}^2}
{m_t^{{\rm pole}\,2}}\right)\,,
\nonumber \\
X_{tb}&\equiv &
-\frac{3}{16\pi^2}\left(
|h_t|^2 \ln\frac{M_{\tilde{t}}^2}{m_t^{{\rm pole}\,2}}-
t_\beta^2 |h_b|^2 \ln\frac{M_{\tilde{b}}^2}{m_t^{{\rm pole}\,2}}\right)\,.
\end{eqnarray}
We  note that the  phase of  the $\mu$  parameter, $\Phi_\mu$,  is not
renormalized.

Once  $|\mu|^2$  is  found,  ${\rm  Re}\,\overline{m}_{12}^2$  can  be
obtained   from   $T_{\phi_1}=0$   or   $T_{\phi_2}=0$.   With   ${\rm
Re}\,\overline{m}_{12}^2$ known, the charged Higgs-boson pole mass can
be obtained by solving the following equation iteratively:
\begin{equation}
\left(M^{{\rm pole}}_{H^\pm}\right)^2=
\frac{{\rm Re}\,\overline{m}_{12}^2}{s_\beta c_\beta}+
\frac{1}{2}\overline\lambda_4 v^2-
{\rm Re}\,\widehat\Pi_{H^+H^-}(\sqrt{s}=M^{\rm pole}_{H^\pm})\,.
\end{equation}
For  the   explicit  form  of  $\widehat\Pi_{H^+H^-}$,   we  refer  to
Ref.~\cite{HiggsPole}.   We note  that, for  large  $\tan\beta$, ${\rm
Re}\,\overline{m}_{12}^2/s_\beta             c_\beta            \simeq
M_{H_d}^2-M_{H_u}^2-M_Z^2$ at the  tree level. Finally, after imposing
the CP-odd  tadpole condition ${\rm Im}\,(B\mu)=0$,  we use $B\mu={\rm
Re}\,\overline{m}_{12}^2$       to       calculate      the       2HDM
contribution~(\ref{2HDMapprox}),   by  noting  $H_uH_d=-\Phi_1^\dagger
\Phi_2$.

\end{document}